\documentclass{article}

\usepackage{arxiv}
\usepackage[utf8]{inputenc} 
\usepackage[T1]{fontenc}    
\usepackage{hyperref}       
\usepackage{url}            
\usepackage{booktabs}       
\usepackage{amsmath,amssymb,amsfonts}%
\usepackage{amsthm}%
\usepackage{mathrsfs}%
\usepackage{nicefrac}       
\usepackage{microtype}      
\usepackage{graphicx}
\usepackage[numbers]{natbib}
\usepackage{doi}
\usepackage{wrapfig}

\usepackage[font=footnotesize,labelfont=bf]{caption}
\usepackage{subcaption}
\usepackage{colortbl}
\usepackage[capitalise,nameinlink]{cleveref}
\usepackage{rotating}
\usepackage{array}
\usepackage{ragged2e}
\usepackage{tabularray}
\usepackage{makecell}
\usepackage{multirow}%
\usepackage{authblk}

\newcommand{\lb}[1]{\left( #1 \right)}

\title{A Two-stage Bayesian Small Area Estimation Approach for Proportions}

\author{
\href{https://orcid.org/0000-0002-4666-5900}{\includegraphics[scale=0.06]{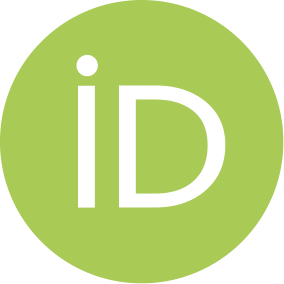}\hspace{1mm} James Hogg$^1$}\thanks{\href{mailto:james.hogg@hdr.qut.edu.au}{james.hogg@hdr.qut.edu.au}} 
\hspace{5mm}%
\href{https://orcid.org/0000-0002-8161-0358}{\includegraphics[scale=0.06]{plots/orcid.png}\hspace{1mm}Jessica Cameron$^{2,4}$} 
\hspace{5mm}%
\href{https://orcid.org/0000-0001-9041-9531}{\includegraphics[scale=0.06]{plots/orcid.png}\hspace{1mm}Susanna Cramb$^3$} 
\hspace{5mm}%
\href{https://orcid.org/0000-0001-8576-8868}{\includegraphics[scale=0.06]{plots/orcid.png}\hspace{1mm}Peter Baade$^{1,4}$} 
\hspace{5mm}%
\href{https://orcid.org/0000-0001-8625-9168}{\includegraphics[scale=0.06]{plots/orcid.png}\hspace{1mm}Kerrie Mengersen$^{1,2}$}}
\date{%
    $^1$Centre for Data Science, Queensland University of Technology\\%
    $^2$School of Mathematical Sciences, Queensland University of Technology\\%
    $^3$Australian Centre for Health Services Innovation, School of Public Health and Social Work, Queensland University of Technology\\%
    $^4$Viertel Cancer Research Centre, Cancer Council Queensland\\[2ex]%
}

\renewcommand{\shorttitle}{TSLN model}

\hypersetup{
colorlinks,
allcolors=blue,
pdftitle={A Two-stage Bayesian Small Area Estimation Method for Proportions},
pdfsubject={},
pdfauthor={Hogg, Cameron, Cramb, Baade, Mengersen},
pdfkeywords={Bayesian statistics, sample surveys, small area estimation, area level model, individual level model},
}

\begin{document}
\maketitle

\begin{abstract}
	With the rise in popularity of digital Atlases to communicate spatial variation, there is an increasing need for robust small-area estimates. However, current small-area estimation methods suffer from various modeling problems when data are very sparse or when estimates are required for areas with very small populations. These issues are particularly heightened when modeling proportions. Additionally, recent work has shown significant benefits in modeling at both the individual and area levels. We propose a two-stage Bayesian hierarchical small area estimation approach for proportions that can: account for survey design; reduce direct estimate instability; and generate prevalence estimates for small areas with no survey data. Using a simulation study we show that, compared with existing Bayesian small area estimation methods, our approach can provide optimal predictive performance (Bayesian mean relative root mean squared error, mean absolute relative bias and coverage) of proportions under a variety of data conditions, including very sparse and unstable data. To assess the model in practice, we compare modeled estimates of current smoking prevalence for 1,630 small areas in Australia using the 2017-2018 National Health Survey data combined with 2016 census data.
\end{abstract}


\newpage
\section{Introduction}\label{sec1}

Although the popularity in using digital health atlases \cite{RN113, RN26} to effectively communicate spatial patterns continues to rise, data for most health outcomes are generally not available for the entire population. Instead, researchers must rely on large surveys to generate estimates for small geographical areas. However, this often results in small sample sizes for each small area. A popular remedy is to employ statistical methods of small area estimation (SAE) which use the concordance between survey and census data to generate estimates of small area characteristics \cite{RN28, pratesi2016, morales2021course}. 

The two common frameworks for SAE are direct and model-based estimators. Direct estimators (e.g. the Horvitz-Thompson estimator \cite{horvitz1952}), that use only sampled individuals, yield estimates with low mean squared error (MSE) when sample sizes are large, but high MSE when sample sizes are small \cite{RN28}. Unlike direct estimators, model-based methods, such as the seminal individual level \cite{RN48} or area level models \cite{RN54}, can borrow statistical strength across areas and thus can provide more efficient estimates \cite{RN82}. 

However, these existing model-based frameworks are increasingly put to the test as the availability of large-scale surveys fails to match the escalating need for high-resolution estimates. New model-based methods must be developed to address the issue of data sparsity, which refers to settings where area level sample sizes are very small. 

Although model-based methods were initially developed for continuous outcomes and are generally less affected by sparsity, proportions of binary or categorical outcomes such as smoking status, are often more useful in health settings \cite{RN461}. Unfortunately, modeling proportions with highly sparse data gives rise to unique methodological challenges that neither individual level nor area level models can address. 

Area level models require access to direct proportion estimates and their sampling variances \cite{RN400, RN408, liu2014}. Under data sparsity, the consistency of these estimates can present a considerable limitation as they can be highly fluctuating, frequently collapsing to zero or one. In this work, we define these areas as \emph{unstable}. As sparsity increases, the likelihood of instability also dramatically increases. In conjunction with sparsity, instability is exacerbated when the characteristic is either very rare or common, or when the sample design is very informative. Furthermore, unstable direct estimates give invalid sampling variances, rendering them inapplicable in standard area level models \cite{RN533}. 

While unstable sampling variances can be imputed using generalized variance functions \cite{RN430}, some solutions to unstable direct estimates include: small conditional or unconditional perturbations prior to modeling \cite{liu2014}, use of zero-or-one inflated models \cite{RN402, RN525}, assuming a distributional form for the strata-specific proportions \cite{RN476, RN497}, dropping them from analysis altogether \cite{RN408} or modeling at a higher administrative level (i.e. using larger areas). Some of these solutions may be considered ad hoc, while others may compromise the resolution of the estimates.

Unlike area level models, individual level logistic models do not suffer from instability issues. Instead, they require covariate data for the entire population of interest \cite{RN404, RN146}. This requirement can be avoided by using only area level covariates \cite{chen2012, liu2017} if the assumption of within-area exchangeability is reasonable. Alternatively, if all the individual level covariates are categorical, multilevel regression with poststratification (MrP) can be used \cite{RN32, RN403, RN404}. However, given MrP's reliance on population counts, it can create issues related to covariate choice and data accessibility \cite{RN467, RN462}. In both cases, to ensure concordance between survey and census covariates, less flexible models must be adopted, and this sacrifices predictive performance. 

Recent research has seen the emergence of two-stage approaches to small area estimation \cite{RN376, gao2023_sma, Das2022}. For instance, Gao and Wakefield \cite{gao2023_sma} introduced a smoothed model-assisted approach, where estimates from an individual level model were input into an area level model. This concept of multi-stage or double smoothing is also explored by Das \emph{et al.} \cite{Das2022}, who modeled data cross-sectionally before using a temporal model. These approaches hint at, but don't fully explore, the potential benefits of two-stage approaches. We suggest that employing both individual and area level models mitigates their individual weaknesses and leverages their strengths.

\subsection{Motivation}

Although two-stage approaches hold broad applicability, the motivation for this study lies in the challenges associated with data sparsity and proportion estimation in the context of Australia. Australia's population density is highly variable, about 80\% of Australians live in east coast cities, and there are huge inland areas that are sparsely populated \cite{RN87}. Unless a survey has been designed specifically for SAE, sample sizes for small areas can be prohibitively small, and remote areas may be excluded altogether \cite{RN113, RN27}. A review of SAE papers revealed some international studies using area level sample sizes in excess of 25 \cite{gao2023_sma}, with most using sample sizes greater than 50 \cite{RN147, RN44}. 

\begin{wrapfigure}{r}{0.3\textwidth}
    \begin{center}
        \includegraphics[trim=35mm 0mm 35mm 0mm, clip]{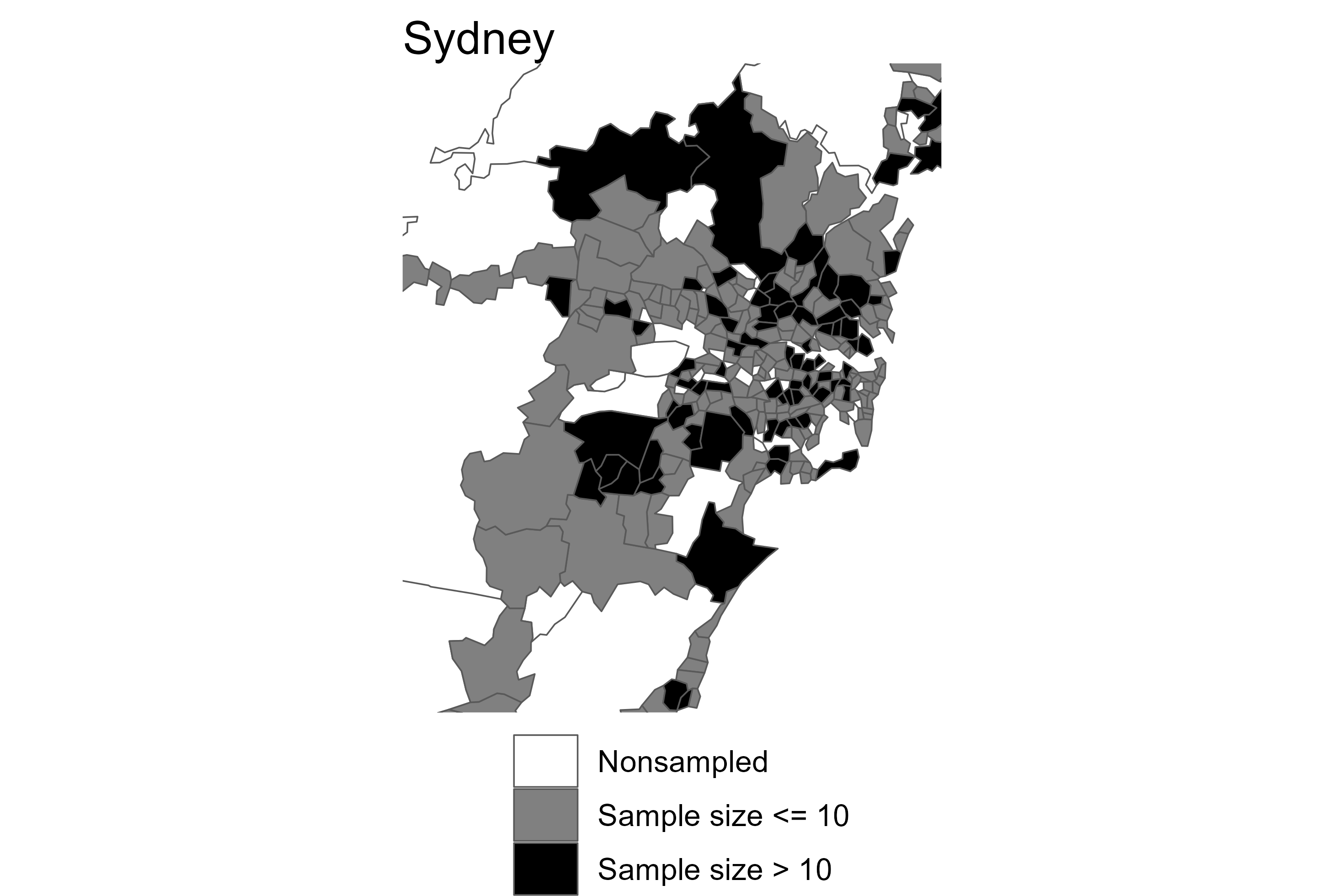}
    \end{center}
    \caption{Map of 282 small areas in and around Sydney on the east coast of Australia. Each area is coloured according to the sample size (greater or less than 10) or sample status of the area (sampled or nonsampled) in the 2017-2018 National Health Survey.}
    \label{fig:ss_map_Sydney}
\end{wrapfigure}

In contrast, the 2017-2018 National Health Survey (NHS), the focus of this work, has area level sample sizes ranging from 5 to 13 (see \cref{fig:ss_map_Sydney}). The consequence of these very small sample sizes (i.e. data sparsity) is that roughly 50\% of the direct estimates, such as those for current smoking, are either missing or unstable, making standard area level models inappropriate. While modeling for larger areas may solve the data sparsity issue, as in the Social Health Atlases of Australia \cite{RN113}, this practice risks concealing critical regional differences due to increased within-area heterogeneity \cite{RN75}. 

While census microdata are not easily accessible in Australia, publicly available aggregated census data are available. However, cross-tabulations of these aggregated data, as required for MrP, are limited to a maximum of four demographic variables (e.g. area, sex, age and education) with no available health-related covariates (such as obesity or chronic illness). The flexibility and predictive capabilities of individual level logistic models are compromised without population level data at a finer granularity. 



The aim of this work is to conceptualize and demonstrate a method for modeling proportions that can provide superior estimates for sparse SAE applications, such as when using the NHS. Our proposed solution, called the two-stage logistic-normal (TSLN) approach, will address the methodological and applied issues described above by using both individual and area level models in a two-stage approach, thus leveraging the strengths of each while mitigating their respective limitations. We adopt a Bayesian perspective and estimate both models using Markov chain Monte Carlo (MCMC) methods \cite{RN576}. Initially, the TSLN approach stabilizes direct estimates through an individual-level logistic model. Subsequently, it inputs the first stage's posterior draws into an area-level hierarchical linear model, thereby propagating uncertainty.

The paper is structured as follows. First, we describe the two Bayesian models of our proposed approach and how these link together. Then we describe alternative two-stage and one-stage approaches for proportions. Next, we describe the simulation study conducted to assess the performance of the TSLN approach compared with four alternative models. Finally, we describe a case study where we generated small area level prevalence estimates of current smoking on the east coast of Australia using the 2017-18 NHS. 

\section{Proposed method} \label{sec:tsln}

\subsection{Notation} \label{sec:notation}

We define a finite population where each individual resides in one of $M$ small geographical areas. Allow $N_i$ individuals in each area $i = 1, \dots, M$, such that $\sum_{i=1}^M N_i = N$. We are interested in a binary characteristic $y_{ij} \in \{0,1\}$ which is equal to $1$ if individual $j$ in area $i$ has the characteristic and $0$ otherwise. We wish to generate estimates, $\hat{\boldsymbol{\mu}} = \left( \hat{\mu}_1, \dots, \hat{\mu}_M \right)$, of the true proportion of the population with the characteristic, $\boldsymbol{\mu} = \left( \mu_1, \dots, \mu_M \right)$.  

Without loss of generality, assume that we have samples from the first $m$ areas, $m<M$, and no samples for the remaining $M - m$ areas. Denote the $n_i$ sampled individuals in area $i$ by $j \in r_i$ and the $N_i - n_i$ nonsampled individuals by $j \in r_i^C$ where $r_i^C$ is the complement set of $r_i$. 

Generally, the survey data used in applications of SAE are collected according to a specified survey design. Similar to Vandendijck \emph{et al.} \cite{RN147}, we assume that in secondary analysis we do not have sufficient details or data to create weights and instead rely on the sampling weights, $w_{ij}^{\text{raw}}$, provided by the data custodians. Using these weights, design-unbiased population estimates can be derived \cite{RN28}. 

Throughout this work, $\hat{\mu}_i^D, \hat{\mu}_i, \mu_i$ will denote a direct estimate (using the Hajek \cite{RN571}), a model-based estimate and the unknown true small area proportion, respectively. Hence,  

\begin{equation}
    \hat{\mu}_i^D = \frac{\sum_{j \in r_i} w_{ij} y_{ij}}{n_i} . \label{eq:hajek}
\end{equation}

By assuming $w_{ij} = w_{ij}^{\text{raw}} n_i \lb{\sum_{j \in r_i} w_{ij}^{\text{raw}}}^{-1}$ and that all the sampling fractions, $n_i/N_i$, are sufficiently small, we use the following approximation to the sampling variance of the direct proportion estimator \cite{RN147, RN552},

\begin{eqnarray}
        \psi_i^D = \widehat{\text{v}}\left( \hat{\mu}_i^D \right)  = \frac{1}{n_i} \left( 1 - \frac{n_i}{N_i} \right) \left( \frac{1}{n_i - 1} \right) \nonumber
        \\
        \sum_{j \in r_i} \left( w_{ij}^2 \left( y_{ij} - \hat{\mu}_i^D  \right)^2 \right). \label{eq:ht_var}
\end{eqnarray}

Direct estimators have low variance and are design-unbiased for $\mu_i$ when $n_i$ is large, but high variance when $n_i$ is small \cite{RN37}. Although, model-based methods can be biased, they reduce variance, resulting in lower MSEs \cite{RN10}.

\subsection{TSLN approach}

The proposed two-stage logistic-normal (TSLN) approach involves two models/stages. In the first stage, an individual level logistic model is fitted to the survey data, where individual level predictions are used to generate area level estimates \cite{RN534}. In the second stage, these area estimates serve as data in an area level Fay and Herriot \cite{RN54} (FH) model to further smooth and impute estimates for nonsampled areas \cite{gao2023_sma}. 

For the sake of clarity, we describe and investigate our approach in its most generalized form, encompassing both individual and area level models of a similarly generic nature. It is worth noting that more sophisticated modeling techniques --- such as non-linear regression or spatial smoothing \cite{RN458, RN394} --- are not only feasible but also advisable. Moreover, benchmarking approaches \cite{RN15, RN131} could be utilized after fitting the stage 2 model or incorporated directly into the stage 2 model using fully Bayesian benchmarking \cite{RN30}. We employ some of these sophisticated modeling techniques in \cref{sec:case_study}.

\subsubsection{Stage 1} \label{sec:stage1_model}

The goal of the stage 1 model is to stabilize the area level direct estimates and sampling variances whilst simultaneously reducing their bias and accommodating the sample design. This is achieved by fitting a Bayesian logistic mixed model to the individual level binary outcomes, $y_{ij}$. 

For design-consistent estimation of the model parameters we use pseudo-likelihood, which can be described as the Horvitz-Thompson estimator of the population-level log-likelihood \cite{RN500, RN501} (see Supplemental Materials B \cite{self_cite}). Following the notation of Parker \emph{et al.} \cite{RN44,Parker2023}, we represent the pseudo-likelihood for a probability density function, $p(.)$, as $p(y_{ij})^{\tilde{w}_{ij}}$, where $\tilde{w}_{ij}$ are the sample scaled weights, $\tilde{w}_{ij} = w_{ij}^{\text{raw}} n \lb{\sum_{i = 1}^m \sum_{j = 1}^{n_i} w_{ij}^{\text{raw}}}^{-1}$. The rescaling of the weights is recommended by Savitsky and Toth \cite{RN501} in their work on Bayesian pseudo-likelihoods.   

The stage 1 model, similar to Model 1 in Parker \emph{et al.} \cite{RN44}, is, 

\begin{eqnarray}
    y_{ij} &\sim& \text{Bernoulli}(p_{ij})^{\tilde{w}_{ij}} \label{eq:tsln_F_s1_log} 
    \\
    \text{logit}(p_{ij}) &=& \mathbf{x}_{ij} \boldsymbol{\beta} + e_i \nonumber 
    \\
    e_i &\sim& N(0, \sigma_e^2) \nonumber
\end{eqnarray}

\noindent where $j = 1, \dots, n_i; i = 1, \dots, m$. In \eqref{eq:tsln_F_s1_log}, $\mathbf{x}$ is the design matrix, which includes the fixed effects (individual level survey-only and area level fixed effects), and $\boldsymbol{\beta}$ the corresponding regression coefficients. 

Noninformative and diffuse priors (e.g. $\boldsymbol{\beta} \propto 1$ and $\sigma_e \propto 1$) are common choices in Bayesian SAE \cite{RN28, RN150}. Recently however, weakly informative priors, which consider the scale of the data, are preferred to these flat priors \cite{RN524}. We used $N(0,2)$ priors for the fixed regression coefficients $\boldsymbol{\beta}$ and a student-$t(df = 3, \mu = 0, \sigma = 1)$ prior for the intercept (implied in $\mathbf{x}_{ij} \boldsymbol{\beta}$). A weakly informative $N(0,1)^{+}$ prior was used for $\sigma_e$.  

The stage 1 model will be referred to as the TSLN-S1 model hereafter.

\subsubsection{Stage 1 (S1) estimates} \label{sec:stage1_est}

To derive area level estimates suitable for area level modeling, we collapse the individual level data to area level data by calculating S1 proportion estimates, $\hat{\mu}_i^{\text{S1}}$, and their corresponding sampling variances, $\psi_i^{\text{S1}}$. These values are derived using the posterior distributions for the individual level probabilities, $p_{ij}$ instead of the observed binary outcomes, $y_{ij}$. 

Smoothing the individual level data in this way, which is similar to the work by Gao and Wakefield \cite{RN536}, permits one to view the TSLN-S1 model as a model-based smoothing method. However, unlike their work \cite{gao2023_sma}, the S1 estimates are aggregates of the sampled individuals only. This highlights the distinction between our proposed approach and the model-assisted framework \cite{RN457}. Note that using S1 estimates guarantees that the values are stable: all $\hat{\mu}_i^{\text{S1}} \in (0,1)$ and $\psi_i^{\text{S1}}>0$. 

To accommodate the uncertainty of the TSLN-S1 model, we approximate the posterior distribution for the S1 estimates. To do so, all the steps to aggregate the individual level predictions are done for each posterior draw, which are indexed by $t = 1, \dots, T$.

The S1 estimate for area $i$ and posterior draw $t$ is,  

\begin{eqnarray}
    \hat{\mu}^{\text{S1},t}_i & = & \frac{\sum_{j \in r_i} w_{ij} p_{ij}^t}{n_i} \label{eq:tsln_mu_S1}
    \\
    \psi_i^{\text{S1},t} & = & \widehat{\text{v}}\left( \hat{\mu}^{\text{S1},t}_i \right) = \widehat{\text{v}}\left( \hat{\mu}^D_i \right) + \widehat{\text{v}}\left( \hat{B}_i^t \right) \label{eq:tsln_psi_S1}
    \\
    \hat{B}_i^t & = & \frac{\sum_{j \in r_i} w_{ij} \left( p_{ij}^t - y_{ij} \right)}{n_i} \label{eq:tsln_bias}
\end{eqnarray}

\noindent where $\hat{B}_i^t$ quantifies the level of smoothing achieved by using the TSLN-S1 model (See Supplemental Materials A \cite{self_cite} for details). The formula used to calculate both $\widehat{\text{v}} \lb{.}$ terms is given in \eqref{eq:ht_var}. 

To accommodate the constraints on the S1 estimates in the stage 2 model, we use the common empirical logistic transformation \cite{RN476, RN552}, 

\begin{eqnarray}
    \hat{\theta}_i^{\text{S1},t} & = & \text{logit}\left( \hat{\mu}_i^{\text{S1},t} \right) \label{eq:tsln_thetaD}
    \\
    \gamma_i^{\text{S1},t} & = & \psi_i^{\text{S1},t} \left[ \hat{\mu}_i^{\text{S1},t} \left( 1 - \hat{\mu}_i^{\text{S1},t} \right) \right]^{-2} \label{eq:tsln_gammaD}
\end{eqnarray}

\noindent where $\hat{\theta}_i^{\text{S1},t}$ is the log-odds of the S1 estimate and $\gamma_i^{\text{S1},t}$ the corresponding sampling variance for area $i$ and posterior draw $t$. The logistic transformation permits the use of a Gaussian likelihood in the stage 2 model, which improves computation and avoids the limitations of the Beta distribution (see Supplemental Materials B \cite{self_cite}).

\subsubsection{Assessing smoothing properties} \label{sec:smooth}

Whilst multi-stage or double smoothing of survey data has been shown to provide optimal predictive performance \cite{gao2023_sma, Das2022}, the degree of smoothing should be monitored. Predictions from the TSLN-S1 model should provide acceptable approximations to the observed values, whilst still ensuring a level of generalisability. Severely overfit TSLN-S1 models will yield minimal smoothing and stability gains, while poor-fitting models can exhibit very high sampling variances, biased S1 estimates and inaccurate estimates for nonsampled areas. Therefore, the stability gained from smoothing the individual level data must be balanced by ensuring adequate agreement (and comparable variability) of the S1 and direct estimates. 

To assess the level of agreement, we regress the posterior median of $\lb{ \hat{\theta}_i^{\text{S1},t}, \dots, \hat{\theta}_i^{\text{S1},T} }$ on $\hat{\theta}_i^D$ with weights $\frac{1}{\psi_i^D}$. The estimated slope from this regression is defined as the area linear comparison (ALC), with values closer to 1 indicating less smoothing and more agreement between the S1 and direct estimates.  

We are yet to provide a theoretical threshold for when the ALC measure may indicate over-smoothing. However, the simulation study described in Supplemental Materials E suggested that optimal performance could be achieved approximately when $0.55 < ALC < 0.75$, with performance atrophy outside these bounds. We found that values at the lower end of these bounds generally provided narrower and more reliable uncertainty intervals.  

\subsubsection{Stage 2} \label{sec:stage2_model}

To accommodate the posterior draws from the stage 1 model, we derive the following summaries of the S1 estimates. Let $\bar{\gamma}_i^{\text{S1}}$ be the empirical mean of $\lb{ \gamma_i^{\text{S1},1:T} }$, $\widehat{\text{v}}\lb{\hat{\theta}_i^{\text{S1}}}$ be the empirical variance of $\lb{ \hat{\theta}_i^{\text{S1},1:T} }$ and $\hat{\boldsymbol{\theta}}_i^{\text{S1}}$ be a random subset from $\lb{ \hat{\theta}_i^{\text{S1},1:T} }$ of size $\widetilde{T}$. We use a random subset to reduce the computational burden, whilst simultaneously accommodating the uncertainty of the stage 1 model. In this work we use $\widetilde{T} = T/2$ (i.e. 50\% of the posterior draws). A more detailed description and some alternative approaches to accommodate the uncertainty of the stage 1 model are given in Supplemental Materials A \cite{self_cite}.

The stage 2 model, referred to as the TSLN-S2 model hereafter, is composed of: a measurement error model \cite{RN549}, 

\begin{equation}
    \hat{\boldsymbol{\theta}}_i^{\text{S1}} \sim N\lb{ \hat{\bar{\theta}}_i, \widehat{\text{v}}\lb{\hat{\theta}_i^{\text{S1}}} }, \label{eq:tsln_F_s2_me}
\end{equation}

\noindent which accommodates some of the uncertainty of the stage 1 model; a sampling model,  

\begin{equation}
    \hat{\bar{\theta}}_i \sim N\lb{ \hat{\theta}_i, \bar{\gamma}_i^{\text{S1}} }, \label{eq:tsln_F_s2_sampling}
\end{equation}

\noindent which accommodates the sampling variance; and a linking model, 

\begin{eqnarray}
    \hat{\theta}_i & = & \mathbf{Z}_{i} \boldsymbol{\lambda} + v_i \label{eq:tsln_F_s2_linking}
    \\
    v_i & \sim & N(0, \sigma_v^2), \nonumber
\end{eqnarray}

\noindent which relates the modeled values, $\hat{\theta}_i$, to a series of covariates and area level random effects $\mathbf{v} = \left( v_1, \dots, v_M \right)$ for $i=1, \dots , M$. We used $N(0,2)$ priors for the fixed regression coefficients $\boldsymbol{\lambda}$ and a student-$t(df = 3, \mu = 0, \sigma = 1)$ prior for the intercept (implied in $\mathbf{Z}_{i} \boldsymbol{\lambda}$). A weakly informative $N(0,2)^{+}$ prior was used for $\sigma_v$.  

Note that $\mathbf{Z}$ is the design matrix, which should include area level covariates that are available for all $M$ areas, and $\boldsymbol{\lambda}$ is the corresponding regression coefficients. Although S1 estimates are not available for areas $m+1,\dots,M$, Tzavidis \emph{et al.} \cite{RN14} show how estimates can be obtained by combining the areas' known covariate values (i.e. $\mathbf{Z}$) and random draws from the priors. To ensure that posterior uncertainty remains unaffected by the choice of $\widetilde{T}$, we downscale the likelihood contribution in \eqref{eq:tsln_F_s2_me} by $1/\widetilde{T}$.

The parameter of interest (i.e. proportion) is the posterior distribution of $\hat{\mu}_i = \text{logit}^{-1}( \hat{\theta}_i)$, which can be described by deriving summary quantities (such as means, variances and highest density intervals) of the posterior draws of $\hat{\mu}_i$ \cite{RN36, RN37, RN28}. 


\subsubsection{Generalized variance functions (GVF)} \label{sec:gvf}

FH models require valid direct estimates and sampling variances for \emph{all} sampled areas. As discussed previously, for very sparse survey data, many sampled areas may give unstable estimates ($\hat{\mu}_i^D = 0$ or $1$) and thus sampling variances that are highly fluctuating, undefined or exactly zero. The reader is reminded that an area is defined as unstable according to its direct estimate only.

The stage 1 model provides more consistent area level estimates and sampling variances, reducing the need for pre-smoothing, a standard practice in area level modeling \cite{RN28, RN137}. That said, S1 sampling variances do exhibit undesirable limit properties. This is because they are derived from probabilities rather than binary random variables. For example, consider a single area $i$ with a sample size of 4 and assume that $w_{ij}=1, \forall j$. While it is possible for $\hat{\mu}^{\text{S1}}_i = 0.01$, to obtain $\hat{\mu}^D_i = 0.01$ one would require 25 times the sample size. In other words, for a fixed sample size, S1 estimates are able to be much closer to zero (or one), without being exactly zero (or one), than direct estimates. The result of this phenomenon is unrealistically low S1 sampling variances for unstable areas. 

Our solution is to use generalized variance functions (GVF), which relate the sampling variance of a direct estimate to a set of covariates, often including the sample sizes and direct estimates, with necessary transformations \cite{RN430, RN28, RN137}. To ensure that S1 sampling variances for unstable areas do not inadvertently affect the fit of the FH model in \eqref{eq:tsln_F_s2_sampling}, we impute the $m-m_s$ S1 sampling variances using GVF, where $m_s < m < M$ is the number of stable areas. 

In this work, we generalize the approach used by Das \emph{et al.} \cite{Das2022} to a fully Bayesian framework by implementing the GVF within our model. By fitting the GVF and stage 2 model jointly \cite{RN536} we ensure the uncertainty of the imputations is appropriately taken into account. 

Unlike standard practice where GVFs are applied to all direct sampling variances \cite{RN28, RN137}, our approach only requires this for unstable areas. We use the stable S1 sampling variances, $\bar{\gamma}_i^{\text{S1}}$, to estimate the parameters of the GVF and then use the fitted GVF to impute the S1 sampling variances for the unstable areas only. Note that S1 sampling variances were not imputed for non-sampled areas. For more details on the GVF used in this work, see Supplemental Materials A \cite{self_cite}. 

\section{Existing methods} \label{sec:existing_methods}

\begin{sidewaystable*}
    \centering
    \resizebox{\columnwidth}{!}{\begin{tabular}{m{2cm}m{9.5cm}lm{1.5cm}} 
    Name & Model & Limitations & Reference \\ 
    \hline
    Logistic model (LOG) & \begin{eqnarray*}
         y_{ij} & \sim & \text{Bernoulli}(p_{ij})^{\tilde{w}_{ij}} \quad\quad j =1, \dots, n_i; i = 1, 
        \dots, m
        \\
        \text{logit}(p_{ij}) & = & \mathbf{x}_{ij} \boldsymbol{\beta} + e_i \quad\quad j =1, \dots, N_i; i = 1, 
        \dots, M
        \\
        \hat{\mu}_i & = & \frac{1}{N_i} \sum_{j = 1}^{N_i} p_{ij} \quad\quad i=1, \dots, M
    \end{eqnarray*} & Requires access to covariate data for the entire population & \RaggedRight \cite{RN501, RN44} \\ 
    \hline
    Binomial model (BIN) & \begin{eqnarray*}
        \sum_{j \in r_i} y_{ij} & \sim &  \text{Binomial}(n_i, \hat{\mu}_i) \quad\quad i=1, \dots, m
        \\
        \text{logit}(\hat{\mu}_i) & = & \mathbf{Z}_{i} \boldsymbol{\lambda} + v_i \quad\quad i=1, \dots, M
    \end{eqnarray*} & Does not accommodate sample design & \cite{RN461, RN460} \\ 
    \hline
    Beta model (BETA) & \begin{eqnarray*}
        \hat{\mu}^D_i & \sim & \text{Beta} \lb{ \hat{\mu}_i \phi_i, \phi_i - \hat{\mu}_i \phi_i } \quad\quad i=1, \dots, m
        \\
        \phi_i & = & \frac{\hat{\mu}_i (1- \hat{\mu}_i)}{\psi^D_i} - 1 \quad\quad i=1, \dots, m
        \\
        \text{logit}(\hat{\mu}_i) & = & \mathbf{Z}_{i} \boldsymbol{\lambda} + v_i \quad\quad i=1, \dots, M
    \end{eqnarray*} & \begin{tabular}[c]{@{}l@{}}
    Requires $\hat{\mu}_i^D \in (0,1)$ and $\psi^D_i \in (0,0.25)$, $\forall i$\\
    Possible bimodal posterior\\
    Requires GVFs for unstable \emph{and} missing areas\\
    Areas with very large $\psi^D_i$, give $\hat{\mu}_i$ values close to 0.5\end{tabular} & \cite{RN408, liu2014}\\ 
    \hline
   Empirical-logistic normal model (ELN) & \begin{eqnarray*}
        \text{logit} \left( \hat{\mu}_i^D \right) & \sim & N \left( \text{logit}\lb{ \hat{\mu}_i }, \gamma_i^D \right) \quad\quad i=1, \dots, m
        \\
        \gamma_i^D & = & \psi_i^D \left[ \hat{\mu}_i^D \left( 1 - \hat{\mu}_i^D \right) \right]^{-2} \quad\quad i=1, \dots, m
        \\
        \text{logit}\lb{ \hat{\mu}_i } & = & \mathbf{Z}_i \boldsymbol{\lambda} + v_i \quad\quad i=1, \dots, M
    \end{eqnarray*} & \begin{tabular}[c]{@{}l@{}}
    Requires $\hat{\mu}_i^D \in (0,1)$, $\forall i$\\
    Requires GVFs for unstable areas\end{tabular} &  \cite{RN476, RN552, liu2014}\\
    \hline
    \end{tabular}
    }
    \caption{\small Description of the individual and area level comparison models with their corresponding model identifiers (e.g. ELN, LOG, etc). Model details can be found in both \cref{sec:prop_models} and Supplemental Materials B \cite{self_cite}, while generalized variance functions (GVFs) are described in \cref{sec:gvf}. Note that $e_i \sim N(0, \sigma_e^2)$ and $v_i \sim N(0, \sigma_v^2)$ and weakly informative priors are used for the remaining parameters (see \cref{sec:stage1_model,sec:stage2_model,sec:sim}).}
    \label{table:comp_models}
\end{sidewaystable*}

\subsection{Two-stage approaches}

Our two-stage approach builds on the similar intuition of Gao and Wakefield \cite{gao2023_sma} (hereafter named the GW model) and Das \emph{et al.} \cite{Das2022}, but with several improvements. Firstly, unlike previous approaches that treat predictions from the first stage as fixed, the TSLN-S2 model accommodates the uncertainty in the model parameters and predictions from the TSLN-S1 model. Gao and Wakefield \cite{gao2023_sma} warn against overconfident estimates, as their stage 1 LGREG estimator, fit using frequentist inference, ignores model parameter uncertainty. By using Bayesian inference at both the individual and area level, our estimates inherit the uncertainty from model fitting.

Secondly, our approach relaxes some covariate requirements. While the GW model assumes that the stage 1 model provides ``unbiased'' predictions for all individuals and areas this assumption can be restrictive. Specifically, the GW model requires individual-level covariates for all population individuals and prohibits the use of survey-only covariates. Fortunately, the TSLN approach has no such requirement or prohibition. Note that Gao and Wakefield \cite{gao2023_sma} acknowledge the limitations of requiring access to covariate information for the entire population. Their solution involved redefining individuals as very fine spatial grids, for which population data were available.

Thirdly, the GW approach requires access to first and second-order inclusion probabilities in order to calculate the sampling variance of their first stage LGREG estimates. In most cases, first-order inclusion probabilities are not provided by data custodians, and further, it is very rare to have access to second-order probabilities. Our TSLN approach only assumes access to sample weights, which are typically provided with survey data. Finally, unlike the model of Das \emph{et al.} \cite{Das2022}, which operates solely at the area level, the TSLN approach can provide more efficient estimates by smoothing the data at both the individual and area levels. 

Note that we do not use the GW model as a comparison model in \cref{sec:sim} as it uses a frequentist approach for the stage 1 model and requires pairwise sampling probabilities which we assume are inaccessible. Nor do we compare our approach to that of Das \emph{et al.} \cite{Das2022} model as it requires temporal data. 

\subsection{Models for proportions} \label{sec:prop_models}

As with the TSLN approach, those developed by Gao and Wakefield \cite{gao2023_sma} and Das \emph{et al.} \cite{Das2022} are composed of basic component models such as the logistic and FH. \cref{table:comp_models} gives an overview of four different model-based techniques for small area estimation of proportions that serve as comparison models in the simulation experiment described in \cref{sec:sim}. However, now we briefly address some of the challenges and limitations of these and other alternative methodologies (see Supplemental Materials B \cite{self_cite} for more details), as well as how the TSLN approach offers some solutions. In addition, we provide justification for the four comparison models used in this work. 


\textbf{BIN model} Despite their common use \cite{RN28}, standard FH models are generally not suitable to model sparse binary data due to unstable direct estimates. Instability is alleviated by considering an aggregation of the binary outcomes via a binomial model. However, these models do not generally accommodate the sampling methodology. In contrast, our approach provides stable direct estimates and accommodates the survey design. In this study, we deliberately selected a binomial model that ignores the sampling methodology as a baseline (crude) model against which the TSLN approach was evaluated in terms of design consistency.  

\textbf{BETA model} Although the Beta distribution is a natural choice when modeling proportions \cite{RN58,RN400, RN408, liu2014}, it has several statistical and computational limitations that arise when modeling sparse survey data (see \cref{table:comp_models} and Supplemental Materials B \cite{self_cite}). These include: problematic constraints on the $\hat{\mu}_i$'s which depend on the $\psi^D_i$'s; the necessity to impute sampling variances for non-sampled areas; bimodal behaviour which causes significant MCMC convergence issues; and undefined likelihoods for unstable direct estimates. Nevertheless, the FH Beta model is a popular solution for modeling proportions \cite{liu2014}; thus, it was chosen as one of the comparison models.

\textbf{ELN model} Other area level models for proportions assume Gaussian distributions for the direct estimates, allowing the use of typical FH models \cite{RN400}. Although this method is effective for applications with large area sample sizes, in the situation of sparsity, the Gaussian approximation proves insufficient \cite{liu2014}. Instead, researchers use the empirical logistic transformation before using an FH model. This is the technique used by Mercer \emph{et al.} \cite{RN476} and Cassy \emph{et al.} \cite{RN552}, and by us. Liu \emph{et al.} \cite{liu2014} present a similar model (Model 3), which utilizes the design effect to compute sampling variances. Since the ELN model is most similar to ours, we included it in our comparisons to evaluate the advantages of using S1 over direct estimates.

\textbf{LOG model} Individual level models for binary data offer significant advantages over the area level models discussed above. Most notably, working at the individual level eliminates the need for direct estimates, and allows for greater model flexibility. However, these models require covariate data for \emph{all} persons in the population \cite{RN404, RN146}, which is often sourced from census microdata. This introduces further restrictions on covariate choice due to the necessary concordance between survey and census covariates. Covariates collected in a census may be less predictive of health-related outcomes, whilst individual level survey-only covariates may be more useful. The TSLN-S1 model does not predict values for the whole population, hence easing the requirement for survey-census concordance and permitting the inclusion of survey-only covariates.

Another limitation of individual level models is that they do not accommodate known sample weights and are susceptible to bias under informative sampling. A simple solution is to include the sample weights as covariates, but this requires imputation for nonsampled individuals \cite{RN147}. Bayesian pseudo-likelihood \cite{RN500} is an alternative used by Parker \emph{et al.} \cite{RN44,Parker2023} (see Supplemental Materials B \cite{self_cite}). By construction, our TSLN approach accommodates the sample design at both stages; pseudo-likelihood in \eqref{eq:tsln_F_s1_log} and the weighted estimators in \eqref{eq:tsln_mu_S1}. Comparing our TSLN approach to a pseudo-likelihood LOG model (similar to MrP \cite{RN32, RN403, RN404}) enables us to assess the effectiveness of utilizing survey-only covariates and the utility of two-stage methods in general.

\section{Simulation Study} \label{sec:sim}

To evaluate the performance of the TSLN approach, we undertook a simulation experiment based on the approach used by Buttice and Highton \cite{RN442}. In contrast to conventional simulation studies where data are generated directly from the target model, the two-stage structure of our approach required a reverse method. We generated a census by first simulating the area level proportions and then generating individual and area level covariates based on these, implicitly allowing for specified levels of association and random effects. For simplicity, no spatial autocorrelation was imposed between areas. From the fixed census, we repeatedly drew unique surveys using a multi-stage informative design.

First, we generated individual level census data for $100$ small areas with populations ranging from $500$ and $3000$. The simulated census included the binary outcomes, $\mathbf{y}$, two individual-level covariates (one survey-only categorical covariate, $\mathbf{x}^{\text{survey}}$ with three groups, and one continuous covariate, $\mathbf{x}^{\text{census}}$ available in both the census and survey) and a single area level covariate ($\mathbf{Z}$). To explore how the association of the survey-only covariate affected the overall performance of our approach, we used two levels; high and average (see \cref{table:sim_specs}). 

Keeping the census, and thus true proportions, fixed, we then drew 500 unique (repetitions) surveys, fitting the TSLN and four comparison models (\cref{table:comp_models}) to each. We used a sampling fraction of 0.4\% and only sampled $60$ areas for each repetition. Following Hidiroglou and You \cite{RN137}, we devised the sampling method so that individuals with a binary value of 0 were more likely to be sampled and constructed sample weights to reflect this. Extensive details of the simulation algorithm can be found in Supplemental Materials C \cite{self_cite}. 

The median sample size, $n$, population size, $N$, and area sample size, $n_i$ were 771, 177263, and 8 respectively. Note that it is relatively rare for simulation experiments in the SAE literature to use such small area level sample sizes; generally $n_i > 50$ \cite{gao2023_sma, RN111, RN44, RN56}, which is less relevant in the Australian context. Unstable direct estimates for all comparison models were stabilized via the simple perturbation method \cite{liu2014}; add or subtract a value of $0.001$ to any $\hat{\mu}_i^D = 0$ or $1$, respectively, prior to modeling. 

Overall, we wished to explore how the TSLN approach performed for sparse survey data. In addition, we wanted to assess performance for varying degrees of prevalence and for the inclusion of more predictive individual level survey-only covariates. These objectives translated to the six simulation scenarios given in \cref{table:sim_specs}. Note that Sc1, Sc3, and Sc5 ensure that the survey-only covariate is much more predictive of the outcome than the census-available covariate. Given the informative sample design, the rare scenarios (Sc3 and Sc4) will provide a high number of unstable direct estimates (see \cref{table:sim_summary}).  

\begin{table}
  \centering
    \begin{tabular}{lllll}
     &  & Predictive $\mathbf{x}^\text{survey}$ & $L$ & $U$ \\ 
    \hline
    \multirow{2}{*}{50-50} & Sc1 & High & \multirow{2}{*}{0.35} & \multirow{2}{*}{0.65} \\
     & Sc2 & Average &  &  \\ 
    \hline
    \multirow{2}{*}{Rare} & Sc3 & High & \multirow{2}{*}{0.1} & \multirow{2}{*}{0.4} \\
     & Sc4 & Average &  &  \\ 
    \hline
    \multirow{2}{*}{Common} & Sc5 & High & \multirow{2}{*}{0.6} & \multirow{2}{*}{0.9} \\
     & Sc6 & Average &  &  \\
    \hline
    \end{tabular}
    \caption{Summary of the six simulation scenarios. $L$ and $U$ give the lower and upper bounds of the true proportions, respectively (see Supplemental Materials C \cite{self_cite} for more details).}
    \label{table:sim_specs}
\end{table}

As described in \cref{sec:existing_methods} we compared the performance of our approach to the four model-based SAE methods summarized in \cref{table:comp_models}. Both models of the TSLN approach and the BETA, LOG and ELN models were fit using \texttt{rstan}, leveraging Hamiltonian Monte Carlo (HMC) \cite{RN452}. The BIN model was fit using JAGS via \texttt{rjags} \cite{RN447}. The covariates used in the models are given in \cref{table:cov_specs}. Only models deemed to have converged were used in the simulation results. Convergence was assessed using $\widehat{R}$ \cite{RN499}. Any analysis for which at least one parameter of interest had $\widehat{R} > 1.02$ was discarded. We used $2,000$ post-warmup and $6,000$ post-burnin draws for each of four chains for the Stan and JAGS models, respectively. \texttt{R} code to run the simulations, with the necessary Stan and JAGS code is available \href{https://github.com/JamieHogg-depo/two_stage_saemodel_proportions}{here}.

\begin{table}
  \centering
    \begin{tabular}{rrlll}
    \multicolumn{1}{l}{} & \multicolumn{1}{l}{} & $\mathbf{x}^\text{survey}$ & $\mathbf{x}^\text{census}$ & $\mathbf{Z}$ \\ 
    \hline
    \multirow{2}{*}{Individual level} & TSLN-S1 & \checkmark & \checkmark & \checkmark \\
     & LOG &  & \checkmark & \checkmark \\ 
    \hline
    \multirow{4}{*}{Area level} & TSLN-S2 &  &  & \checkmark \\
     & ELN &  &  & \checkmark \\
     & BETA &  &  & \checkmark \\
     & BIN &  &  & \checkmark \\
    \hline
    \end{tabular}
    \caption{Illustrates which of the simulated covariates, described in \cref{sec:sim}, were used in which models in the simulation study. Ticks denote that the model specified in the row used the covariate (shown in the column) as a fixed effect. The estimated regression coefficients and variance terms varied across models, scenarios and repetitions (see Supplemental Materials C \cite{self_cite}).}
    \label{table:cov_specs}
\end{table}

Similar to the TSLN approach, all fixed regression parameters in the comparison models were given a $N(0,2)$ prior, whilst intercept terms were given a student-$t(df = 3, \mu = 0, \sigma = 1)$ prior. Standard deviation parameters in the TSLN-S1, LOG and BIN models were given a weakly informative $N(0,1)^{+}$ prior, while standard deviation parameters in the BETA, ELN and TSLN-S2 models were given a $N(0,2)^{+}$ prior. Note that the GVF described in \cref{sec:gvf} was necessarily adjusted for the Beta model (See Supplemental Materials B \cite{self_cite} for details). All design matrices were mean-centered prior to model fitting and the popular QR decomposition was used \cite{RN572}. 


\subsection{Performance metrics} \label{sec:sim_perform}

To compare the models we use Bayesian performance metrics which are calculated separately depending on whether an area was sampled or nonsampled in repetition $d$. Unlike common frequentist metrics (see Supplemental Materials C \cite{self_cite}), Bayesian metrics use all the appropriate posterior samples during calculation. Thus, they generally favor posteriors with smaller variance, whilst still penalizing inaccuracy. While Bayesian coverage is summed over all areas and repetitions, the remaining metrics are calculated independently for each repetition, $d$. The simulations provide $500 \times T$ posterior draws for each area, model and scenario. Let $\hat{\mu}_{idt}$ be the $t$th posterior draw for repeat $d$ in area $i$ and $\mu_i$ be the true proportion in area $i$. Also let $\hat{\mu}_{id}^{\text{(L)}}$ and $\hat{\mu}_{id}^{\text{(U)}}$ denote the lower and upper bounds, respectively, of the posterior 95\% highest density interval (HDI) for area $i$ and repeat $d$. 
\\
\noindent Absolute relative bias: ARB$_{id}$
\begin{equation}
    \left| \frac{ \frac{1}{T} \sum_{t=1}^T \lb{\hat{\mu}_{idt}-\mu_i} }{\mu_i} \right| \label{met:b_arb}
\end{equation}
\noindent Relative root mean square error: RRMSE$_{id}$
\begin{equation}
    \frac{\sqrt{  \frac{1}{T} \sum_{t=1}^T \lb{\hat{\mu}_{idt}-\mu_i}^2  }}{\mu_i}  \label{met:b_rrmse}
\end{equation}
\noindent Coverage
\begin{equation}
    \frac{1}{MD} \sum_{i=1}^M \sum_{d=1}^D \mathbb{I}\left( \hat{\mu}_{id}^{\text{(L)}} < \mu_i < \hat{\mu}_{id}^{\text{(U)}} \right) \label{met:f_coverage}
\end{equation}

\begin{figure*}
    \centering
    \includegraphics[width=\columnwidth]{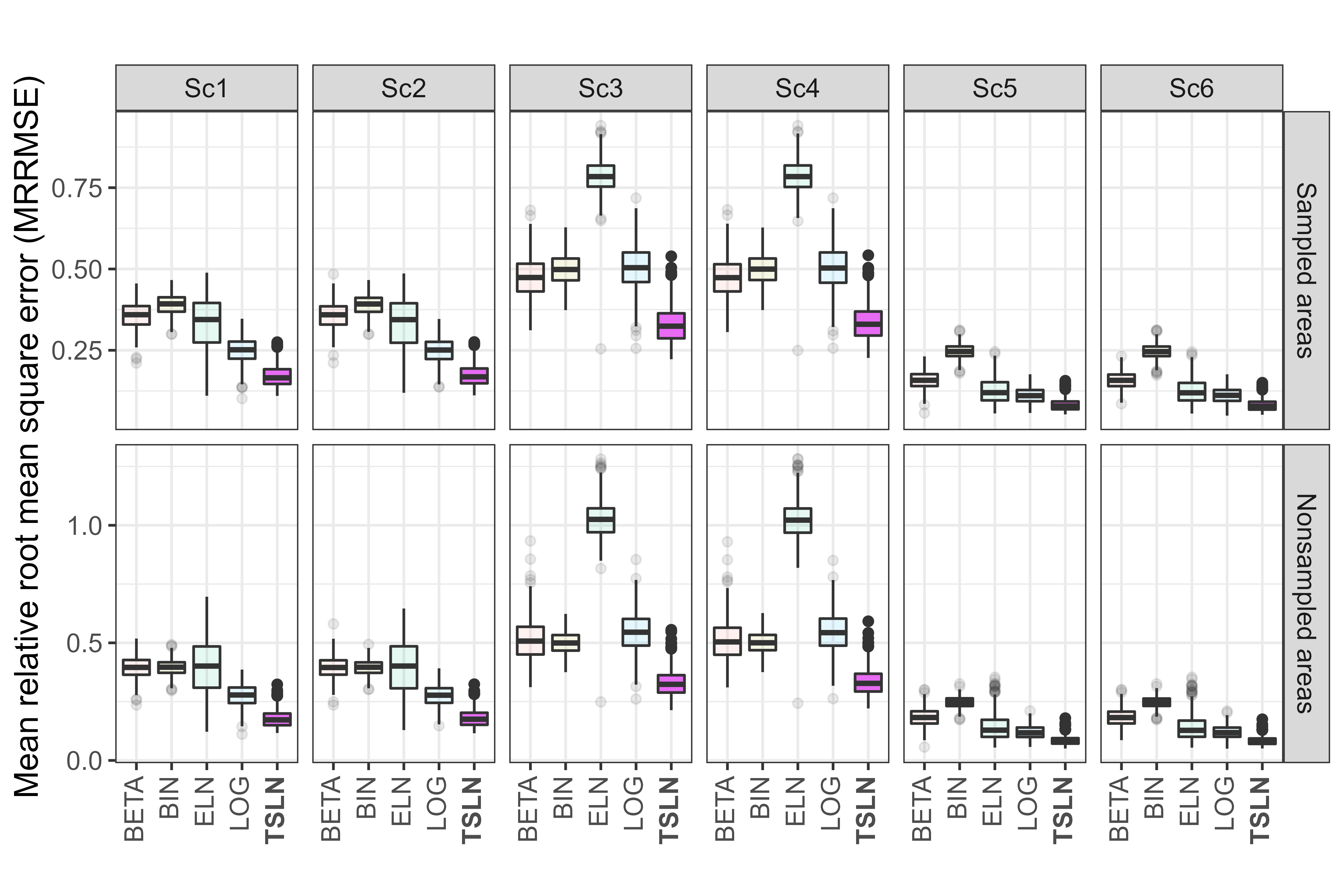}
    \caption{Boxplots of the 500 MRRMSE values. The medians of each boxplot are available in \cref{table:bay_mse}.}
    \label{fig:bay_MRRMSE}
\end{figure*}

\begin{figure*}
    \centering
    \includegraphics[width=\columnwidth]{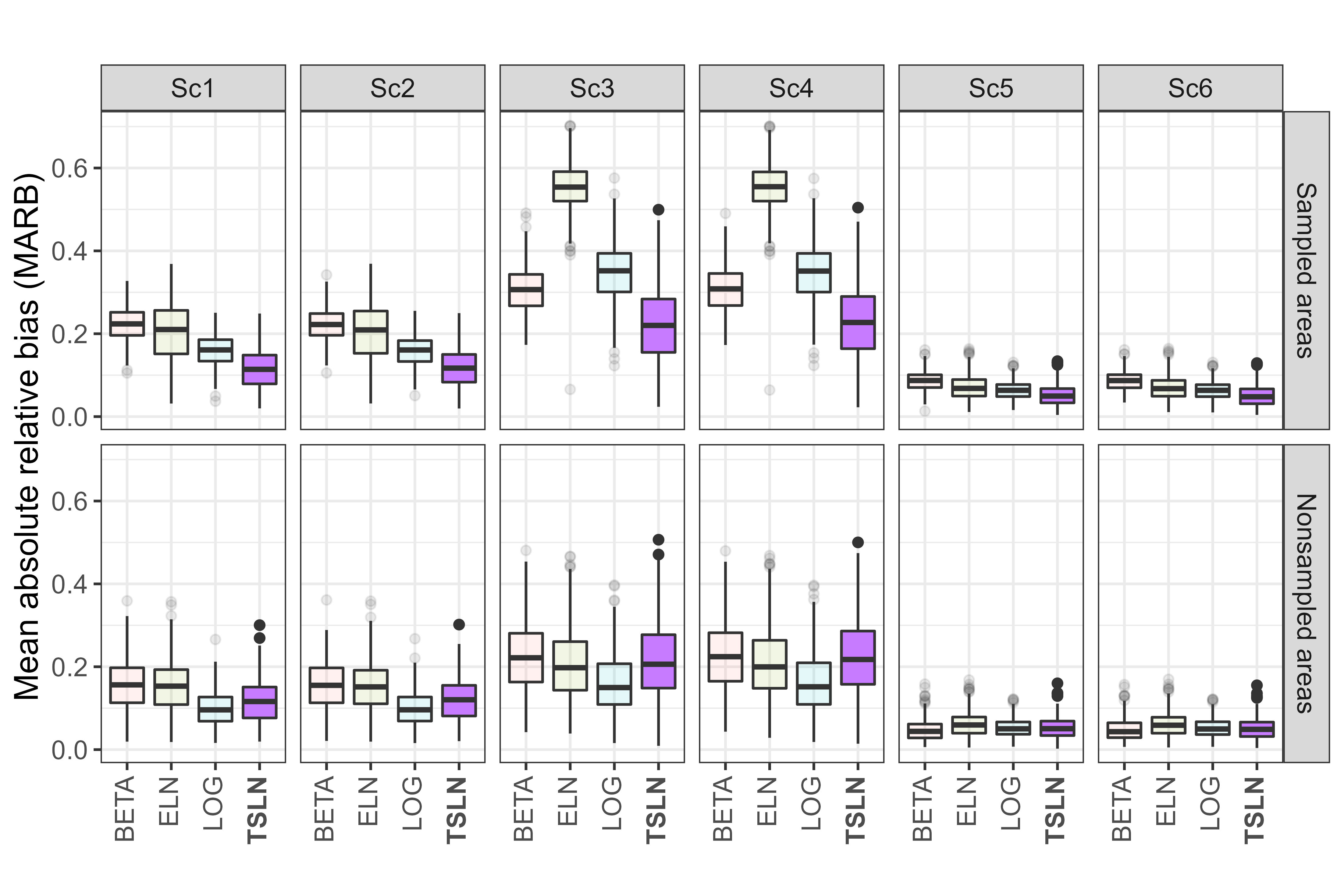}
    \caption{Boxplots of the 500 MARB values. The medians of each boxplot are available in \cref{table:bay_mse}. We have omitted the BIN model; its very large bias distorts the plot.}
    \label{fig:bay_MARB}
\end{figure*}

We take the mean across all areas to derive the mean RRMSE (MRRMSE) and mean ARB (MARB) for each repetition. While the distribution of these 500 MRRMSE and MARB values are visualized in \cref{fig:bay_MRRMSE} and \ref{fig:bay_MARB}, respectively, summaries are given in \cref{table:bay_mse}.

\subsection{Results}


\begin{table*}
\centering
\begin{tabular}{rrrrrr}
 &  & \makecell[r]{Percent of\\sampled areas\\unstable (\%)} & ALC & \makecell[r]{Percent\\increase in\\sampling\\variance (\%)} & \makecell[r]{Percent\\reduction in\\MAB (\%)}\\
 \hline\hline
50-50 & Sc1 & 3.3 (1.7, 5.0) & 0.69 (0.66, 0.73) & 56 (27, 91) & 25 (22, 28)\\
 & Sc2 & 3.3 (1.7, 5.0) & 0.65 (0.60, 0.68) & 74 (40, 107) & 30 (27, 34)\\
 \hline
Rare & Sc3 & 25.0 (21.7, 28.3) & 0.78 (0.73, 0.82) & 79 (36, 140) & 24 (21, 28)\\
 & Sc4 & 25.0 (21.7, 28.3) & 0.76 (0.71, 0.80) & 101 (53, 162) & 27 (24, 31)\\
 \hline
Common & Sc5 & 1.7 (0.0, 3.3) & 0.66 (0.62, 0.70) & 66 (29, 108) & 30 (26, 34)\\
 & Sc6 & 1.7 (0.0, 3.3) & 0.59 (0.55, 0.64) & 79 (36, 123) & 38 (34, 42)\\
  \hline\hline
\end{tabular}
\caption{Overview of the simulation and smoothing properties of the TSLN-S1 model (see Supplemental Materials C \cite{self_cite}), including the area linear comparison (ALC) metric (\cref{sec:smooth}), and mean absolute bias (MAB) along with the interquartile bounds in brackets. Column one is derived by calculating the percentage of sampled areas with unstable direct estimates for each repetition, before taking the median (and IQ bounds) of these 500 percentages. Column two is derived by taking the median (and IQ bounds) of the 500 ALC values. The third column is derived by first calculating the median of the area-specific ratios of the stage 1 sampling variances to the direct sampling variances for each repetition, before taking the median (and IQ bounds) of the 500 values (details are given in Supplemental Materials C \cite{self_cite}). Large values indicate large increases in the S1 sampling variances above that of the direct, which is undesirable. The final column is derived by first calculating the ratio between the MABs for the direct and S1 estimates for each repetition, before taking the median (and IQ bounds) of these 500 ratios (details are given in Supplemental Materials C \cite{self_cite}). Large values indicate larger reductions in MAB when using the S1 estimates, which is preferable.}
\label{table:sim_summary}
\end{table*}

\begin{sidewaystable*}
\centering
\begin{tabular}{lllllllllll}
 &  &  & Sampled areas &  &  &  & Nonsampled areas &  &  & \\
 \hline\hline
 &  &  & MRRMSE & MARB & CI width & Coverage & MRRMSE & MARB & CI width & Coverage\\
  \hline\hline
50-50 & Sc1 & BETA & 0.36 \textcolor[rgb]{0.502,0.502,0.502}{(2.17)} & 0.22 \textcolor[rgb]{0.502,0.502,0.502}{(1.96)} & 0.46 & 0.80 & 0.40 \textcolor[rgb]{0.502,0.502,0.502}{(2.29)} & 0.16 \textcolor[rgb]{0.502,0.502,0.502}{(1.35)} & 0.61 & 1.00\\
 &  & BIN & 0.39 \textcolor[rgb]{0.502,0.502,0.502}{(2.38)} & 0.38 \textcolor[rgb]{0.502,0.502,0.502}{(3.35)} & \textbf{0.18} & 0.03 & 0.40 \textcolor[rgb]{0.502,0.502,0.502}{(2.30)} & 0.38 \textcolor[rgb]{0.502,0.502,0.502}{(3.29)} & 0.20 & 0.03\\
 &  & ELN & 0.34 \textcolor[rgb]{0.502,0.502,0.502}{(2.08)} & 0.21 \textcolor[rgb]{0.502,0.502,0.502}{(1.84)} & 0.43 & 0.91 & 0.40 \textcolor[rgb]{0.502,0.502,0.502}{(2.32)} & 0.15 \textcolor[rgb]{0.502,0.502,0.502}{(1.32)} & 0.64 & 1.00\\
 &  & LOG & 0.25 \textcolor[rgb]{0.502,0.502,0.502}{(1.52)} & 0.16 \textcolor[rgb]{0.502,0.502,0.502}{(1.41)} & 0.32 & 0.90 & 0.28 \textcolor[rgb]{0.502,0.502,0.502}{(1.61)} & \textbf{0.10 \textcolor[rgb]{0.502,0.502,0.502}{(0.83)}} & 0.47 & 1.00\\
 &  & TSLN & \textbf{0.17 \textcolor[rgb]{0.502,0.502,0.502}{(1.00)}} & \textbf{0.11 \textcolor[rgb]{0.502,0.502,0.502}{(1.00)}} & 0.22 & \textbf{0.94} & \textbf{0.17 \textcolor[rgb]{0.502,0.502,0.502}{(1.00)}} & 0.12 \textcolor[rgb]{0.502,0.502,0.502}{(1.00)} & 0.23 & \textbf{0.95}\\
 \hline
 & Sc2 & BETA & 0.36 \textcolor[rgb]{0.502,0.502,0.502}{(2.13)} & 0.22 \textcolor[rgb]{0.502,0.502,0.502}{(1.90)} & 0.46 & 0.80 & 0.40 \textcolor[rgb]{0.502,0.502,0.502}{(2.25)} & 0.16 \textcolor[rgb]{0.502,0.502,0.502}{(1.29)} & 0.62 & 1.00\\
 &  & BIN & 0.39 \textcolor[rgb]{0.502,0.502,0.502}{(2.33)} & 0.38 \textcolor[rgb]{0.502,0.502,0.502}{(3.25)} & \textbf{0.18} & 0.03 & 0.40 \textcolor[rgb]{0.502,0.502,0.502}{(2.26)} & 0.38 \textcolor[rgb]{0.502,0.502,0.502}{(3.17)} & 0.20 & 0.03\\
 &  & ELN & 0.34 \textcolor[rgb]{0.502,0.502,0.502}{(2.04)} & 0.21 \textcolor[rgb]{0.502,0.502,0.502}{(1.79)} & 0.43 & 0.91 & 0.40 \textcolor[rgb]{0.502,0.502,0.502}{(2.29)} & 0.15 \textcolor[rgb]{0.502,0.502,0.502}{(1.26)} & 0.64 & 1.00\\
 &  & LOG & 0.25 \textcolor[rgb]{0.502,0.502,0.502}{(1.49)} & 0.16 \textcolor[rgb]{0.502,0.502,0.502}{(1.37)} & 0.32 & 0.90 & 0.28 \textcolor[rgb]{0.502,0.502,0.502}{(1.58)} & \textbf{0.10 \textcolor[rgb]{0.502,0.502,0.502}{(0.80)}} & 0.47 & 1.00\\
 &  & TSLN & \textbf{0.17 \textcolor[rgb]{0.502,0.502,0.502}{(1.00)}} & \textbf{0.12 \textcolor[rgb]{0.502,0.502,0.502}{(1.00)}} & 0.22 & \textbf{0.93} & \textbf{0.18 \textcolor[rgb]{0.502,0.502,0.502}{(1.00)}} & 0.12 \textcolor[rgb]{0.502,0.502,0.502}{(1.00)} & 0.23 & \textbf{0.95}\\
  \hline
Rare & Sc3 & BETA & 0.47 \textcolor[rgb]{0.502,0.502,0.502}{(1.46)} & 0.31 \textcolor[rgb]{0.502,0.502,0.502}{(1.39)} & 0.21 & 0.66 & 0.51 \textcolor[rgb]{0.502,0.502,0.502}{(1.57)} & 0.22 \textcolor[rgb]{0.502,0.502,0.502}{(1.08)} & 0.34 & \textbf{0.97}\\
 &  & BIN & 0.50 \textcolor[rgb]{0.502,0.502,0.502}{(1.54)} & 0.48 \textcolor[rgb]{0.502,0.502,0.502}{(2.16)} & \textbf{0.12} & 0.08 & 0.50 \textcolor[rgb]{0.502,0.502,0.502}{(1.54)} & 0.47 \textcolor[rgb]{0.502,0.502,0.502}{(2.29)} & \textbf{0.13} & 0.08\\
 &  & ELN & 0.78 \textcolor[rgb]{0.502,0.502,0.502}{(2.42)} & 0.55 \textcolor[rgb]{0.502,0.502,0.502}{(2.52)} & 0.39 & 0.68 & 1.03 \textcolor[rgb]{0.502,0.502,0.502}{(3.17)} & 0.20 \textcolor[rgb]{0.502,0.502,0.502}{(0.96)} & 0.77 & 1.00\\
 &  & LOG & 0.50 \textcolor[rgb]{0.502,0.502,0.502}{(1.55)} & 0.35 \textcolor[rgb]{0.502,0.502,0.502}{(1.60)} & 0.28 & 0.82 & 0.55 \textcolor[rgb]{0.502,0.502,0.502}{(1.68)} & \textbf{0.15 \textcolor[rgb]{0.502,0.502,0.502}{(0.73)}} & 0.43 & 1.00\\
 &  & TSLN & \textbf{0.32 \textcolor[rgb]{0.502,0.502,0.502}{(1.00)}} & \textbf{0.22 \textcolor[rgb]{0.502,0.502,0.502}{(1.00)}} & 0.19 & \textbf{0.86} & \textbf{0.32 \textcolor[rgb]{0.502,0.502,0.502}{(1.00)}} & 0.21 \textcolor[rgb]{0.502,0.502,0.502}{(1.00)} & 0.20 & 0.88\\
  \hline
 & Sc4 & BETA & 0.47 \textcolor[rgb]{0.502,0.502,0.502}{(1.43)} & 0.31 \textcolor[rgb]{0.502,0.502,0.502}{(1.36)} & 0.21 & 0.66 & 0.50 \textcolor[rgb]{0.502,0.502,0.502}{(1.54)} & 0.22 \textcolor[rgb]{0.502,0.502,0.502}{(1.03)} & 0.34 & \textbf{0.97}\\
 &  & BIN & 0.50 \textcolor[rgb]{0.502,0.502,0.502}{(1.51)} & 0.48 \textcolor[rgb]{0.502,0.502,0.502}{(2.10)} & \textbf{0.12} & 0.09 & 0.50 \textcolor[rgb]{0.502,0.502,0.502}{(1.53)} & 0.47 \textcolor[rgb]{0.502,0.502,0.502}{(2.18)} & \textbf{0.14} & 0.09\\
 &  & ELN & 0.78 \textcolor[rgb]{0.502,0.502,0.502}{(2.38)} & 0.55 \textcolor[rgb]{0.502,0.502,0.502}{(2.44)} & 0.39 & 0.68 & 1.02 \textcolor[rgb]{0.502,0.502,0.502}{(3.12)} & 0.20 \textcolor[rgb]{0.502,0.502,0.502}{(0.92)} & 0.77 & 1.00\\
 &  & LOG & 0.50 \textcolor[rgb]{0.502,0.502,0.502}{(1.52)} & 0.35 \textcolor[rgb]{0.502,0.502,0.502}{(1.55)} & 0.28 & 0.82 & 0.54 \textcolor[rgb]{0.502,0.502,0.502}{(1.66)} & \textbf{0.15 \textcolor[rgb]{0.502,0.502,0.502}{(0.70)}} & 0.43 & 1.00\\
 &  & TSLN & \textbf{0.33 \textcolor[rgb]{0.502,0.502,0.502}{(1.00)}} & \textbf{0.23 \textcolor[rgb]{0.502,0.502,0.502}{(1.00)}} & 0.19 & \textbf{0.86} & \textbf{0.33 \textcolor[rgb]{0.502,0.502,0.502}{(1.00)}} & 0.22 \textcolor[rgb]{0.502,0.502,0.502}{(1.00)} & 0.19 & 0.88\\
  \hline
Common & Sc5 & BETA & 0.16 \textcolor[rgb]{0.502,0.502,0.502}{(2.01)} & 0.09 \textcolor[rgb]{0.502,0.502,0.502}{(1.76)} & 0.31 & 0.89 & 0.18 \textcolor[rgb]{0.502,0.502,0.502}{(2.25)} & 0.04 \textcolor[rgb]{0.502,0.502,0.502}{(0.87)} & 0.43 & 1.00\\
 &  & BIN & 0.25 \textcolor[rgb]{0.502,0.502,0.502}{(3.13)} & 0.24 \textcolor[rgb]{0.502,0.502,0.502}{(4.80)} & 0.19 & 0.05 & 0.25 \textcolor[rgb]{0.502,0.502,0.502}{(3.04)} & 0.23 \textcolor[rgb]{0.502,0.502,0.502}{(4.64)} & 0.21 & 0.08\\
 &  & ELN & 0.12 \textcolor[rgb]{0.502,0.502,0.502}{(1.52)} & 0.07 \textcolor[rgb]{0.502,0.502,0.502}{(1.37)} & 0.26 & 0.98 & 0.13 \textcolor[rgb]{0.502,0.502,0.502}{(1.58)} & 0.06 \textcolor[rgb]{0.502,0.502,0.502}{(1.18)} & 0.31 & 1.00\\
 &  & LOG & 0.11 \textcolor[rgb]{0.502,0.502,0.502}{(1.40)} & 0.06 \textcolor[rgb]{0.502,0.502,0.502}{(1.28)} & 0.24 & \textbf{0.97} & 0.12 \textcolor[rgb]{0.502,0.502,0.502}{(1.45)} & \textbf{0.05 \textcolor[rgb]{0.502,0.502,0.502}{(0.99)}} & 0.29 & 1.00\\
 &  & TSLN & \textbf{0.08 \textcolor[rgb]{0.502,0.502,0.502}{(1.00)}} & \textbf{0.05 \textcolor[rgb]{0.502,0.502,0.502}{(1.00)}} & \textbf{0.16} & 0.98 & \textbf{0.08 \textcolor[rgb]{0.502,0.502,0.502}{(1.00)}} & 0.05 \textcolor[rgb]{0.502,0.502,0.502}{(1.00)} & 0.17 & \textbf{0.99}\\
  \hline
 & Sc6 & BETA & 0.16 \textcolor[rgb]{0.502,0.502,0.502}{(2.02)} & 0.09 \textcolor[rgb]{0.502,0.502,0.502}{(1.81)} & 0.31 & 0.89 & 0.18 \textcolor[rgb]{0.502,0.502,0.502}{(2.26)} & 0.04 \textcolor[rgb]{0.502,0.502,0.502}{(0.88)} & 0.43 & 1.00\\
 &  & BIN & 0.25 \textcolor[rgb]{0.502,0.502,0.502}{(3.15)} & 0.24 \textcolor[rgb]{0.502,0.502,0.502}{(4.96)} & 0.19 & 0.05 & 0.25 \textcolor[rgb]{0.502,0.502,0.502}{(3.05)} & 0.23 \textcolor[rgb]{0.502,0.502,0.502}{(4.78)} & 0.21 & 0.08\\
 &  & ELN & 0.12 \textcolor[rgb]{0.502,0.502,0.502}{(1.53)} & 0.07 \textcolor[rgb]{0.502,0.502,0.502}{(1.41)} & 0.26 & 0.98 & 0.13 \textcolor[rgb]{0.502,0.502,0.502}{(1.58)} & 0.06 \textcolor[rgb]{0.502,0.502,0.502}{(1.20)} & 0.31 & 1.00\\
 &  & LOG & 0.11 \textcolor[rgb]{0.502,0.502,0.502}{(1.43)} & 0.06 \textcolor[rgb]{0.502,0.502,0.502}{(1.32)} & 0.24 & \textbf{0.97} & 0.12 \textcolor[rgb]{0.502,0.502,0.502}{(1.47)} & 0.05 \textcolor[rgb]{0.502,0.502,0.502}{(1.02)} & 0.30 & 1.00\\
 &  & TSLN & \textbf{0.08 \textcolor[rgb]{0.502,0.502,0.502}{(1.00)}} & \textbf{0.05 \textcolor[rgb]{0.502,0.502,0.502}{(1.00)}} & \textbf{0.16} & 0.98 & \textbf{0.08 \textcolor[rgb]{0.502,0.502,0.502}{(1.00)}} & \textbf{0.05 \textcolor[rgb]{0.502,0.502,0.502}{(1.00)}} & 0.16 & \textbf{0.98} \\
 \hline\hline
\end{tabular}
\caption{Median of the 500 MRRMSE and MARB values. The table also includes the width of 95\% posterior credible interval (CI width) and coverage across the 500 repetitions. Bold numbers represent the lowest value in each column, scenario and whether the metric is for sampled or nonsampled areas. For coverage, bold numbers denote the value closest to 0.95. Gray numbers in brackets give the ratio of the value to that of the TSLN approach.}
\label{table:bay_mse}
\end{sidewaystable*}

\subsubsection{Stage 1 results}

\cref{table:sim_summary} summarises the simulated data and provides details on the smoothing properties of the TSLN-S1 model. As anticipated with more predictive individual level covariates (e.g. Sc1 versus Sc2), the ALC is closer to 1. However, the differences are subtle. The biggest effect of having more predictive individual level covariates is the significantly smaller increase in sampling variance. However, this comes at the cost of a lower reduction in the mean absolute bias between the S1 or direct estimates and the true values. This pattern is expected because the S1 estimates will collapse to the direct estimates when the TSLN-S1 model ``overfits'' the survey data.  


\subsubsection{Overall modeling results}


\textbf{Accuracy} Overall, the TSLN approach provides MRRMSE that are between 40\% to 52\% smaller than those for the next smallest MRRMSE (\cref{fig:bay_MRRMSE} and \cref{table:bay_mse}). Across all scenarios and for sampled areas, at least, the TSLN approach provides a 28\% to 41\% smaller MARB than the next smallest MARB; a clear pattern in \cref{fig:bay_MARB}. However, for nonsampled areas, the TSLN approach does not outperform the alternative models, with MARB values ranging from 11\% to 30\% bigger than the best-performing approach (often the LOG model). These findings for the MARB (nonsampled) were expected given the additional smoothing enforced under the TSLN approach. That said, the between-simulation variability of the MARB estimates (e.g. the sizes of the boxes in \cref{fig:bay_MARB}) suggest that the MARB for the TSLN approach is at least comparable to the other models for nonsampled areas.

Table 1 in Supplemental Materials C \cite{self_cite} provides frequentist MSE results \cite{RN497}. Across all scenarios and for sampled areas we found that the TSLN approach had MSE values ranging from 78\% to 194\% smaller than the next smallest. For nonsampled areas, the LOG and BETA models had smaller MSE than that of the TSLN approach, apart from in the rare scenarios, where the TSLN approach outperformed all comparison models (5\% to 14\% smaller MSE than the BETA model). 

\textbf{Uncertainty} Similar to Gomez-Rubio \emph{et al.} \cite{RN35}, we found that the credible intervals for all comparison models were too wide for nonsampled areas, but generally too narrow for sampled areas. The TSLN approach had, in most cases, coverage closer to the nominal 95\% level than the comparison models. 

By excluding the BIN model due to its excessive bias, our simulation study found that the TSLN approach consistently provided the smallest CI widths, with the improvement most notable for nonsampled areas. For sampled areas, posterior credible intervals were consistently between 11\% and 50\% smaller than that of the other models. For nonsampled areas the CI widths ranged from 71\% to 104\% smaller. Since the TSLN-S1 model pre-smooths estimates, the TSLN-S2 model has considerably less variance to accommodate than the ELN and BETA models. This manifests in a smaller $\sigma_v$ (see Table 2 in Supplemental Materials C \cite{self_cite}), which results in less posterior variance (i.e. smaller credible intervals). Interestingly when the perturbation applied to unstable direct estimates is less extreme (e.g. setting any $\hat{\mu}_i^D = 0.01$ instead of $0.001$), the size of $\sigma_v$ and thus the uncertainty can be reduced, in some cases even halved (results not shown).  

\textbf{Covariates} We found no discernible performance improvements when we used more predictive individual level covariates in the TSLN-S1 model (e.g. comparing Sc1 to Sc2). 

\textbf{Rare outcomes}
Given our interest in instability, the simulation results for the rare scenarios (Sc3 and Sc4) were particularly important. Although the ELN model and TSLN approach are relatively similar in construction, our approach gave superior results under high instability. The TSLN approach had lower median MRRMSE, smaller average CI widths, and better coverage than the ELN model. We believe this is primarily due to the substantially larger random effect variance (see Supplemental Materials C \cite{self_cite}); in Sc3 $\sigma_v = 2.11, 0.17$ for the ELN model and TSLN approach, respectively. 

While the alternative area level model (BETA model) provided better coverage (97\%) for non-sampled areas, it gave very poor coverage (66\%) for sampled areas. Although individual level models (e.g. LOG model) should provide optimal prediction in the unstable setting, we found that the TSLN approach remained superior in terms of median MRRMSE, average CI width and coverage. For the rare scenarios and for nonsampled areas, the LOG model gave the lowest bias; approximately 30\% smaller than the TSLN approach. However, the between-simulation variability of the MARB (see \cref{fig:bay_MARB}) suggests this difference is not significant.

\newpage
\section{Application: Prevalence of current smoking on the east coast of Australia} \label{sec:case_study}

To illustrate the benefits of the two-stage logistic normal (TSLN) model in practice, we generate small area estimates of current smoking prevalence in Australia using the TSLN, ELN and LOG models. Neither the BIN model (poor performance in the simulation study) nor the BETA model (has various limitations, see Supplemental Materials B \cite{self_cite}) were considered. Unless otherwise stated, the model specifications used in this application are identical to those described in the preceding sections.  

\subsection{Data}

\begin{wrapfigure}{r}{0.3\textwidth}
    \begin{center}
        \includegraphics[trim=45mm 0mm 45mm 0mm, clip]{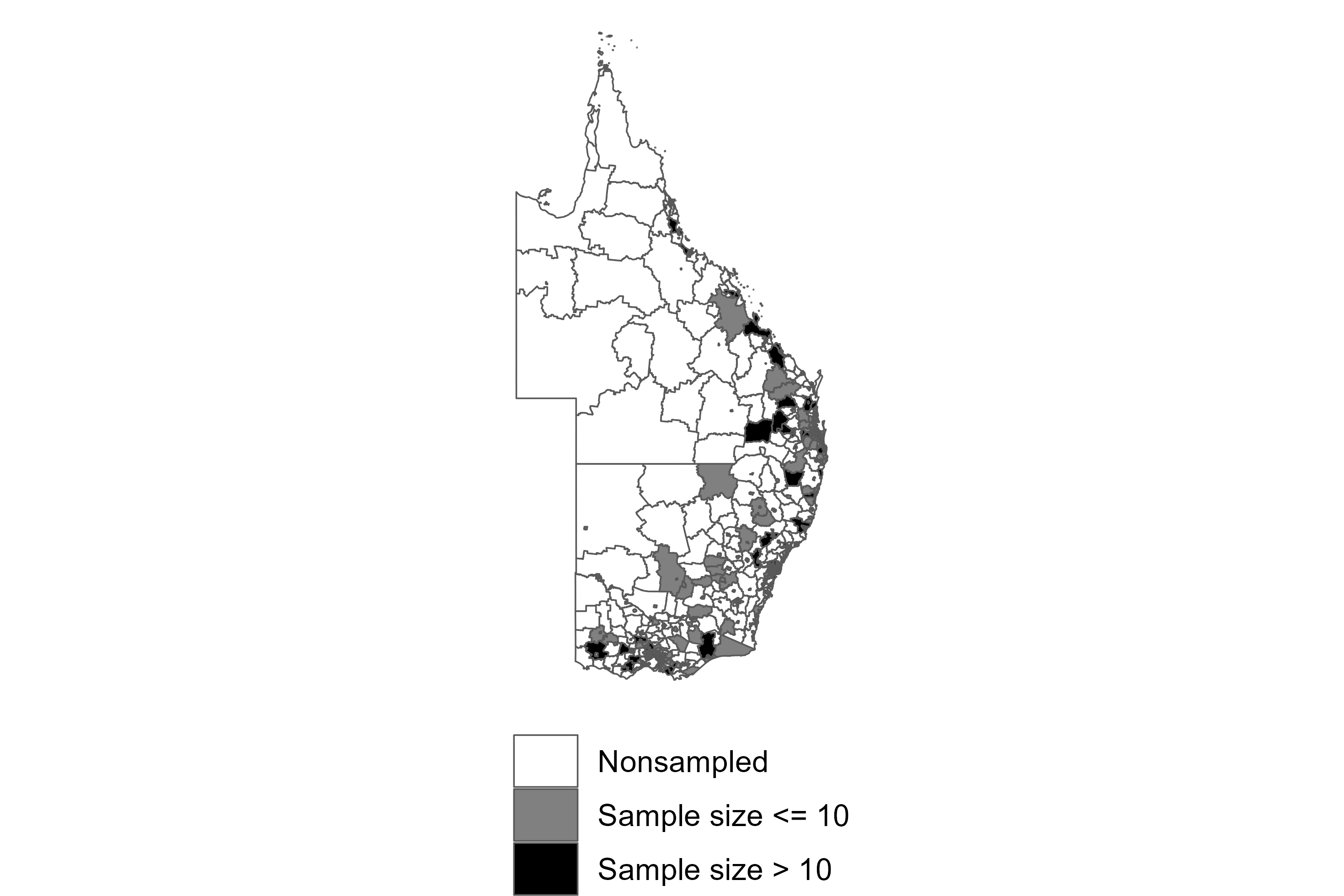}
    \end{center}
    \caption{Map of the 1630 statistical area level 2 (SA2s) on the east coast of Australia. Each SA2 is colored according to the sample size (greater or less than 10) or sample status of the SA2 (sampled or nonsampled).}
    \label{fig:ss_map}
\end{wrapfigure}

The individual level survey data were obtained from the 2017-18 National Health Survey (NHS), which is an Australia-wide population-level health survey conducted every 3-4 years by the Australian Bureau of Statistics (ABS) \cite{RN478, RN598}. The survey aimed to collect a variety of health data on one adult and one child (where possible) in each selected household. Households were selected using a complex multistage design \cite{RN478}. Trained ABS interviewers conducted personal interviews with selected persons in each of the sampled households. To allow researchers to accommodate the complex sample design, the ABS provides survey weights, which we use in this analysis after applying the necessary rescalings (see \cref{sec:notation}). For the area level auxiliary data, we use data from the 2016 Australian census, represented as proportions. We obtained the Estimated Resident Population (ERP) stratified by age (15 years and above), sex and small area for both 2017 and 2018. In this study, the population counts were derived by averaging across the two years. Although the 2017-18 NHS collected data across Australia, we reduced our analysis to just those states on the east coast, both to ease the interpretation of visualizations and reduce the computation time.

We classify current smoking as daily, weekly, or less-than-weekly smokers, like the Social Health Atlas \cite{RN113}. We additionally enforce that all current smokers must have smoked at least 100 cigarettes in their lifetime. After exclusions, we have data for 10,918 respondents aged 15 years and above. The overall weighted prevalence of current smoking in our study region is 14.7\%. 

The goal of this analysis is to generate prevalence estimates for 1,630 small areas, of which 1,262 (77\%) were sampled. Of the sampled areas, 781 (62\%) gave stable direct estimates. Area level sample sizes range from 1 to 140, with a median of 7.

The small areas we use are derived from the 2016 Australia Statistical Geography Standard (ASGS), which is the geographic standard maintained by the ABS \cite{RN348}. The ASGS splits Australia into a hierarchical structure of areas that completely cover the country. We generate prevalence estimates at the statistical areas level 2 (SA2), which is the lowest level of the ASGS hierarchy for which detailed census population characteristics are publicly available. 

\subsection{Model details}

Variable selection is an intricate step of our approach in practice. Not only do we have two models for which variable selection must be performed, but variable selection decisions made at the second stage are dependent on the model fit at the first stage; an issue we do not tackle in the current paper. 

Given the extensive range of covariates available in the 2017-18 NHS, we used two phases of variable selection to mitigate the computational burden. First, as recommended by Goldstein \cite{RN553}, we employed frequentist inference methods (i.e. the \texttt{lme4} package \cite{RN573}) to identify an initial set of candidate fixed effects. Models were evaluated based on the AIC, BIC, and ALC, with a preference for lower AIC and BIC values. This frequentist stage served as an effective initial filter for variable selection.

Second, final decisions on covariates and random effects were made using Bayesian leave-one-out cross-validation (LOOCV) \cite{RN116}, which provided a Bayesian assessment of model fit with higher values preferred. The use of the \texttt{loo} package necessitated an approximation that disregarded the uncertainty from the stage 1 model.

Where possible the fixed and random effects for all the models were chosen according to the TSLN approach. That is, to enable fair comparisons, fixed and random effects for all models were as similar as possible. See Supplemental Materials B \cite{self_cite} for further details of model selection.  

\subsubsection{Individual level models} 
We used the following individual level categorical covariates in the TSLN-S1 model: age, sex and their interaction; registered marital status; high school completion status; Kessler psychological distress score; educational qualifications;  self-assessed health; and labor force status. We also used the following household level categorical covariates: number of daily smokers, tenure type, and whether there were Indigenous Australian household members. Along with the individual and household level covariates, we used some SA2-level contextual covariates including state, the Index of Relative Socio-Economic Disadvantage (IRSD) from the Socio-Economic Indexes for Areas (SEIFA) \cite{RN560}, and the following SA2-level demographic variables as proportions: occupation, Indigenous Australian status, income, unemployment and household composition. 

In addition to the individual and area level fixed effects, we introduced a random effect on a new categorical covariate, constructed from every unique combination of sex, age, and four binary-coded individual level risk factors (see Table 4 in the Supplemental Materials). The resulting covariate had 274 groups, with the number of participants in each category ranging from 1 to 299, with a median of 20. 

To further improve the predictive accuracy of the TSLN-S1 model and increase the ALC, the final component of the model was an individual level residual error term with a fixed standard deviation of 0.5. The chosen TSLN-S1 model had $\text{ALC} = 0.67$. 

Since census microdata was unavailable, we could not mirror the TSLN-S1 model complexity in the LOG model. We omitted the risk factor categorical random effect and were restricted to just three individual level covariates: age, sex and marital status. These three covariates gave a poststrata dataset with 146,700 rows. Note that we also omitted the individual level residual error term from the LOG model as, unlike the TSLN-S1 model, the LOG model must prioritize generalisability in order to perform well. 

Further details and definitions for the covariates used in the individual level models can be found in Supplemental Materials B \cite{self_cite}. 

\subsubsection{Area level models} 
For both the TSLN-S2 and ELN models, we utilized the following SA2-level covariates: IRSD, state and the first six principal components derived from SA2-level census proportion data. Similar to \cref{sec:gvf} we use GVFs where the log of the SA2-level sample sizes was the only predictor. Details on variable selection and the covariates used in the area level models can be found in Supplemental Materials D \cite{self_cite}.  

\subsubsection{Spatial random effects} \label{sec:sre}
Unlike in the simulation study where data were not generated with any spatial autocorrelation, we expect smoking prevalence to exhibit spatial clustering as smoking is generally higher in areas of lower socioeconomic status which can be geographical neighbors \cite{RN204}. There is a considerable collection of research that develops spatial SAE models \cite{RN91, chandra2018, chandra2017_gwr, Guhu2023}. Further, it has been shown that accommodating spatial structure in model-based SAE methods can provide considerable efficiency gains \cite{RN458, RN147, RN343, gao2023_sma, RN533}. 

To adjust for the spatial autocorrelation between areas and enforce global \emph{and} local smoothing, we use the BYM2 prior \cite{RN394} at the SA2-level for the TSLN-S2, ELN and LOG models (see Supplemental Materials D \cite{self_cite} for details). Although others have used conditional autoregressive (CAR) or simultaneous autoregressive (SAR) priors only, Gomez-Rubio \emph{et al.} \cite{RN35} conclude that models with just the CAR prior generally over-estimate the small area estimates and argue that including a structured and unstructured random effect provides a useful compromise between producing accurate small area estimates and their corresponding variances.

\begin{figure*}
    \centering
    \includegraphics[width=\columnwidth]{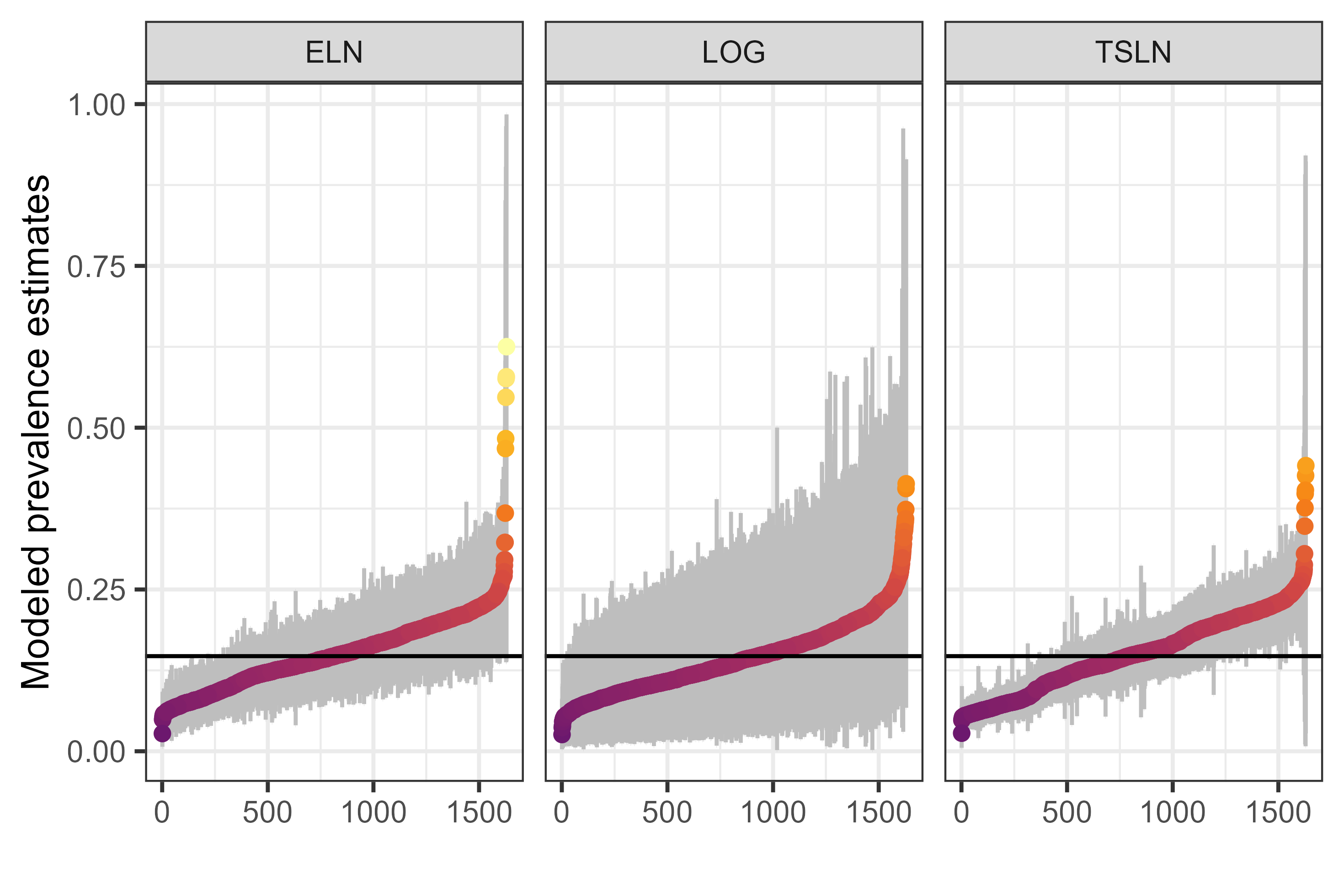}
    \caption{Comparison of the modeled SA2 level current smoking prevalence estimates from the three models. The caterpillar plots display the posterior medians and 95\% highest density intervals (grey bars) for all 1630 SA2s, ordered by their magnitude. The point colors mirror the $y$-axis and the vertical black line is the overall current smoking prevalence.}
    \label{fig:cat_prev_medianci}
\end{figure*}

\subsubsection{Priors and computation}
Most of the priors used in this case study mirror those specified in \cref{sec:sim}. We utilized a relatively informative prior for $\rho$, the mixing parameter of the BYM2 spatial prior which controls the amount of spatially structured as opposed to unstructured variation, where $\rho = 1$ gives a scaled intrinsic CAR prior \cite{RN394}. We use $\rho \sim $Beta$\lb{\text{shape} = 3.05, \text{rate} = 1.65}$, a prior which places roughly 45\% of density above $\rho = 0.7$. In this application, areas with survey data may have many neighbors but few with survey data, thus this informative prior ``encouraged'' the models to borrow information locally. The median number of neighbors is 7, whereas the median number that has sampled data is 5. This informative prior on $\rho$ slightly improved both model fit and predictive accuracy. The posterior median of $\rho$ was 0.89 under this informative prior, but 0.5 under a Uniform$(0,1)$ prior. 

We used $5000$ post-warmup draws for each of four chains in Stan \cite{RN452}, feeding a random subset of 500 posterior draws from the TSLN-S1 to the TSLN-S2 model. For storage reasons we thinned the draws by four, resulting in $5000$ useable posterior draws. Convergence of the models was assessed using $\widehat{R}$ \cite{RN499}, where a $\widehat{R} < 1.02$ was used as the cutoff for convergence for the parameters of interest, namely $\boldsymbol{\mu}$. We also explored trace plots and autocorrelation plots to verify convergence. All proportion parameters, $\boldsymbol{\mu}$, had effective sample sizes $>$500. 

\subsubsection{Benchmarking} \label{sec:bench}
To help validate our small area estimates, we utilized state-level estimates as internal benchmarks \cite{RN15}. There are four states on the east coast of Australia, with a median sample and population size of 3100 and 4.6 million, respectively. At this level of aggregation, the direct estimates are reliable. We employed inexact fully Bayesian benchmarking \cite{RN30}, which acts as a soft constraint on the model by penalizing discrepancies between the modeled state estimates and the direct state estimates. Unlike previous approaches that use posterior point estimates \cite{RN37}, Bayesian benchmarking directly includes the benchmarks in the joint posterior distribution, which accounts for benchmarking-induced uncertainty. We use Bayesian benchmarking in the TSLN-S2 and ELN models and exact benchmarking for the LOG model. Full details and a performance comparison with and without benchmarking is given in Supplemental Materials D \cite{self_cite}. 

\begin{figure*}
    \begin{subfigure}[h]{0.3\linewidth}
    \includegraphics[width=\linewidth]{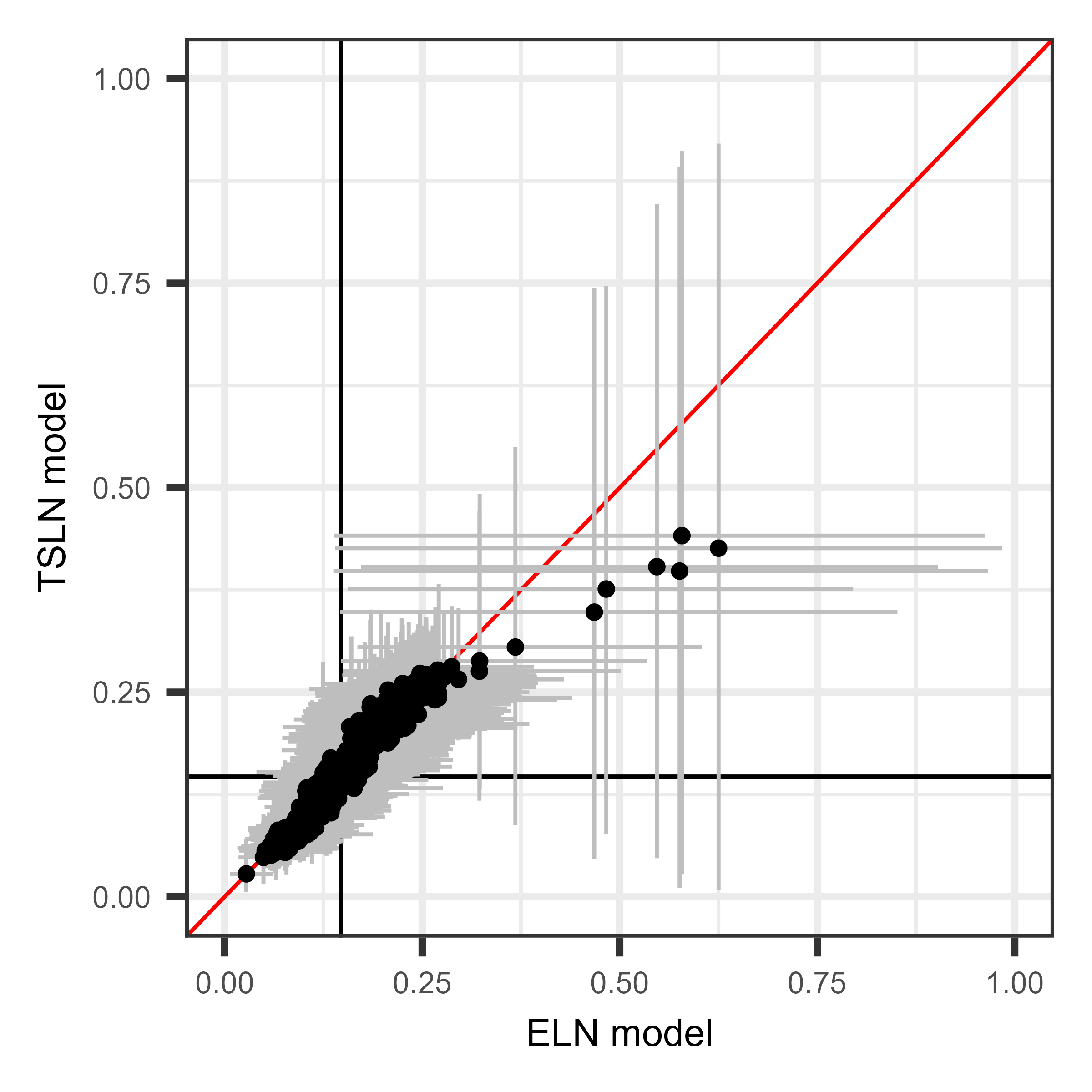}
    \caption{}
    \end{subfigure}
    \hfill
    \begin{subfigure}[h]{0.3\linewidth}
    \includegraphics[width=\linewidth]{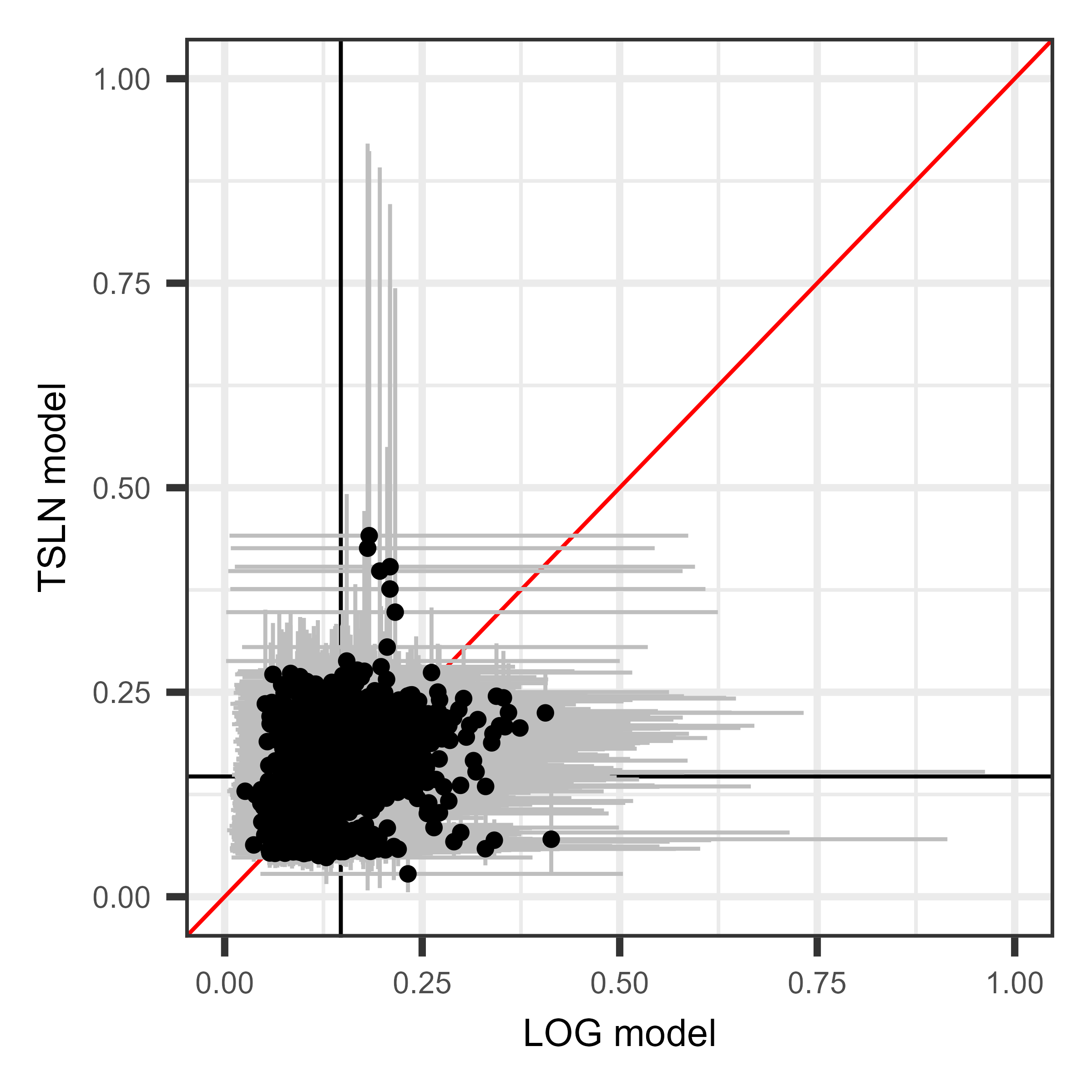}
    \caption{}
    \end{subfigure}
    \hfill
    \begin{subfigure}[h]{0.3\linewidth}
    \includegraphics[width=\linewidth]{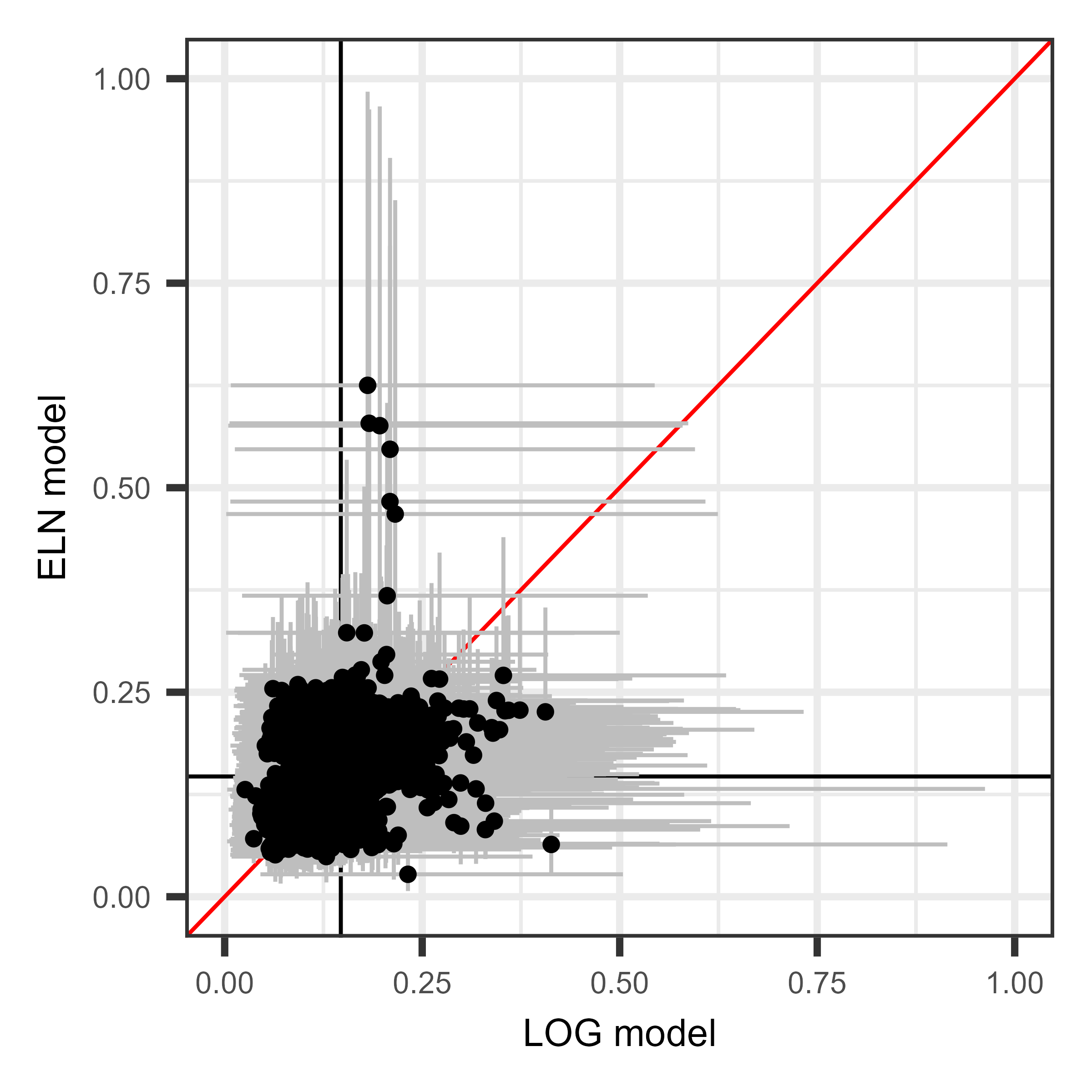}
    \caption{}
    \end{subfigure}%
    \caption{Comparison of the modeled estimates from the TSLN, LOG and ELN models. The scatter plots display the posterior medians (black points) and 95\% highest density intervals (gray bars) of the modeled SA2 level current smoking prevalence estimates, with the red lines denoting equivalence. The vertical and horizontal black lines are the overall current smoking prevalence.}
    \label{fig:scatter_TSLNvs}
\end{figure*}

\subsubsection{Visualisations}

We map both absolute and relative measures of current smoking. The absolute measures are the posterior medians and highest density intervals (HDIs) of the estimated prevalence from the models, while the relative measures rely on odds ratios (ORs). We derive ORs as follows,

\begin{equation}
    \widehat{\text{OR}}_i = \frac{\hat{\mu}_i/(1-\hat{\mu}_i)}{\hat{\mu}^D/(1-\hat{\mu}^D)}, \label{eq:or}
\end{equation}

\noindent where $\hat{\mu}^D$ is the overall direct estimate of current smoking prevalence. By deriving $\widehat{\text{OR}}_i$ for all posterior draws, we can map their posterior medians and HDIs. To quantify whether an OR is significantly different to 1, we use the exceedance probability (EP) \cite{RN115, RN460}, where $t = 1, \dots, T$ indexes the MCMC draws. 

\begin{equation}
    EP_i = \lb{\frac{1}{T} \sum_t \mathbb{I} \lb{\widehat{\text{OR}}_{it} > 1}} \label{eq:dpp}
\end{equation}

A high (low) $EP_i$ is interpreted as a high level of evidence that the OR for SA2 $i$ is significantly higher (lower) than 1. Generally an $EP_i$ above 0.8 or below 0.2 is considered significant \cite{RN396}.

\subsection{Results}

\begin{figure*}
    \centering
    \includegraphics[width=\columnwidth]{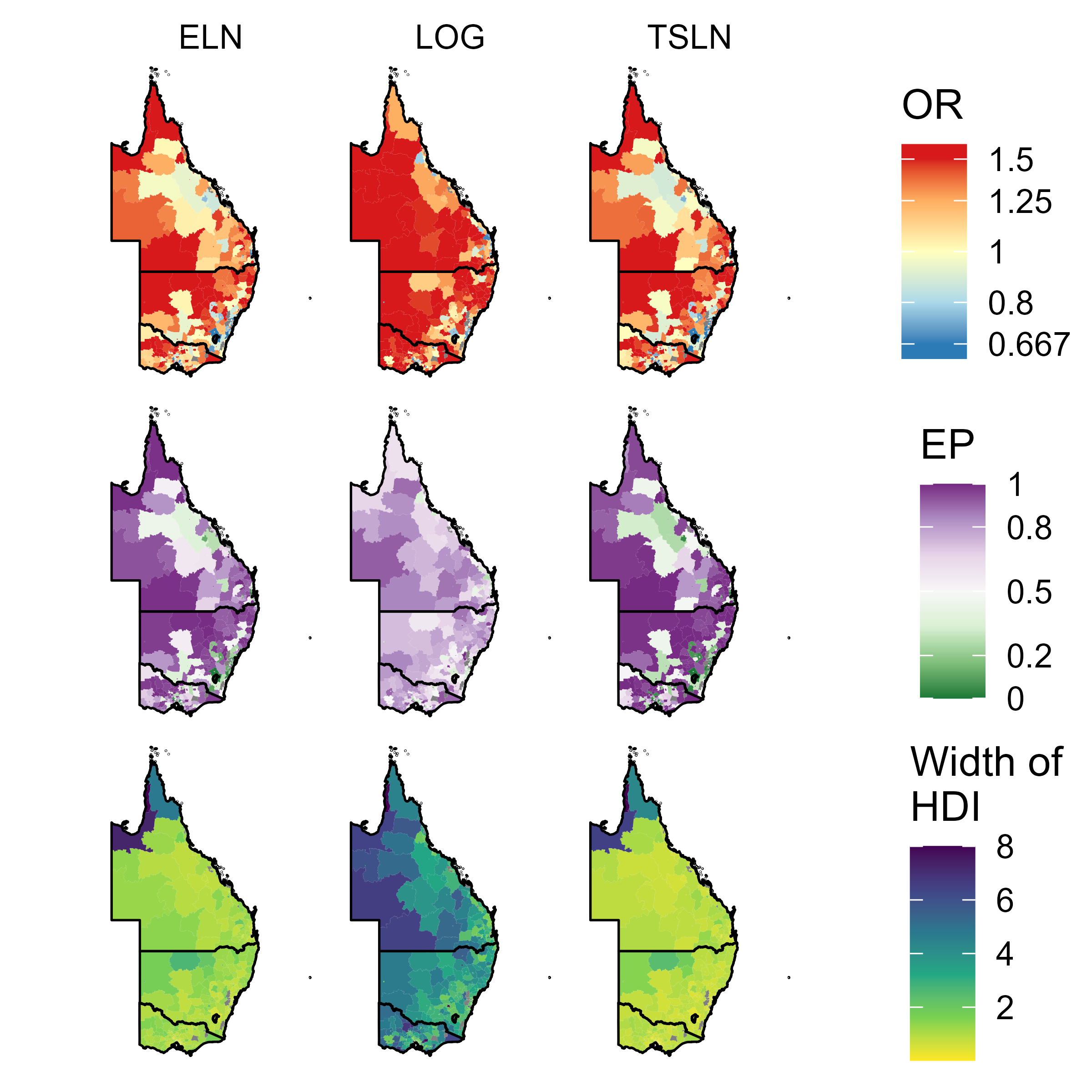}
    \caption{Choropleth maps displaying the modeled ORs for smoking prevalence in 1630 SA2s on the east coast of Australia. For each model, this figure displays the posterior medians (top row), exceedance probabilities ($EP$s) (middle row) and width of the 95\% HDI (bottom row) for the ORs. Note that some values are lower than the range of color scales shown --- for these values, the lowest color is shown. Gray areas were excluded from estimation due to having 2016-2017 ERPs smaller than or equal to 10. Black lines represent the boundaries of the four states on the east coast of Australia.}
    \label{fig:map_or}
\end{figure*}

\cref{fig:cat_prev_medianci} gives separate caterpillar plots for the SA2-level prevalence estimates and HDIs for the three models. Supporting this plot, \cref{fig:scatter_TSLNvs} compares the estimates from the TSLN approach to the ELN and LOG models. Both figures show the similarities in the modeled estimates from the two area level models and the superior interval sizes of the TSLN approach. The LOG model provides estimates with little correspondence to those from the TSLN or the ELN models. 

For areas with high prevalence the TSLN approach provides more conservative estimates than those from the ELN model; a result of the two stages of smoothing applied when using the TSLN approach. Given that the sparsity of the survey data results in very noisy and unstable direct estimates, in this case study we prefer estimates that are slightly over-smoothed rather than undersmoothed.  

The six outlying points visible in (a) of \cref{fig:scatter_TSLNvs} are SA2s in the Cape York (Northern section) region of Queensland. The high level of uncertainty for these SA2s is reasonable as they are remote, have small populations and are far from areas with survey data.  


In \cref{fig:map_or} we map the relative measures for all 1630 SA2s, including those without survey data. The figure displays posterior median ORs, HDIs and corresponding $EP$s for the three models. Observe that because the area level models (TSLN and ELN) generally provide more certainty in the estimates, the $EP$s are more extreme than those for the LOG model. The prevalence of current smoking is significantly higher than the overall prevalence in the western and northern parts of the region, while significantly lower in urban centers, such as Sydney (the most populous city on the east coast of Australia), Melbourne and Brisbane. \cref{fig:map_muonly_BriSydMel} gives the absolute prevalence estimates for these three cities. See Supplemental Materials D \cite{self_cite} for a plot stratifying the modeled estimates by socioeconomic status and maps of the absolute prevalence estimates from the models.  

By treating the direct estimates at a higher aggregation level as the truth, we compare the direct and modeled estimates at the statistical area level 4 using RRMSE, ARB, coverage and interval overlap. The details and possible limitations of these comparative performance results are given in Supplemental Materials D \cite{self_cite}. We found that the TSLN approach provided superior MRRMSE and interval overlap. Furthermore, the performance metrics for the TSLN approach show the smallest changes when we benchmark, providing more support for the validity of the estimates from the TSLN approach. Prevalence estimates from the LOG model, on the other hand, saw considerable changes when benchmarking was applied. We posit that the LOG model is providing poor estimates given the restricted set of individual level census covariates that were available, given the requirements of the LOG model.  

\begin{figure*}
    \centering
    \includegraphics[width=\columnwidth]{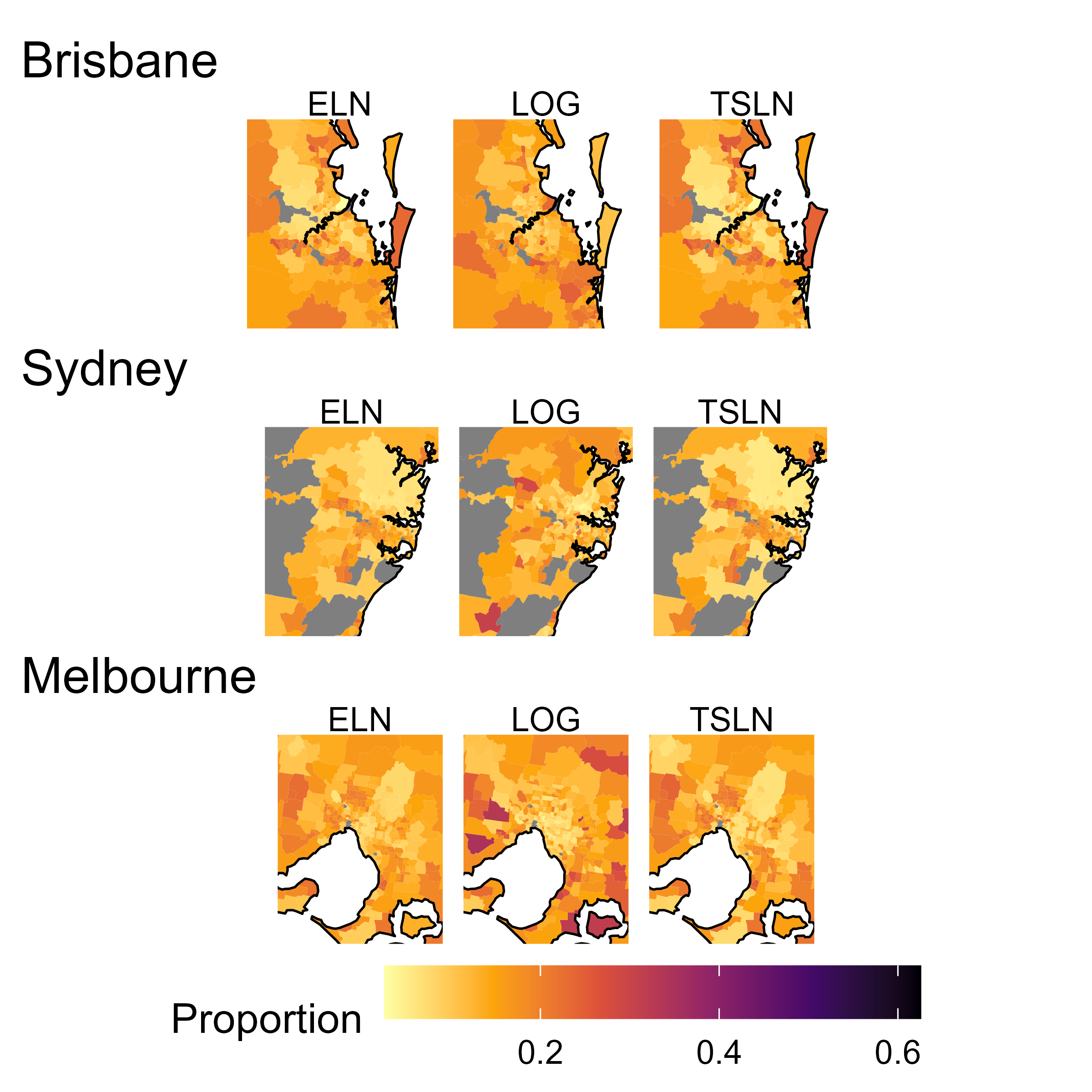}
    \caption{Choropleth inset maps displaying the modeled estimates of the proportion of current smokers in and around Sydney, Melbourne and Brisbane; three capital cities on the east coast of Australia. Gray areas were excluded from estimation due to having 2016-2017 ERPs smaller than or equal to 10. Most of these excluded areas are National parks, industrial areas, airports or cemeteries.}
    \label{fig:map_muonly_BriSydMel}
\end{figure*}

\section{Discussion}

We have proposed a new Bayesian model-based SAE approach to estimate proportions for small areas from sparse survey data. The TSLN approach is able to model these data by using both individual and area level models to reduce instability, alleviate covariate restrictions, and generate estimates for nonsampled areas. We have shown that the TSLN approach can provide superior proportion estimates for sampled and nonsampled areas compared to some alternatives, with similar or slightly more bias but much smaller variance; resulting in consistently smaller MRRMSEs and credible intervals. Compared with other available approaches, the TSLN approach appears to be the best option in sparse settings, such as when using the NHS in Australia. Moreover, we have demonstrated, along with others \cite{gao2023_sma, Das2022}, that the multi-stage or double smoothing properties of modeling at both the individual and area level are beneficial across a broad range of SAE applications.  

    
Although the simulation results and case study illustrate the potential benefits of the TSLN approach, some limitations motivate future theoretical and applied research in two-stage SAE. One pressing need is the development of statistical theory to motivate the use of double smoothing and two-stage approaches in general.

Our approach follows the independence assumptions proposed by Gao and Wakefield \cite{gao2023_sma}, but the potential for correlation among the S1 estimates warrants further investigation. Das \emph{et al.} \cite{Das2022} found that incorporating the covariances between area predictions did little to improve the bias or variance of their final estimates. Our own simulation study corroborated the low correlation between S1 estimates. We provide further details in Supplemental Material A. Future research could assess the validity and implications of this independence assumption. One option is to create a specialized MCMC algorithm, which would eliminate the need for these assumptions and allow one to fit the TSLN approach in a single step. Alternatively, our TSLN approach could be conceptualized as a modular model, by leveraging Bayesian cut distributions \cite{RN530}.  

We recognize that the GVF applied exclusively to unstable S1 sampling variances represents a unique solution to the undesirable properties mentioned in \cref{sec:gvf}. Thus, another area of exploration is the development of non-GVF solutions. Alternatively, methods to adjust for the instability of direct estimates could be formally compared. On a related note, while coverage was relatively stable for the TSLN approach, one could explore methods of conformal prediction \cite{RN531}, which can be used to guarantee frequentist coverage in small area estimation \cite{RN532}.

The Gaussian distribution assumption in the stage 2 model, although consistent with existing literature \cite{gao2023_sma}, presents another limitation. The empirical logistic transformation of direct estimates yields a small set of possible values if area level sample sizes are small. This issue is alleviated when using our S1 estimates, which aggregate probabilities rather than binary variables. Although the smoothing introduced by the stage 1 model makes the Gaussian assumption more plausible, especially under data sparsity, further theoretical work is required to determine when the assumption becomes inappropriate. We initially explored an alternative approach using a Beta likelihood, but encountered issues with MCMC convergence and unidentifiability, warranting future research. 

Another consideration is substantial clustering of the sampling, which results in the effective sample size of the survey being significantly lower than the nominal sample size. While the current study does not delve into the specific impacts of clustering, the stage 1 model is inherently flexible and is therefore capable of accommodating clustering effects similar to standard individual-level logistic models.

The TSLN approach is generic in its component models, which within an applied context allows researchers to extend our approach by using more flexible classes of models (such as semi or non-parametric models or even machine learning algorithms), as long as uncertainty can be captured. Performance could be improved by leveraging robust SAE for the component models, such as those developed by Chambers \emph{et al.} \cite{chambers2014outlier} or Liu and Lahiri \cite{liu2017}. 

Alternatively, the TSLN approach could be extended to the multivariate setting by using univariate stage 1 models, followed by a multivariate stage 2 model \cite{Guhu2023, arima2017multivariate}. Finally, further work is required to develop model selection tools for both stages of the TSLN approach, with a focus on the flow-on effect of model choices from the first stage to the second and how LOOCV can be better utilized in this setting. 

\section{Conclusion}\label{sec13}

Given that the need for higher resolution area level estimates is increasing faster than funding for larger surveys, methods of SAE must be capable of tackling sparsity issues. In this work, we have developed a solution to SAE for severely sparse data by leveraging both area and individual level models. Similar to other work \cite{gao2023_sma, Das2022}, this research represents another important step in continuing the positive narrative surrounding two-stage approaches and highlights their benefits and future avenues of research in SAE. As expressed by Fuglstad \emph{et al.} \cite{RN461}, ``$\dots$ the goal of the analysis should determine the approach, and different goals may call for different approaches.'' We envisage that our approach will allow practitioners to set more ambitious goals for their small area estimates in the future. 

\subsection*{Acknowledgments}

This study has received ethical approval from the Queensland University of Technology Human Research Ethics Committee (Project ID: 4609) for the project entitled ``Statistical methods for small area estimation of cancer risk factors and their associations with cancer incidence''. 

We thank the Australian Bureau of Statistics (ABS) for designing and collecting the National Health Survey data and making it available for analysis in the DataLab. The views expressed in this paper are those of the authors and do not necessarily reflect the policy of QUT, CCQ or the ABS.

\subsection*{Competing interests}
The authors declare that they have no competing interests.

\subsection*{Funding}
JH was supported by the Queensland University of Technology (QUT) Centre for Data Science and Cancer Council QLD (CCQ) Scholarship. SC receives salary and research support from a National Health and Medical Research Council Investigator Grant (\#2008313).

\subsection*{Supplemental materials}
The supplemental material mentioned throughout this work can be found on at the end of this document. This additional material includes further details, plots and results to accompany Sections 2-5 of this paper.

\bibliographystyle{unsrtnat}
\bibliography{ref}  

\end{document}


\maketitle

\setcounter{section}{0}
\setcounter{subsection}{0}
\renewcommand{\thesection}{\Alph{section}}
\renewcommand{\thesubsection}{\thesection.\arabic{subsection}}

\section{Proposed approach} \label{supp:A}

\subsection{Definition for S1 sampling variance} \label{supp:var_psi}


The sampling variance for the S1 estimate is lower bounded by the direct estimate sampling variance. Initially let, 

\begin{equation*}
    \frac{\sum_{j \in r_i} w_{ij} y_{ij}}{n_i} = \frac{\sum_{j \in r_i} w_{ij} y_{ij}}{n_i} + \frac{\sum_{j \in r_i} w_{ij} p_{ij}}{n_i} - \frac{\sum_{j \in r_i} w_{ij} p_{ij}}{n_i},
\end{equation*}

which can be rewritten as, 

\begin{equation*}
    \hat{\mu}_i^{\text{S1}} = \hat{\mu}_i^D + \hat{B}_i,
\end{equation*}

where 

\begin{equation*}
    \hat{B}_i = \frac{\sum_{j \in r_i} w_{ij} (p_{ij}-y_{ij})}{n_i},
\end{equation*}

is the difference between $\hat{\mu}_i^{\text{S1}}$ and $\hat{\mu}_i^D$. Thus, by treating both $y_{ij}$ and $w_{ij}$ as fixed quantities and by definition assuming $\text{cov}\left( \hat{\mu}_i^D, \hat{B}_i \right) = 0$, the sampling variance of the S1 estimator is given by,

\begin{equation*}
    \widehat{\text{v}}\left( \hat{\mu}_i^{\text{S1}} \right) \leq \widehat{\text{v}}\left( \hat{\mu}_i^D \right) + \widehat{\text{v}}\left( \hat{B}_i \right).
\end{equation*}

The $\leq$ symbol is used as empirical results suggested that without assuming $y_{ij}$ as fixed the $\text{cov}\left( \hat{\mu}_i^D, \hat{B}_i \right) < 0$. 

The specification for $\widehat{\text{v}}\left( \hat{\mu}_i^{\text{S1}} \right)$ ensures that poorly specified logistic models will give large sampling variances, because $\widehat{\text{v}}\left( \hat{B}_i \right)$ will be large. On the other hand, as $p_{ij}$ tends to $y_{ij}$, $\widehat{\text{v}}\left( \hat{\mu}_i^{\text{S1}} \right) \approx \widehat{\text{v}}\left( \hat{\mu}_i^D \right)$.  

\subsection{Generalized variance functions (GVF)} \label{supp:gvf}

By letting $f(.)$ be an appropriate link function, $\mathbf{L}$ an area level design matrix, and $\boldsymbol{\omega}$ the corresponding regression coefficients, 

\begin{equation}
    f \left( \bar{\gamma}_i^{\text{S1}} \right) \sim N\left( \mathbf{L}_i \boldsymbol{\omega}, \sigma_{\text{gvf}}^2 \right), \label{eq:gvf_norm}
\end{equation}

\noindent where $i=1,\dots, m_s$. We used a $\text{Cauchy}\lb{0,2}^{+}$ prior for $\sigma_{\text{gvf}}$ and $N(0,2)$ prior for $\boldsymbol{\omega}$. 

The S1 sampling variances for $i=m_s+1,\dots, m$ are imputed by

\begin{equation}
    \bar{\gamma}_i^{\text{S1}} = f^{-1}\left( \mathbf{L}_i \boldsymbol{\omega} \right). \label{eq:gvf_pred}
\end{equation}

The general form specified in \eqref{eq:gvf_norm} can be adopted according to the scale of the sampling variances and covariates. Note that when $f(x) = \text{log}(x)$, we use the following bias correction of Das \emph{et al.} \cite{Das2022} when back-transforming, $\bar{\gamma}_i^{\text{S1}} = \text{exp}\left( \mathbf{L}_i \boldsymbol{\omega} + 0.5 \sigma_{\text{gvf}}^2 \right)$. For the TSLN-S2 model, we set $f(x) = \text{log}\left(\sqrt{x}\right)$ and use $\text{log}(n_i)$ as the single covariate. 

\subsection{Accommodating uncertainty of stage 1 model} \label{supp:alt_inf}

The specification given in Section 2.2.4 of the main paper can be viewed as a type of FH model where the variance of the sampling model is an independent combination of sampling and measurement (i.e. model) error. Thus, Eq. 9 and 10 from the main paper can be collapsed into 

\begin{equation*}
    \hat{\boldsymbol{\theta}}^{\text{S1}, \text{its}}_{i} \sim N\left( \hat{\theta}_i, \bar{\gamma}_i^{\text{S1}} + \widehat{\text{v}}\left( \hat{\theta}_i^{\text{S1}} \right) \right).
\end{equation*}

There are simpler and more complex approaches to estimating the parameters of our TSLN approach. The solution of Das \emph{et al.} \cite{RN479}\cmmnt{textcite} and Gao and Wakefield \cite{gao2023_sma}\cmmnt{textcite} was to treat the posterior means for $\hat{\theta}_i^{\text{S1}}$ and $\gamma_i^{\text{S1}}$ from the stage 1 model as known quantities in the stage 2 model. However, this approach neglects the uncertainty in the first-stage logistic model. In a more complex model setup, we would also add $\boldsymbol{\gamma}^{\text{S1}, \text{its}}_{i} \sim N\left( \bar{\gamma}_i^{\text{S1}}, \tilde{\text{v}}\left( \gamma_i^{\text{S1}} \right) \right)$, rather than treating $\bar{\gamma}_i^{\text{S1}}$ as a fixed quantity. 

Further extensions could incorporate the correlation structure between posterior draws for $\hat{\theta}_i^{\text{S1}}$ and $\gamma_i^{\text{S1}}$. One could initially treat the diagonal elements of the variance-covariance matrices, $\sum \in \mathbb{R}^{m \times m}$, as fixed, but estimate a constant correlation between all areas (i.e. use an exchangeable correlation structure). A further step would be to consider an unstructured correlation matrix, by using the empirical variance-covariance matrices, $\widetilde{\sum} \in \mathbb{R}^{m \times m}$, and a multivariate normal distribution. In this case, 

\begin{equation*}
    \hat{\boldsymbol{\theta}}^{\text{S1}, \text{its}}_{t} = \lb{\hat{\theta}^{\text{S1}, \text{its}}_{it},\dots, \hat{\theta}^{\text{S1}, \text{its}}_{Mt}} \sim \text{MVN}\left( \hat{\bar{\boldsymbol{\theta}}}, \widetilde{\sum} \right), 
\end{equation*}

\noindent where $\hat{\bar{\boldsymbol{\theta}}} = \left( \hat{\bar{\theta}}_1,\dots, \hat{\bar{\theta}}_m \right)$. Note that Das \emph{et al.} \cite{RN479}\cmmnt{textcite} found that incorporating the covariances between area predictions did little to improve the bias or variance of their final estimates. In our simulation study, we found little correlation between the posterior draws for different areas. 

\newpage
\section{Existing methods} \label{supp:B}

\subsection{Seminal model-based methods}

\paragraph{Fay-Herriot (FH) model} \label{supp:fh}

Let $\hat{\theta}^D_i$ and $\gamma^{D}_i$ be the known (and fixed) area level direct estimates and sampling variances for areas $i = 1,\dots, m$ and $\mathbf{Z} \in \mathbb{R}^{M \times (q^a + 1)}$ be the area level design matrix with corresponding regression coefficients, $\boldsymbol{\lambda} \in \mathbb{R}^{(q^a + 1) \times 1}$, for the $q^a$ area level covariates. A fully Bayesian FH model is given by, 

\begin{eqnarray}
    \hat{\theta}^D_i & \sim & N(\hat{\theta}_i, \gamma^D_i) \quad\quad i = 1,\dots, m \label{eq:fh_sampling}
    \\
    \hat{\theta}_i & = & \mathbf{Z}_{i} \boldsymbol{\lambda} + v_i \quad\quad i = 1,\dots, M \label{eq:fh_linking}
    \\
    v_i & \sim & N(0, \sigma_v^2) \quad\quad i = 1,\dots, M \nonumber
\end{eqnarray}

where independent and uninformative priors are used for the other model parameters ($\boldsymbol{\lambda}, \sigma_v$). The small area estimate for area $i$ is given by the marginal posterior of $\theta_i$. By construction, the FH model gives $\hat{\theta_i}$, the synthetic estimate in \eqref{eq:fh_linking}, when $\gamma_i^D$ is very large, but $\hat{\theta}_i^D$ when $\gamma_i^D$ is very small. 

\paragraph{Nested error (NER) model} \label{supp:ner}

With access to detailed individual level continuous outcome data, $\delta_{ij}$, individual level models become another avenue for SAE \cite{RN48}. The Bayesian nested error (NER) model is \cite{RN28},

\begin{eqnarray}
    \delta_{ij} & \sim & N\lb{\mathbf{x}_{ij} \boldsymbol{\beta} + e_i, \sigma_r^2} \quad\quad j = 1,\dots, n_i; i = 1,\dots, m \label{eq:ner}
    \\
    e_i & \sim & N\lb{0, \sigma_e^2} \quad\quad i = 1,\dots, M. \nonumber
\end{eqnarray}

Independent and uninformative priors are used for the model parameters. By assuming that $N_i$ is large \cite{RN150}, the empirical best linear unbiased predictor (EBLUP) is calculated as, 

\begin{equation*}
    \hat{\theta}_i = \bar{\mathbf{X}}_i \boldsymbol{\beta} + e_i, \quad\quad i = 1,\dots, M \nonumber
\end{equation*}

where we distinguish between the design matrix, $\mathbf{x} \in \mathbb{R}^{n \times (q^u + 1)}$, for the $q^u$ individual level covariates and the area level means for the same covariates, $\bar{\mathbf{X}} \in \mathbb{R}^{M \times (q^u + 1)}$. Note that $n$ is the sample size of the survey.  

Unlike the Bayesian FH model (\eqref{eq:fh_sampling}), the NER \eqref{eq:ner}, does not incorporate the individual level sampling weights and is thus not design-unbiased \cite{RN138}. Although individual level models have been shown to outperform area level models \cite{RN137}, the core limitation of \eqref{eq:ner} is that one must have access to the individual level and known population means for all $q^u$ covariates; restricting one to census covariates only.  

\subsection{Area level models for proportions} \label{supp:area}

\paragraph{Normal-logit model}

A simple adaption of the FH model (\eqref{eq:fh_sampling}) for proportions is achieved by accommodating the bounds of the direct proportion estimates in the link function. Liu, Lahiri, and Kalton \cite{RN400}\cmmnt{textcite} describe this approach as a normal-logit model. 

\begin{eqnarray}
    \hat{\mu}^D_i & \sim & N(\hat{\mu}_i, \psi^D_i) \quad\quad i = 1,\dots, M \label{eq:nl}
    \\
    \text{logit}(\hat{\mu}_i) & = & \mathbf{Z}_i \boldsymbol{\lambda} + v_i \quad\quad i = 1,\dots, M \nonumber
    \\
    v_i & \sim & N(0, \sigma_v^2) \quad\quad i=1,\dots, M \nonumber
\end{eqnarray}

Liu, Lahiri, and Kalton \cite{RN400}\cmmnt{textcite} argue that \eqref{eq:nl} is applicable only when $n_i$ is sufficiently large, resulting in small sample variances, $\psi^D_i$, and approximate normality of $\hat{\mu}^D_i$. 

\paragraph{Empirical logistic-normal model}

By targeting the boundary issues with the normal-logit model (\eqref{eq:nl}), Mercer \emph{et al.} \cite{RN476}\cmmnt{textcite} and then Cassy \emph{et al.} \cite{RN552}\cmmnt{textcite} used an empirical logit transformation of the direct proportion estimates.

\begin{eqnarray}
    \text{logit} \left( \hat{\mu}_i^D \right) & \sim & N \lb{ \text{logit} \left( \hat{\mu}_i \right), \gamma_i^D }  \quad\quad i=1,\dots, m \label{eq:eln}
    \\
    \text{logit} \left( \hat{\mu}_i \right) & = & \mathbf{Z}_i \boldsymbol{\lambda} + v_i \quad\quad i=1,\dots, M \nonumber
    \\
    \gamma_i^D  & = & \psi_i^D \left[ \hat{\mu}_i^D \lb{1-\hat{\mu}_i^D} \right]^{-2} \quad\quad i=1,\dots, m \nonumber
    \\
    v_i & \sim & N(0, \sigma_v^2) \quad\quad i=1,\dots, M \nonumber
\end{eqnarray}

The small area estimate for area $i$ is given by $\hat{\mu}_i$. 

\paragraph{Binomial model}

An alternative approach to the normal logit model is to model the area level sample counts, $\widehat{Y}_{i} = \sum_{j \in s_i} y_{ij}$, with a binomial distribution \cite{RN461, RN87, RN460}, 

\begin{equation*}
    \widehat{Y}_{i} \sim \text{Binomial}(n_i, \hat{\mu}_i),
\end{equation*}

\noindent where the linear predictor and random effects are the same as those in \eqref{eq:nl}.  

Although the binomial model does not automatically accommodate the sample design, interested readers are referred to Vandendijck \emph{et al.} \cite{RN147}\cmmnt{textcite} and Chen, Wakefield, and Lumely \cite{RN497}\cmmnt{textcite} for binomial models that do. These authors adjust both $\widehat{Y}_{i}$ and ${n}_i$ according to the sample design to derive the effective number of counts $\widetilde{Y}_{i}$ and effective sample size, $\tilde{n}_i$. In practice however their method results in non-integer values for $\widetilde{Y}_{i}$ and $\tilde{n}_i$. Consequently, to implement this model within a Bayesian framework, particularly in probabilistic programming languages like Stan, it becomes necessary to extend the traditional discrete binomial distribution to accommodate these non-integer values.

\paragraph{Beta model} \label{supp:beta}

The Beta distribution, which is naturally bounded between 0 and 1, is a favorable choice for modeling small area proportions \cite{RN58,RN400, RN408}. Although formally parameterized by two parameters (i.e. $\kappa^{(1)}$ and $\kappa^{(2)}$) that control the shape and scale, it is common to adopt the following mean-precision parameterization \cite{RN401} where $\kappa^{(1)} = \phi \mu$ and $\kappa^{(2)} = \phi - \phi \mu = \phi - \kappa^{(1)}$. The parameter $\mu \in (0,1)$ is the mean of the bounded outcome, whilst $\phi > 0$ can be interpreted as a precision parameter. Hence for $Z \in (0,1)$, note that

\begin{eqnarray*}
    \text{E}[Z] & = & \frac{\kappa^{(1)}}{\kappa^{(1)}+\kappa^{(2)}} = \mu
    \\
    \text{Var}[Z] & = & \frac{\kappa^{(1)} \kappa^{(2)}}{(\kappa^{(1)} + \kappa^{(2)})^2 (\kappa^{(1)} + \kappa^{(2)} +1)} = \frac{\mu (1- \mu)}{\phi + 1}.
\end{eqnarray*}

Following the specification of the FH model \cite{RN54}, $\phi_i$ is considered known. 

\begin{equation}
     \phi_i = \frac{\hat{\mu}_i (1-\hat{\mu}_i)}{\psi^D_i} - 1
\end{equation}

The Bayesian area level FH Beta model follows directly,

\begin{eqnarray}
    \hat{\mu}^D_i & \sim & \text{Beta} \left( \kappa^{(1)}_i = \hat{\mu}_i \phi_i, \kappa^{(2)}_i =  \phi_i - \hat{\mu}_i \phi_i \right) \quad\quad i=1,\dots, m \label{eq:fh_beta}
    \\
    \text{logit}(\hat{\mu}_i) & = & \mathbf{Z}_{i} \boldsymbol{\lambda} + v_i \quad\quad i=1,\dots, M \nonumber
    \\
    v_i & \sim & N(0, \sigma_v^2) \quad\quad i=1,\dots, M \nonumber,
\end{eqnarray}

where the proportion estimate for area $i$ is given by $\hat{\mu}_i$. 

Unlike other implementations of FH Beta models \cite{RN525, RN408, liu2014}, we specify the shape parameters in terms of $\phi_i$ directly rather than the effective sample size, $\tilde{n}_i = \phi_i + 1$. Below we show how these implementations are equivalent. 

\begin{eqnarray*}
    \hat{v}_{\text{srs}} \left( \hat{\mu}_i^D \right) & = & \frac{\hat{\mu}_i^D \left( 1 - \hat{\mu}_i^D \right)}{n_i} 
    \\
    \text{deff}_i & = & \frac{\psi_i^D}{\hat{v}_{\text{srs}} \left( \hat{\mu}_i^D \right)}
    \\
    \phi_i + 1 & = & \frac{n_i}{\text{deff}_i},
    \\
    \phi_i + 1 & = & \frac{\left(\frac{n_i}{1}\right)}{\left(\frac{\psi_i^D}{\left(\frac{\hat{\mu}_i^D \left( 1 - \hat{\mu}_i^D \right)}{n_i}\right)}\right)} = \frac{n_i \left(\frac{\hat{\mu}_i^D \left( 1 - \hat{\mu}_i^D \right)}{n_i}\right)}{\psi_i^D}
    \\
    \therefore \phi_i & = & \frac{\hat{\mu}_i^D \left( 1 - \hat{\mu}_i^D \right)}{\psi_i^D} - 1
\end{eqnarray*}

Unfortunately, the FH Beta model has several statistical and computational limitations, some of which are summarized in Table 1\cmmnt{cref_table:comp_models} of the main paper. 

The first is a constraint on the mean of the Beta distribution. In order to ensure that the shape parameters remain strictly positive,

\begin{equation}
    \hat{\mu}_i \in \left( \frac{1-\sqrt{1-4 \psi^D_i}}{2}, \frac{1+\sqrt{1-4 \psi^D_i}}{2} \right). \label{eq:beta_mu_bounds}
\end{equation}

Very imprecise or unstable direct estimates can give biased posterior distributions for $\hat{\mu}_i$ since its range tends to $0$ as $\psi^D_i$ tends to $0.25$. 

As a result of the first limitation, the bounds on $\hat{\mu}_i$ must be applied to all $M$ areas to ensure consistency between estimates for both sampled and nonsampled areas. One requires a version of the GVF in \cref{eq:gvf_norm} that can impute sampling variances for all $M$ --- even those areas with no data. Arguably it makes little sense to estimate sampling variances for areas with no sample data, even if this is only for computational reasons. In this work, we assume that $n_i \propto N_i$, and then use $\text{log}(N_i)$ as the single covariate in $\mathbf{L}$. We also set $f(x) = \text{log}\left( \frac{x}{0.25 + x} \right)$, which respects the constraint $\psi_i^D \leq 0.25$ imposed by the Beta distribution. In practice, the selection of covariates for a GVF (of the form in \cref{eq:gvf_norm}) is limited to those available across all areas. This limitation prevents the inclusion of sample sizes and direct estimates, which are anticipated to be predictive of the sampling variances. 

A computational limitation of the Beta model is its possible bimodal behavior; a disastrous affair for standard MCMC methods. A constraint $\kappa^{(1)},\kappa^{(2)}>1$ must be imposed to ensure the Beta distribution remains unimodal. However, in the case of the FH Beta model, this constraint places a very restrictive upper limit on the values of $\psi_i^D$. Any areas with large sampling variances generally produce bimodal Beta distributions and thus cannot be accommodated into the FH Beta model. To improve convergence, we constrain $\hat{\mu}_i \in (0.03, 0.97)$. This is a valid computational constraint and is unlikely to affect model performance.

A final limitation is that the FH Beta model is particularly affected by unstable direct estimates because the likelihood for the Beta distribution becomes undefined if $\hat{\mu}_i^D$ is exactly equal to $0$ or $1$.

\subsection{Individual level models for proportions} \label{supp:unit}

\paragraph{Pseudo-likelihood logistic mixed model}
To extend the NER model to the binary setting and accommodate the sample design, one can use an individual level Bayesian pseudo-likelihood logistic mixed model \cite{RN500, RN44}. 

\begin{eqnarray}
    y_{ij} & \sim & \text{Bernoulli}(p_{ij})^{\tilde{w}_{ij}} \quad\quad j = 1,\dots, n_i, i = 1,\dots, m \label{eq:pseudo_LOG}
    \\
    \text{logit}(p_{ij}) & = & \mathbf{x}_{ij} \boldsymbol{\beta} + e_i \quad\quad j = 1,\dots, N_i, i = 1,\dots, M \nonumber
\end{eqnarray}

Because of the nonlinear link function, the area level proportion estimate is derived by aggregating across the observed and unobserved individuals in each small area,  

\begin{equation}
    \hat{\mu}_i = \frac{1}{N_i} \left( \sum_{j \in r_i} y_{ij} + \sum_{j \in r_i^{C}} p_{ij} \right), \label{eq:LOG_estimator1}
\end{equation}

where $p_{ij}$ is estimated using the posterior distributions of the model parameters, $\boldsymbol{\beta}, \sigma_e$, for $j \in r_i^{C}$.

The motivation for Bayesian pseudo-likelihood is to ensure that the posterior distribution is similar (at least asymptotically) with that from the same specified model fit to the entire population \cite{RN501}. For an arbitrary outcome vector $\mathbf{y}$, sampling weights, $\mathbf{w}$, and model parameters, $\boldsymbol{\theta}$, the posterior distribution is specified as 

\begin{equation*}
    p\left( \boldsymbol{\theta} | \mathbf{y} \right) \propto \left[ \prod_{i=1}^N p\left( y_i | \boldsymbol{\theta} \right)^{w_i} \right] p\left( \boldsymbol{\theta} \right),
\end{equation*}

where $\prod_{i=1}^N p\left( y_i | \boldsymbol{\theta} \right)^{w_i}$ denotes the pseudo likelihood for the sample.

\newpage
\section{Simulation study} \label{supp:C}

\subsection{Algorithm} \label{supp:sim_alg}

Below we give more details on the simulation study described in Section 4\cmmnt{cref_sec:sim}. We generated a census using steps one to four below and then drew $D = 500$ unique samples (repetitions) from this census using step five.

\textbf{Step one } Create a $M$-length vector of area level proportions, $\mathbf{U}$, with values equally spaced between $L$ and $U$ (e.g. $\mathbf{U} = \left( U_1 = 0.1,\dots, U_M = 0.4 \right)$). Sample a random vector of area specific population sizes, $\mathbf{N} = (N_1,\dots, N_M)$ from the set $\{500, 3000\}$. Note that $N = \sum_{i=1}^M N_i$. Next, using a binomial distribution sample the area counts, $Y_i \sim \text{Binomial}(n = N_i, p = U_i)$.  Finally, \emph{uncount} $Y_i$ to create the binary outcome $y_{ij} \in \{0,1\}$ for individual $j$ in area $i$. For example, in area 1 the vector $\mathbf{y}_1 = (y_{11},\dots, y_{1N_1})$, will be composed of $Y_1$ 1's and $N_1 - Y_1$ 0's. 

\textbf{Step two } To simulate individual level covariates (one survey-only categorical covariate, $\mathbf{x}^{\text{survey}}$ with three groups, and one continuous covariate, $\mathbf{x}^{\text{census}}$ available in both the census and survey), first sample two standard normal vectors of length $N$, denoted $\mathbf{e}^{\text{survey}}, \mathbf{e}^{\text{census}}$. Then calculate the following two continuous covariates, $x^{\text{survey}}_{*,ij} = y_{ij} + \alpha^{\text{survey}} e^{\text{survey}}_{ij}$ and $x^{\text{census}}_{*,ij} = y_{ij} + \alpha^{\text{census}} e^{\text{census}}_{ij}$, where $\alpha^{\text{survey}}$ and $\alpha^{\text{census}}$ control the predictive power of the individual level covariates; small values of provide a greater correlation between the outcome and the individual level covariates. Finally, convert $\mathbf{x}^{\text{survey}}_{*}$ into a categorical covariate, $\mathbf{x}^{\text{survey}}$, using appropriate quantiles and standardize $\mathbf{x}^{\text{census}}_{*}$ to create $\mathbf{x}^{\text{census}}$.

\textbf{Step three } To generate an area level covariate first calculate the true area level proportions, 

\begin{equation*}
    \boldsymbol{\mu} = \lb{ \mu_1 = \frac{1}{N_1} \sum_{j=1}^{N_1} y_{1j}, \dots, \mu_{100} = \frac{1}{N_{100}} \sum_{j=1}^{N_{100}} y_{100j} }.
\end{equation*}

Note that $\boldsymbol{\mu}$ is constant for all 500 repetitions. Following a similar method to Step 2, simulate a random standard normal vector of length $M$, denoted $\mathbf{g}$, and calculate a continuous covariate, $Z_{*,i} = \text{logit}(\mu_{i}) + u g_{i}$, where $u$ controls the predictive power of the area level covariate (similar to $\alpha^{\text{survey}}$ and $\alpha^{\text{census}}$ above). As before, we standardize $\mathbf{Z}_{*}$ to create $\mathbf{Z}$, and then expand this vector and include it in the census dataset. The simulated census (of size $N \times 5$) has columns $\mathbf{y}, \mathbf{x}^{\text{survey}}, \mathbf{x}^{\text{census}}, \mathbf{Z}$, and $\mathbf{I}$, where $\mathbf{I} \in \{1,\dots, M\}$ is an area identifier.

\textbf{Step four } We fix the sampling fraction at 0.4\% and $m = 60$, and then calculate the fixed area sample sizes, $n_i = \text{round} \left( \frac{100}{60} \times 0.004 \times N_i \right)$. Next, by following the simulation method used by Hidiroglou and You \cite{RN137}\cmmnt{textcite}, we simulate $z_{ij} = \mathbb{I}\left(y_{ij} = 0 \right) + 0.8 h_{ij}$ for all individuals, where $h_{ij}$ is a random draw from an exponential distribution with rate equal to $1$. The values of $z_{ij}$ are used to determine each individual's sampling probability, $\pi_{ij} = z_{ij} \left( \sum_{j=1}^{N_i} z_{ij} \right)^{-1}$, and sampling weight, $w_{ij} = n_i^{-1} \pi_{ij}^{-1}$, based on the fixed area sample size. The calculation of $\pi_{ij}$ makes individuals with $y_{ij} = 0$ more likely to be sampled (i.e. the sampled design is informative). 

\textbf{Step five } Select 60 out of the 100 areas proportional to their size (i.e. randomly select areas according to $\frac{N_i}{N}$). Within each selected area, draw an informative sample of size $n_i$ based on the sampling probabilities, $\pi_{ij}$. Finally, rescale the sampling weights to ensure that the sum within area $i$ equals $N_i$.

The simulation algorithm was purposely stochastic and thus complete control via the various simulation parameters was not definite. This is particularly true for the covariate effect sizes. We conducted an extensive grid search across simulation parameter values (using crude but fast frequentist methods) to determine the values that gave reasonable model coefficients. In this work, we found that $\alpha^{\text{census}} = 1$ gave reasonable coefficients. Although $N_i$ and $n_i$ were fixed, the areas to be sampled and which individuals were sampled in each selected area were stochastic, resulting in different $n$ for each repetition $d$. Note that $u$, which has an inverse effect on the predictive power of the area level covariate, was set to 0.05 for Sc1-Sc2 and 0.01 for Sc3-Sc6 to ensure $\mathbf{Z}$, the area level covariate, was sufficiently predictive.

\subsection{Table 4 details} \label{supp:sim_summary}

The third column in Table 4\cmmnt{cref_table:sim_summary} is derived as follows. For each repetition, we calculate the median of the per-area ratio of the S1 and direct sampling variances using 
\begin{equation*}
    100\lb{\lb{\frac{\bar{\gamma}_{id}^{\text{S1}} + \tilde{\text{v}}\left( \hat{\theta}_{id}^{\text{S1}} \right)}{\gamma^D_{id}}} - 1}, 
\end{equation*}
which is expected to be greater than 0 given that $\bar{\gamma}_{id}^{\text{S1}} > \gamma^D_{id}$. 

The fourth column is derived as follows. For each repetition, we derive the ratio of the mean absolute bias (MAB),
\begin{eqnarray*}
    \text{MAB}^D_d & = & \frac{1}{M} \sum_i |\hat{\mu}^D_{id}-\mu_i|
    \\
    \text{MAB}^{\text{S1}}_d & = & \frac{1}{M} \sum_i |\bar{\hat{\mu}}^{\text{S1}}_{id}-\mu_i|
    \\
    \text{Ratio}_d & = & 100 \lb{1-\frac{\text{MAB}_d^{\text{S1}}}{\text{MAB}^D_d}}
\end{eqnarray*}

for the direct and S1 estimates, where $\bar{\hat{\mu}}^{\text{S1}}_{id}$ is the posterior median of $\hat{\mu}^{\text{S1}}_{id}$. In Table 4\cmmnt{cref_table:sim_summary} we summarise the 500 $\text{Ratio}_d$ values using the median.  

\subsection{Other performance metrics}


Let $\bar{\hat{\mu}}_{id}$ denote the posterior median of a model's parameter of interest for area $i$ for repeat $d$ \cite{RN3}. The following frequentist metrics were used in work by Chen, Wakefield, and Lumely \cite{RN497}\cmmnt{textcite}, and are summarized in \cref{table:freq_mse}.  

\begin{eqnarray}
    \text{Bias} & = & \frac{1}{100} \sum_{i=1}^{100} \left( \bar{\bar{\hat{\mu}}}_i - \mu_i \right) \text{, where } \bar{\bar{\hat{\mu}}}_i = \frac{1}{500} \sum_{d=1}^{500} \bar{\hat{\mu}}_{id} \nonumber
    \\
    \text{Variance} & = & \frac{1}{100} \sum_{i=1}^{100} \left( \frac{1}{500-1} \sum_{d=1}^{500} \left( \bar{\hat{\mu}}_{id} - \bar{\bar{\hat{\mu}}}_i \right)^2 \right) \nonumber
    \\
    \text{MSE} & = & \left( \text{Bias} \right) ^2 + \text{Variance} \label{met:f_mse}
\end{eqnarray}



\begin{table}
\centering
\begin{tabular}{rrrll}
\hline\hline
&  &  & Sampled areas & Nonsampled areas\\
\hline\hline
50-50 & Sc1 & D & 4.26 \textcolor[rgb]{0.502,0.502,0.502}{(10.39)} & \\
 &  & BETA & 2.03 \textcolor[rgb]{0.502,0.502,0.502}{(4.95)} & 0.92 \textcolor[rgb]{0.502,0.502,0.502}{(2.24)}\\
 &  & BIN & 3.63 \textcolor[rgb]{0.502,0.502,0.502}{(8.85)} & 3.52 \textcolor[rgb]{0.502,0.502,0.502}{(8.59)}\\
 &  & ELN & 1.84 \textcolor[rgb]{0.502,0.502,0.502}{(4.49)} & 0.93 \textcolor[rgb]{0.502,0.502,0.502}{(2.27)}\\
 &  & LOG & 0.96 \textcolor[rgb]{0.502,0.502,0.502}{(2.34)} & \textbf{0.32 \textcolor[rgb]{0.502,0.502,0.502}{(0.78)}}\\
 &  & TSLN & \textbf{0.41 \textcolor[rgb]{0.502,0.502,0.502}{(1.00)}} & 0.41 \textcolor[rgb]{0.502,0.502,0.502}{(1.00)}\\
 \hline
 & Sc2 & D & 4.25 \textcolor[rgb]{0.502,0.502,0.502}{(10.12)} & \\
 &  & BETA & 2.02 \textcolor[rgb]{0.502,0.502,0.502}{(4.81)} & 0.91 \textcolor[rgb]{0.502,0.502,0.502}{(2.12)}\\
 &  & BIN & 3.62 \textcolor[rgb]{0.502,0.502,0.502}{(8.62)} & 3.51 \textcolor[rgb]{0.502,0.502,0.502}{(8.16)}\\
 &  & ELN & 1.82 \textcolor[rgb]{0.502,0.502,0.502}{(4.33)} & 0.92 \textcolor[rgb]{0.502,0.502,0.502}{(2.14)}\\
 &  & LOG & 0.95 \textcolor[rgb]{0.502,0.502,0.502}{(2.26)} & \textbf{0.32 \textcolor[rgb]{0.502,0.502,0.502}{(0.74)}}\\
 &  & TSLN & \textbf{0.42 \textcolor[rgb]{0.502,0.502,0.502}{(1.00)}} & 0.43 \textcolor[rgb]{0.502,0.502,0.502}{(1.00)}\\
 \hline
Rare & Sc3 & D & 3.40 \textcolor[rgb]{0.502,0.502,0.502}{(9.44)} & \\
 &  & BETA & 1.06 \textcolor[rgb]{0.502,0.502,0.502}{(2.94)} & 0.85 \textcolor[rgb]{0.502,0.502,0.502}{(2.36)}\\
 &  & BIN & 1.46 \textcolor[rgb]{0.502,0.502,0.502}{(4.06)} & 1.49 \textcolor[rgb]{0.502,0.502,0.502}{(4.14)}\\
 &  & ELN & 2.61 \textcolor[rgb]{0.502,0.502,0.502}{(7.25)} & 2.03 \textcolor[rgb]{0.502,0.502,0.502}{(5.64)}\\
 &  & LOG & 1.10 \textcolor[rgb]{0.502,0.502,0.502}{(3.06)} & 0.41 \textcolor[rgb]{0.502,0.502,0.502}{(1.14)}\\
 &  & TSLN & \textbf{0.36 \textcolor[rgb]{0.502,0.502,0.502}{(1.00)}} & \textbf{0.36 \textcolor[rgb]{0.502,0.502,0.502}{(1.00)}}\\
 \hline
 & Sc4 & D & 3.39 \textcolor[rgb]{0.502,0.502,0.502}{(8.92)} & \\
 &  & BETA & 1.06 \textcolor[rgb]{0.502,0.502,0.502}{(2.79)} & 0.85 \textcolor[rgb]{0.502,0.502,0.502}{(2.18)}\\
 &  & BIN & 1.46 \textcolor[rgb]{0.502,0.502,0.502}{(3.84)} & 1.49 \textcolor[rgb]{0.502,0.502,0.502}{(3.82)}\\
 &  & ELN & 2.61 \textcolor[rgb]{0.502,0.502,0.502}{(6.87)} & 2.03 \textcolor[rgb]{0.502,0.502,0.502}{(5.21)}\\
 &  & LOG & 1.10 \textcolor[rgb]{0.502,0.502,0.502}{(2.89)} & 0.41 \textcolor[rgb]{0.502,0.502,0.502}{(1.05)}\\
 &  & TSLN & \textbf{0.38 \textcolor[rgb]{0.502,0.502,0.502}{(1.00)}} & \textbf{0.39 \textcolor[rgb]{0.502,0.502,0.502}{(1.00)}}\\
 \hline
Common & Sc5 & D & 2.67 \textcolor[rgb]{0.502,0.502,0.502}{(14.83)} & \\
 &  & BETA & 0.73 \textcolor[rgb]{0.502,0.502,0.502}{(4.06)} & \textbf{0.13 \textcolor[rgb]{0.502,0.502,0.502}{(0.72)}}\\
 &  & BIN & 3.01 \textcolor[rgb]{0.502,0.502,0.502}{(16.72)} & 2.94 \textcolor[rgb]{0.502,0.502,0.502}{(16.33)}\\
 &  & ELN & 0.45 \textcolor[rgb]{0.502,0.502,0.502}{(2.50)} & 0.20 \textcolor[rgb]{0.502,0.502,0.502}{(1.11)}\\
 &  & LOG & 0.32 \textcolor[rgb]{0.502,0.502,0.502}{(1.78)} & 0.16 \textcolor[rgb]{0.502,0.502,0.502}{(0.89)}\\
 &  & TSLN & \textbf{0.18 \textcolor[rgb]{0.502,0.502,0.502}{(1.00)}} & 0.18 \textcolor[rgb]{0.502,0.502,0.502}{(1.00)}\\
  \hline
 & Sc6 & D & 2.67 \textcolor[rgb]{0.502,0.502,0.502}{(15.71)} & \\
 &  & BETA & 0.73 \textcolor[rgb]{0.502,0.502,0.502}{(4.29)} & \textbf{0.14 \textcolor[rgb]{0.502,0.502,0.502}{(0.78)}}\\
 &  & BIN & 3.00 \textcolor[rgb]{0.502,0.502,0.502}{(17.65)} & 2.93 \textcolor[rgb]{0.502,0.502,0.502}{(16.28)}\\
 &  & ELN & 0.43 \textcolor[rgb]{0.502,0.502,0.502}{(2.53)} & 0.20 \textcolor[rgb]{0.502,0.502,0.502}{(1.11)}\\
 &  & LOG & 0.32 \textcolor[rgb]{0.502,0.502,0.502}{(1.88)} & 0.16 \textcolor[rgb]{0.502,0.502,0.502}{(0.89)}\\
 &  & TSLN & \textbf{0.17 \textcolor[rgb]{0.502,0.502,0.502}{(1.00)}} & 0.18 \textcolor[rgb]{0.502,0.502,0.502}{(1.00)}\\
 \hline\hline
\end{tabular}
\caption{Frequentist MSE $(\times 10^{-2})$ (\eqref{met:f_mse}). D denotes the Hajek \cite{RN571}\cmmnt{textcite} direct estimator given in Eq. (1)\cmmnt{eq:hajek} in the main paper. Bold numbers represent the lowest MSE value in each column and scenario for sampled and nonsampled areas. Gray numbers in brackets give the ratio of the value to that of the TSLN approach.}
\label{table:freq_mse}
\end{table}

\begin{table}
\centering
\begin{tabular}{lll|llll|ll|}
 &  &  & \multicolumn{4}{c|}{Individual Level} &  \multicolumn{2}{c|}{Area level}\\
\hline\hline
 &  &  & $\beta_1$ & $\beta_2$ & $\beta_3$ & $\sigma_e$ & $\lambda$ & $\sigma_v$\\
 \hline\hline
50-50 & Sc1 & BETA & & & & & 0.38 & 0.60\\
 &  & BIN & & & & & 0.36 & 0.15\\
 &  & ELN & & & & & 0.49 & 0.84\\
 &  & LOG &  &  & 1.23 & 0.65 & 0.41 & \\
 &  & TSLN-S1 & 1.51 & 3.05 & 1.23 & 0.70 & 0.42 & \\
 &  & TSLN-S2 & & & & & 0.39 & 0.12\\
\hline
 & Sc2 & BETA & & & & & 0.38 & 0.60\\
 &  & BIN & & & & & 0.36 & 0.15\\
 &  & ELN & & & & & 0.49 & 0.84\\
 &  & LOG &  &  & 1.23 & 0.65 & 0.41 & \\
 &  & TSLN-S1 & 0.74 & 1.56 & 1.23 & 0.67 & 0.41 & \\
 &  & TSLN-S2 & & & & & 0.40 & 0.12\\
\hline
Rare & Sc3 & BETA & & & & & 0.46 & 0.64\\
 &  & BIN & & & & & 0.48 & 0.20\\
 &  & ELN & & & & & 0.95 & 2.11\\
 &  & LOG &  &  & 1.21 & 0.94 & 0.57 & \\
 &  & TSLN-S1 & 1.37 & 3.08 & 1.25 & 0.96 & 0.59 & \\
 &  & TSLN-S2 & & & & & 0.61 & 0.17\\
\hline
 & Sc4 & BETA & & & & & 0.46 & 0.63\\
 &  & BIN & & & & & 0.49 & 0.20\\
 &  & ELN & & & & & 0.95 & 2.11\\
 &  & LOG &  &  & 1.21 & 0.94 & 0.58 & \\
 &  & TSLN-S1 & 0.75 & 1.59 & 1.24 & 0.96 & 0.58 & \\
 &  & TSLN-S2 & & & & & 0.60 & 0.17\\
\hline
Common & Sc5 & BETA & & & & & 0.55 & 0.55\\
 &  & BIN & & & & & 0.51 & 0.14\\
 &  & ELN & & & & & 0.56 & 0.36\\
 &  & LOG &  &  & 1.15 & 0.44 & 0.53 & \\
 &  & TSLN-S1 & 1.61 & 3.01 & 1.16 & 0.51 & 0.53 & \\
 &  & TSLN-S2 & & & & & 0.53 & 0.10\\
 \hline
 & Sc6 & BETA & & & & & 0.54 & 0.54\\
 &  & BIN & & & & & 0.51 & 0.14\\
 &  & ELN & & & & & 0.56 & 0.35\\
 &  & LOG &  &  & 1.14 & 0.45 & 0.53 & \\
 &  & TSLN-S1 & 0.81 & 1.56 & 1.15 & 0.46 & 0.54 & \\
 &  & TSLN-S2 & & & & & 0.53 & 0.10\\
  \hline\hline
\end{tabular}
\caption{Median of posterior medians of model parameters across all $D=500$ repetitions by model and scenario. Given the scale of the data is different depending on the model, coefficients cannot be easily compared. The first two individual level coefficients ($\beta_1, \beta_2$), relate to the indicators for groups 1 and 2 of $\mathbf{x}^{\text{survey}}$. By construction, these coefficients are larger for Sc1, Sc3 and Sc5. The coefficient $\beta_3$ relates to the individual level continuous covariate, $\mathbf{x}^{\text{census}}$.}
\label{table:model_coefs}
\end{table}

\newpage
\section{Application: further details} \label{supp:D}

\subsection{Covariates}

\paragraph{TSLN-S1} \label{supp:fe_s1}
The breadth of individual level covariates available in the NHS was enormous. We found an initial set of candidate covariates using pseudo-likelihood and \texttt{lme4} \cite{RN573}. As mentioned in Section 2.1.2\cmmnt{cref_sec:stage1_est}, we preferred models with an area linear comparison value (ALC) closer to 1. 

In addition to those covariates included in the LOG model (sex, age, and marital status), we used a variety of others, listed below. Note that we square root transformed, denoted as sqrt, some of the continuous covariates to reduce their skew. 
\begin{itemize}
    \item Individual level categorical covariates with the number of categories given in brackets. For details see \cref{table:categorial_defins}. 
    \begin{itemize}
        \item High school (6)
        \item Kessler psychological distress score (5)
        \item Qualifications (9)
        \item Self-assessed health (5)
        \item laborforce status (6)
        \item Number of daily smokers in the household (3)
        \item Tenure type of household (5)
        \item Whether indigenous members in household (3)
    \end{itemize}
    \item Area level
        \begin{itemize}
            \item Categorical
            \begin{itemize}
                \item Index of Relative Socio-Economic Disadvantage (IRSD) from the Socio-Economic Indexes for Areas (SEIFA) by the ABS (ABS 2016): 10 decile groups of increasing socio-economic disadvantage (SA2s in group 1 are classified as the most disadvantaged)
                \item State: 4 groups
            \end{itemize}
        \item Continuous
            \begin{itemize}
                \item Occupation: proportion of SA2 who are professionals
                \item Indigenous status (sqrt): proportion of SA2 who identify as Aboriginal and/or Torres Strait Islander
                \item Income: proportion of SA2 with a high weekly personal income (specifically between AUD\$1,500 and AUD\$1,750)
                \item Unemployment rate (sqrt): proportion of persons in the labor force in SA2 who are unemployed
                \item Household composition: proportion of households in SA2 with four people
            \end{itemize}
        \end{itemize}
\end{itemize}

\begin{longtable}{ll}
     & Categories \\ 
    \hline
    Age & 15-19 \\
     & 20-24 \\
     & 25-34 \\
     & 35-44 \\
     & 45-54 \\
     & 55-64 \\
     & 65-74 \\
     & 75-84 \\
     & 85+ \\ 
    \hline
    Sex & Males \\
     & Females \\ 
    \hline
    Registered marital status & Never married \\
     & Widowed \\
     & Divorced \\
     & Separated \\
     & Married \\ 
    \hline
    Qualifications & Postgraduate Degree \\
     & Graduate Diploma/Graduate Certificate \\
     & Bachelor Degree \\
     & Advanced Diploma/Diploma \\
     & Certificate III/IV \\
     & Certificate I/II \\
     & Certificate not further defined \\
     & No non-school qualification \\
     & Level not determined \\ 
    \hline
    laborforce status & Employed, working full-time \\
     & Employed, working part-time \\
     & Unemployed, looking for full-time work \\
     & Unemployed, looking for part-time work \\
     & Unemployed, looking for full-time or part-time work \\
     & Not in the labor force \\ 
    \hline
    High school & Year 12 or equivalent \\
     & Year 11 or equivalent \\
     & Year 10 or equivalent \\
     & Year 9 or equivalent \\
     & Year 8 or below \\
     & Never attended school \\ 
    \hline
    Kessler psychological distress score & Low/moderate level of psychological distress (5-11) \\
     & High/very high level of psychological distress (12-25) \\
     & No applicable \\
     & No asked \\
     & Unable to determine \\ 
    \hline
    Tenure type of household & Owner without a mortgage \\
     & Owner with a mortgage \\
     & Renter \\
     & Other \\
     & Not stated \\ 
    \hline
    Self-assessed health & Excellent \\
     & Very good \\
     & Good \\
     & Fair \\
     & Poor \\ 
    \hline
    Number of daily smokers in the household & Less than 2 \\
     & More than 1 \\
     & Not stated \\ 
    \hline
    Whether indigenous members in household & Non-indigenous only household \\
     & Indigenous only household \\
     & Mixed household \\
    \hline
    \caption{Categories for the individual level covariates used in the TSLN and LOG models for current smoking prevalence. Most of the categories for these covariates were derived by the Australian Bureau of Statistics. For details of the definitions, we refer the reader to publicly available data dictionaries. An exception is the Number of daily smokers in the household variable, which was collapsed from its original form to ensure the variable did not \emph{perfectly} predict current smokers.}
    \label{table:categorial_defins}
\end{longtable}

We also found a significant improvement in model fit by adding a hierarchical prior on a risk factor categorical covariate constructed from every unique combination of sex and age and the following binary risk factors: insufficient physical activity, insufficient fruit and vegetable consumption, overweight, and risky alcohol consumption. All the binary risk factors were defined according to current Australian health guidelines with definitions given in \cref{table:rf_defins}.

\begin{table}
    \centering
    \begin{tabularx}{6in}{lXX} 
    \hline
     & Definition & Notes \\ 
    \hline
    Overweight & \RaggedRight{By using the common cut-offs \cite{RN113, RN510, RN511}, those with a BMI greater or equal to 25 are coded as 1.} & \RaggedRight{This includes those who are obese.}  \\
    Alcohol & \RaggedRight{Those who did not meet the revised 2020 Australian National Health and Medical Research Council (NHMRC) guidelines \cite{RN516} are coded as 1.} & \RaggedRight{The guidelines stipulate that adults should drink no more than 10 standard drinks a week and no more than 4 standard drinks on any one day.} \\
    Physical activity & \RaggedRight{Those who did not meet the 2014 Australian Department of Health Physical Activity guidelines \cite{RN517} were coded as 1.} & \RaggedRight{The NHMRC guidelines, which closely mirror those given by the WHO \cite{RN519}\cmmnt{textcite}, stipulate that each week adults should either do 2.5 to 5 hours of moderate-intensity physical activity or 1.25 to 2.5 hours of vigorous-intensity physical activity or an equivalence combination of both. In addition, the guidelines recommend muscle-strengthening activities at least 2 days each week. This binary variable was based on both their leisure and workplace physical activity levels.} \\
    Diet & \RaggedRight{Those who did not meet the fruit and vegetable 2013 NHMRC Australian Dietary guidelines \cite{RN520} were coded as 1.} & \RaggedRight{The NHMRC guidelines stipulate two servings of fruit and five servings of vegetables per day.} \\
    \hline
    \end{tabularx}
    \caption{Definitions for the binary variables used in the risk factor categorical covariate. The categorical covariate has every unique combination of the four binary variables in this table and age and sex.}
    \label{table:rf_defins}
\end{table}

\paragraph{TSLN-S2}

Deriving variable selection metrics for the TSLN-S2 model requires additional thought; the input data are vectors of posterior draws, making the definition of the ``data'' difficult to determine. To align with previous work where the uncertainty inherent in fitting the first stage model is not considered, we use an approximation to the LOOCV. The \texttt{loo} package requires log-likelihood, $\mathbb{L} \lb{.}$ evaluations for all data points and posterior draws \cite{RN116}. Below we restate \eqref{eq:tsln_me} and \eqref{eq:tsln_samp2} as a reminder,   

\begin{eqnarray*}
    \hat{\boldsymbol{\theta}}^{\text{S1}, \text{its}}_{i} & \sim & N\left( \hat{\bar{\theta}}_i, \widehat{\text{v}}\left( \hat{\theta}_i^{\text{S1}} \right) \right) \quad\quad i=1,\dots, m
    \\
    \hat{\bar{\theta}}_i & \sim & N\left( \theta_i, \bar{\gamma}_i^{\text{S1}} \right) \quad\quad i=1,\dots, m.
\end{eqnarray*}

We take the expectation of $\hat{\bar{\theta}}_i$ and treat it as data. In this way, we use

\begin{equation*}
    \mathbb{L}\lb{\theta_{it}; \mathbb{E}\lb{\hat{\bar{\theta}}_i}, \bar{\gamma}_i^{\text{S1}}} \quad\quad i=1,\dots, M; t = 1,\dots, T,
\end{equation*}
to derive the LOOCV using the \texttt{loo} package. Although this approach to deriving the LOOCV ignores the uncertainty in fitting the first stage model, it aligns with previous two-stage approaches \cite{RN479, gao2023_sma}.

The 2016 Australian census collects a large amount of demographic data which can be used as SA2-level covariates in our models. For a single demographic factor, such as qualification, there are several categories, resulting in numerous proportion variables for qualification alone (e.g. bachelor's degree, postgraduate degree). Instead of relying on a set of single proportion covariates and risk missing important predictors, we used Principal Components Analysis on 84 SA2-level census covariates. We found that the first six principal components had 63\% of the variation and when used as covariates provided superior fit than models using the actual census proportions. Thus, the following eight SA2-level variables were included as fixed effects in the TSLN-S2 and ELN models; IRSD, state and principal components one to six. 

\subsection{Spatial prior} \label{supp:bym2}

The BYM2 prior proposed by Riebler \emph{et al.} \cite{RN394}\cmmnt{textcite} is a linear combination of an intrinsic CAR prior \cite{RN363} and an unstructured normal prior. It places a single variance parameter, $\sigma_\delta^2$, on the combined components with the help of a mixing parameter, $\rho \in (0,1)$, that represents the amount of spatially structured as opposed to unstructured residual variation. The BYM2 prior is 

\begin{eqnarray}
    \delta_i & = & \sigma_\delta \left(s_i \sqrt{\rho/\kappa} + v_i \sqrt{1-\rho}\right) \quad\quad i=1,\dots,M \label{eq:bym2}
    \\
    s_i & \sim & N\left( \frac{\sum_{k=1}^{M} W_{ik} s_k}{\sum_{k=1}^{M} W_{ik}}, \frac{1}{\sum_{k=1}^{M} W_{ik}} \right) \quad\quad i=1,\dots,M \nonumber
    \\
    v_i & \sim & N(0,1) \quad\quad i=1,\dots,M, \nonumber
\end{eqnarray}

where $\mathbf{W} \in \mathbb{R}^{M\times M}$ is the spatial weight matrix, which defines the neighborhood structure of the SA2s. As is common in disease mapping \cite{RN8}, we use the binary contiguous specification where $W_{ik} = 1$ if area $i$ and area $k$ are neighbors and zero otherwise. The parameter $\kappa$ is a known scaling factor, while $\rho$ is generally estimated from the data \cite{RN394}. Following the recommendations by Gomez-Rubio \emph{et al.} \cite{RN35}\cmmnt{textcite}, Mohadjer \emph{et al.} \cite{RN433}\cmmnt{textcite} and Banerjee, Carlin, and Gelfand \cite{RN390}\cmmnt{textcite}, the ICAR prior for $\mathbf{s} = (s_1,\dots, s_M)$ is declared for all areas and thus the $s_i$'s for non-sampled areas are implicitly imputed during MCMC. Following advice by Gomez-Rubio \emph{et al.} \cite{RN35}\cmmnt{textcite}, we also tried fitting an unstructured, ICAR and BYM2 prior at a higher administrative level; specifically the statistical areas level 3 (SA3s) which are constructed from aggregated SA2s. However, we found no discernible improvement in model fit.


\subsection{Benchmarking} \label{supp:bench_log}

Let $\widehat{C}_{k[i]}^D$ and $\widehat{\text{v}}\left( \widehat{C}_{k[i]}^D \right)$ be the direct Hajek \cite{RN571}\cmmnt{textcite} estimate and sampling variance for state $k = 1,\dots, 4$ (see (1) and (2) in Section 1.1). The values of $\widehat{C}_{k[i]}^D$ will be the benchmark values, with the goal that our population-weighted model-based estimate of this quantity

\begin{equation}
    \widetilde{C}_{k[i]} = \frac{ \sum_{i \in S_k} \hat{\mu}_i N_i }{\sum_{i \in S_k} N_i}  \quad\quad k=1,\dots,4 \label{eq:bench}
\end{equation}

agrees at least approximately with $\widehat{C}_{k[i]}^D$, where $\hat{\mu}_i$ is the modeled prevalence estimate for SA2 $i$. Note that $S_k$ is the subset of integers that determines which SA2s are contained within state $k$ (i.e. $\mathbb{I} \left( i \in S_k \right) = 1$ if SA2 $i$ is in state $k$ and zero otherwise). Inexact fully benchmarking takes the form,

\begin{equation}
    \widetilde{C}_{k[i]} \sim N\lb{ \widehat{C}_{k[i]}^D, \lb{\epsilon \times \sqrt{\widehat{\text{v}}\left( \widehat{C}_{k[i]}^D \right)}}^{2} } \quad\quad k = 1,\dots, 4, \label{eq:bay_bench}
\end{equation}

where $0 < \epsilon < 1$ is a discrepancy measure. Setting $\epsilon = 0$ gives exact benchmarking, whilst $\epsilon = 1$ enables the model to accommodate the benchmarks in line with their respective accuracy. We enforced stronger concordance between $\widetilde{C}_{k[i]}$ and $\widehat{C}_{k[i]}^D$ by fixing $\epsilon = 0.3$. With our SA4 level direct estimates, $\epsilon = 0.3$ gives standard deviations in \eqref{eq:bay_bench} that range from 0.002 to 0.003.   

Given that SA2-level estimates from the LOG model are derived via poststratification (e.g. a post-model calculation), we could not use Bayesian benchmarking. Instead, we used an exact ratio-adjusted estimator to benchmark the LOG model estimates \cite{RN37}. We first derived an adjustment factor, $R^B_{k[i]t}$, for the $k$th state and $t$th posterior draw,

\begin{eqnarray*}
    \widetilde{Y}_{k[i]t} & = & \sum_{i \in S_k} \hat{\mu}_{it} N_i
    \\
    \widehat{Y}_{k[i]} & = & \widehat{C}_{k[i]}^D N_k
    \\
    R^B_{k[i]t} & = & \frac{\widetilde{Y}_{k[i]t}}{\widehat{Y}_{k[i]}},
\end{eqnarray*}

where $\widetilde{Y}_{k[i]t}$ and $\widehat{Y}_{k[i]}$ are the modeled and direct estimates of the smoking counts in state $k$, respectively. Note that $N_k$ is the population in state $k$. Then the benchmarked LOG model estimates were calculated as,

\begin{equation}
    \hat{\mu}_{it}^B = \frac{\hat{\mu}_{it}}{R^B_{k[i]t}}. \label{eq:exact_bench} 
\end{equation}

which, by design, ensured that $\sum_{i=1}^M \mathbb{I} \left( i \in S_k \right) \hat{\mu}_{it}^B N_i / N_k = \widehat{C}_{k[i]}^D$ for all posterior draws. Finally, posterior summaries were applied to $\hat{\mu}_i^B$. Note that Zhang and Bryant \cite{RN30}\cmmnt{textcite} found that exact benchmarking provided larger reductions in posterior variance than that of inexact Bayesian benchmarking. 

\subsection{Comparative performance}

To compare models, we derived modeled and direct estimates at a higher administrative level: statistical areas level 4 (SA4). There are 65 SA4's across the east coast of Australia, with a median sample and population size of 148 and 220,000, respectively. 

Our decision to conduct model comparisons at the SA4 level, as opposed to the SA2 level, is informed by the relative accuracy of SA4 level estimates. Approximately 75\% of these direct estimates exhibit coefficients of variation below 25\%, which allows us to tentatively consider them as a proxy for the \emph{truth}. While acknowledging the inherent limitations of this assumption, we note that SA4-level estimates are significantly more precise and stable than their SA2-level counterparts, rendering our results illustrative yet essential for a comprehensive analysis.

The SA4 level modeled and direct estimates are calculated using population-weighted averages of the model-based SA2-level estimates and (1)\cmmnt{eq:hajek}, respectively. We compare the models using RRMSE, ARB, and coverage (see Section 4.0.1\cmmnt{cref_sec:sim_perform}). In addition, given that the SA4-level estimates also have uncertainty, we quantify the overlap of the direct and modeled intervals. Overlap probabilities give the proportion of the modeled interval that is contained within the direct estimate interval. A high probability is preferred and a value of 1 denotes that the modeled interval is entirely within the direct estimate interval. To summarize the overlap probabilities across the 65 SA4s, we take a weighted mean where the weights are the inverse direct estimate standard deviations.  

\begin{figure}[H]
    \centering
    \includegraphics[width=\textwidth]{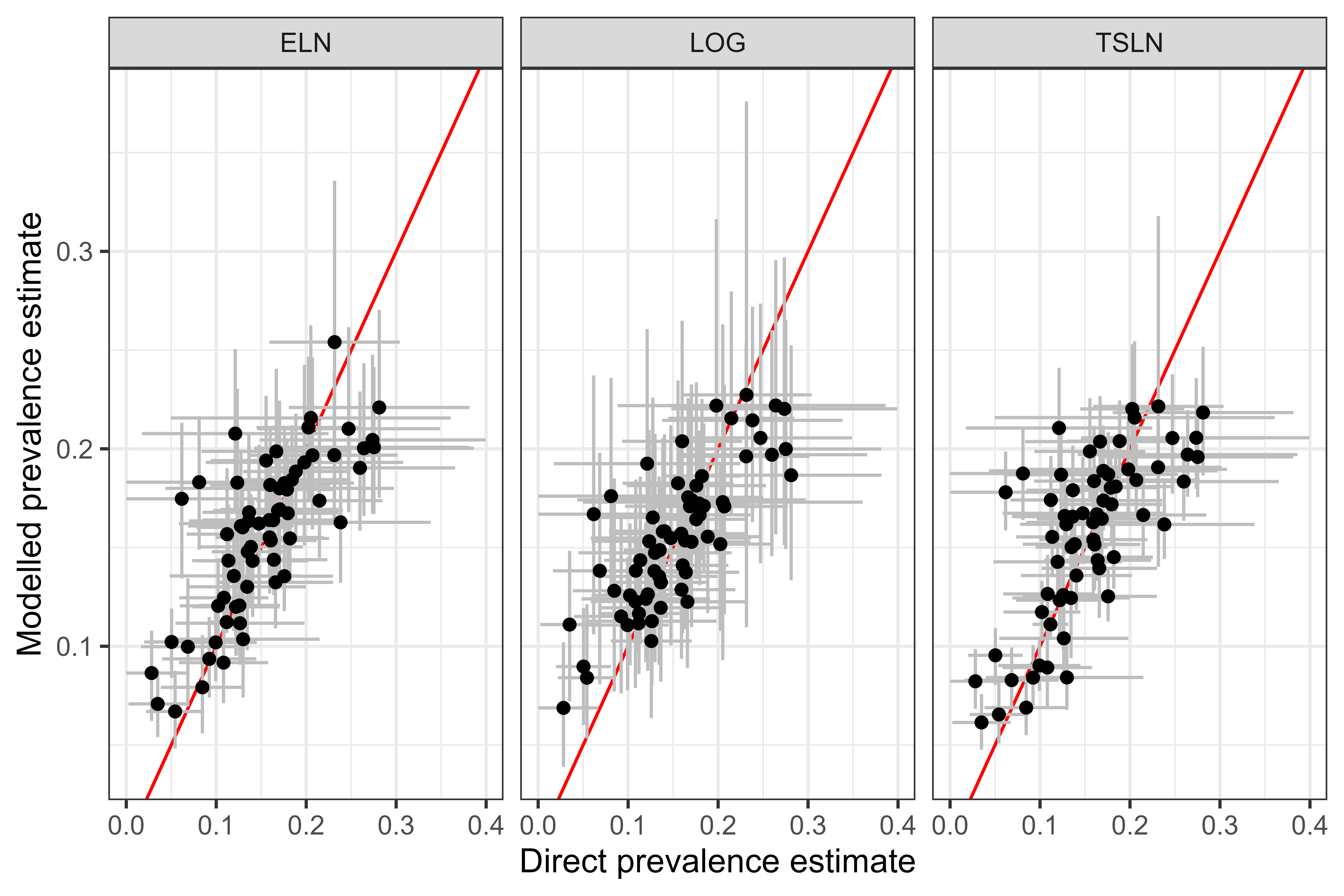}
    \caption{Comparison of the modeled and direct prevalence estimates at the SA4 level. Along with the point estimates, the plot gives the 95\% intervals represented as gray lines. The diagonal red line represents equivalence between the modeled and direct estimates.}
    \label{fig:sa4_directvsmodeled}
\end{figure}

\cref{fig:sa4_directvsmodeled} displays equivalence plots and \cref{table:mse_case_study} compares the mean RRMSE and ARB across all SA4s and summarises the credible interval sizes, coverage and overlap probabilities for the three models. 

Similar to the findings in the simulation study, \cref{table:mse_case_study} shows that the TSLN approach provides smaller MRRMSE and credible interval sizes at the SA4 level. Although the TSLN provides poor Bayesian coverage and higher MARB, our model is preferable in terms of overlap.

\newpage
\subsection{Further plots for case study}

\begin{table}[H]
    \centering
    \begin{tabular}{lllllll}
    &  & MRRMSE & MARB & CI width & Coverage & Overlap \\ 
    \hline
    \multirow{3}{*}{Benchmarked} & ELN & 0.29 & 0.24 & 0.06 & 0.68 & 0.78 \\
    & LOG & 0.35 & 0.27 & 0.09 & 0.83 & 0.73 \\
    & TSLN & 0.29 & 0.26 & 0.04 & 0.48 & 0.80 \\ 
    \hline
    \multirow{3}{*}{Not benchmarked} & ELN & 0.35 & 0.30 & 0.07 & 0.66 & 0.72 \\
    & LOG & 0.55 & 0.28 & 0.22 & 0.98 & 0.52 \\
    & TSLN & 0.29 & 0.25 & 0.05 & 0.55 & 0.75 \\
    \hline
    \end{tabular}
    \caption{Performance metrics for the three models when predicting SA4-level prevalence estimates. The table provides MRRMSE, MARB, median width of 95\% posterior credible interval (CI width), coverage and the weighted mean of overlap probabilities. Results are given separately when benchmarking is and is not used.}
    \label{table:mse_case_study}
\end{table} 

\begin{figure}[H]
    \centering
    \includegraphics[width=\textwidth]{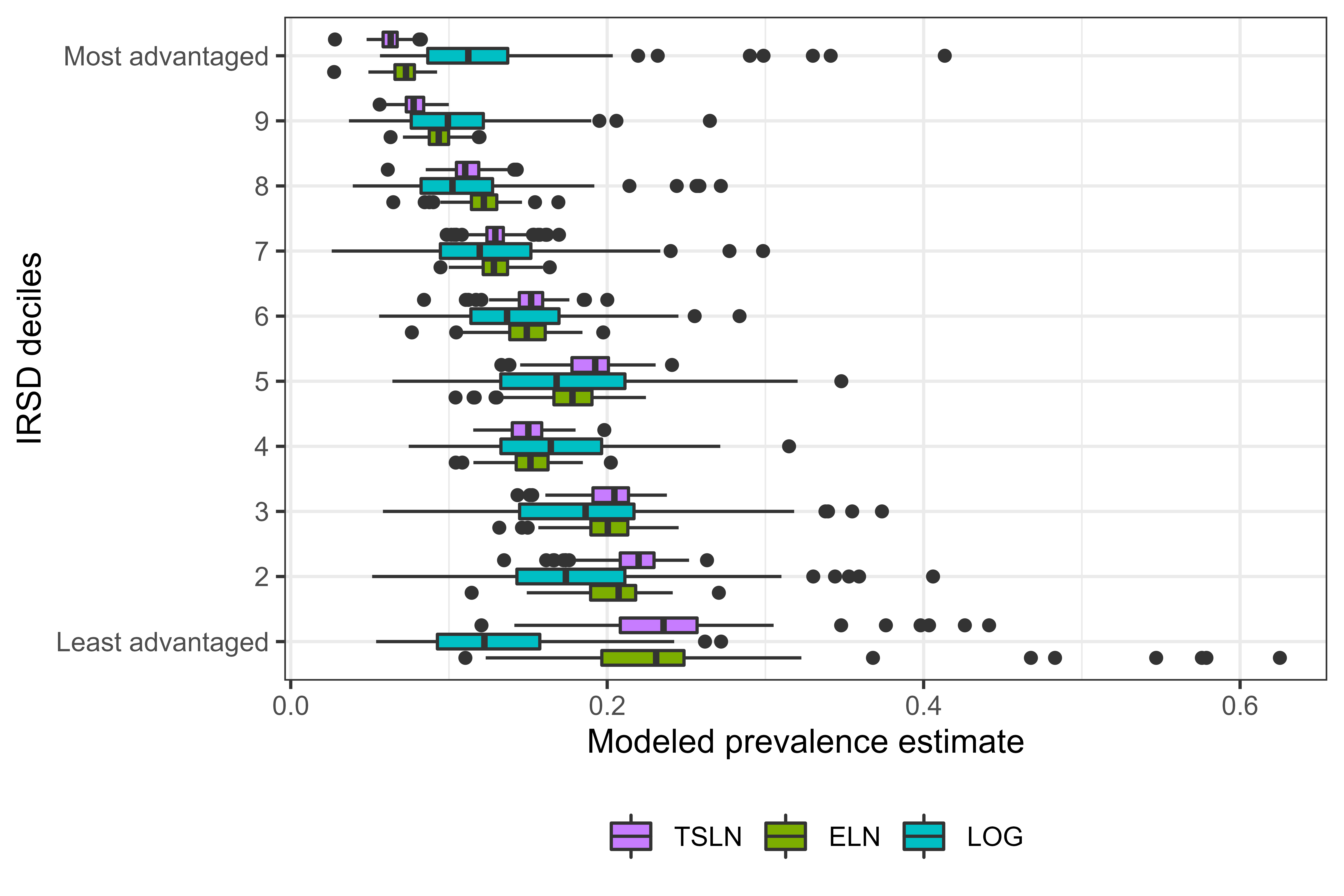}
    \caption{Comparison of the modeled SA2 level current smoking prevalence estimates by IRSD categories. Each boxplot summarises the posterior medians of the prevalence for a specific IRSD deciles and model.}
    \label{fig:boxplot_byseifa}
\end{figure}

\begin{figure}[H]
    \centering
    \includegraphics[width=\textwidth]{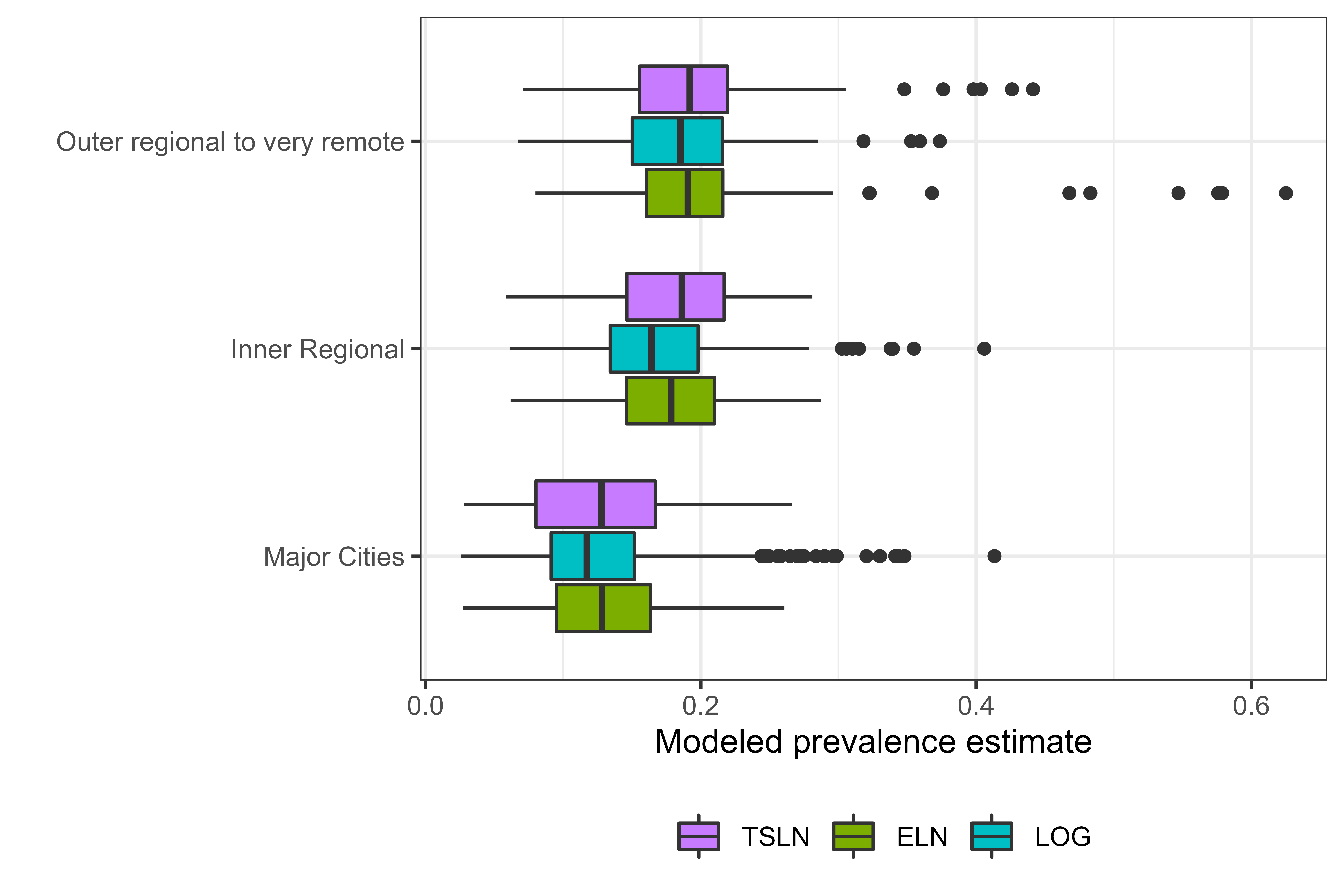}
    \caption{Comparison of the modeled SA2 level current smoking prevalence estimates by remoteness. }
    \label{fig:boxplot_byremoteness}
\end{figure}

\begin{figure}[H]
    \centering
    \includegraphics[width=\textwidth]{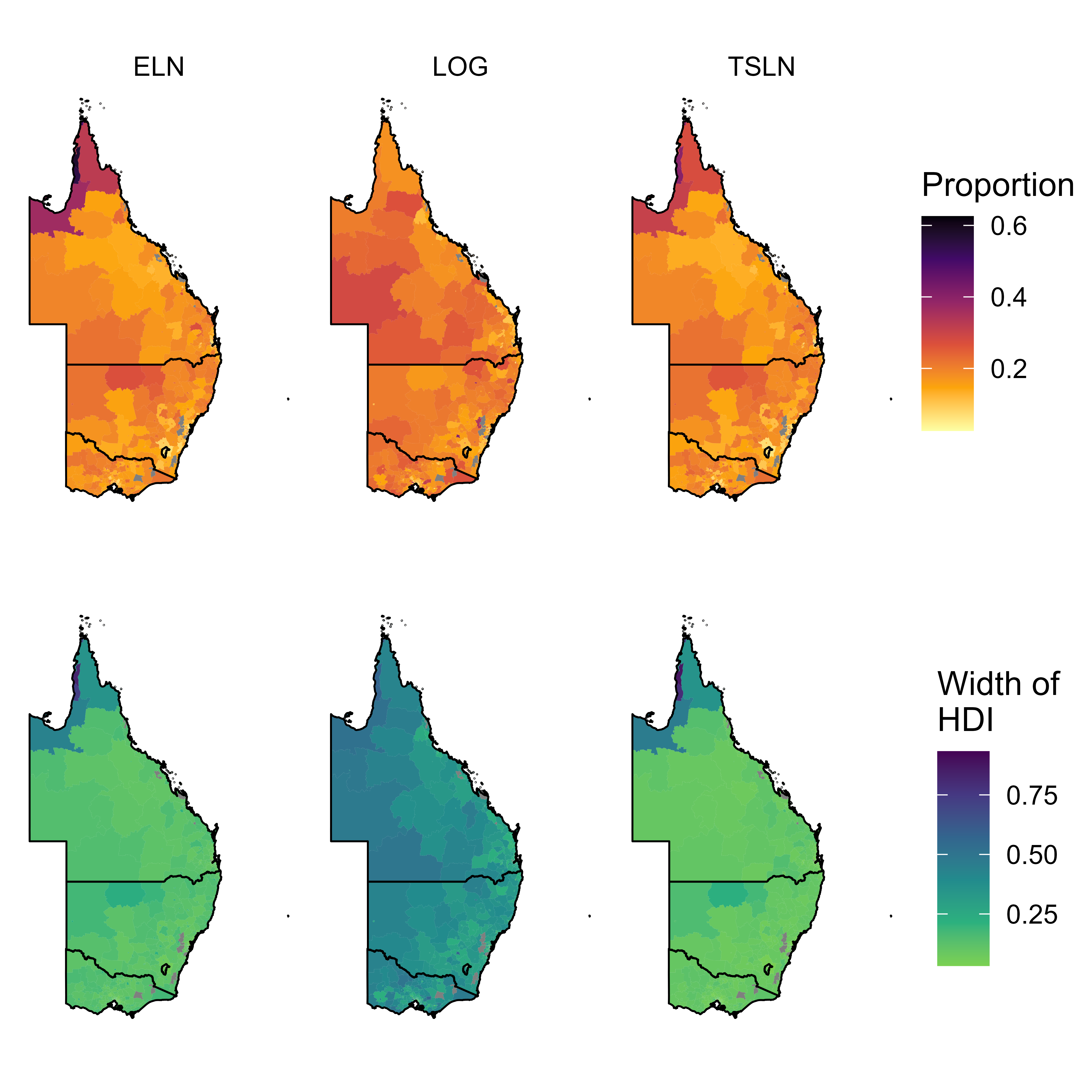}
    \caption{Choropleth maps displaying the modeled estimates of current smoking prevalence for 1630 SA2s on the east coast of Australia. For each model, we mapped the posterior medians and width of the 95\% HDI. Note that some values are lower than the range of color scales shown --- for these values, the lowest color is used.}
    \label{fig:map_mu}
\end{figure}

\begin{figure}[H]
    \centering
    \includegraphics[width=\textwidth]{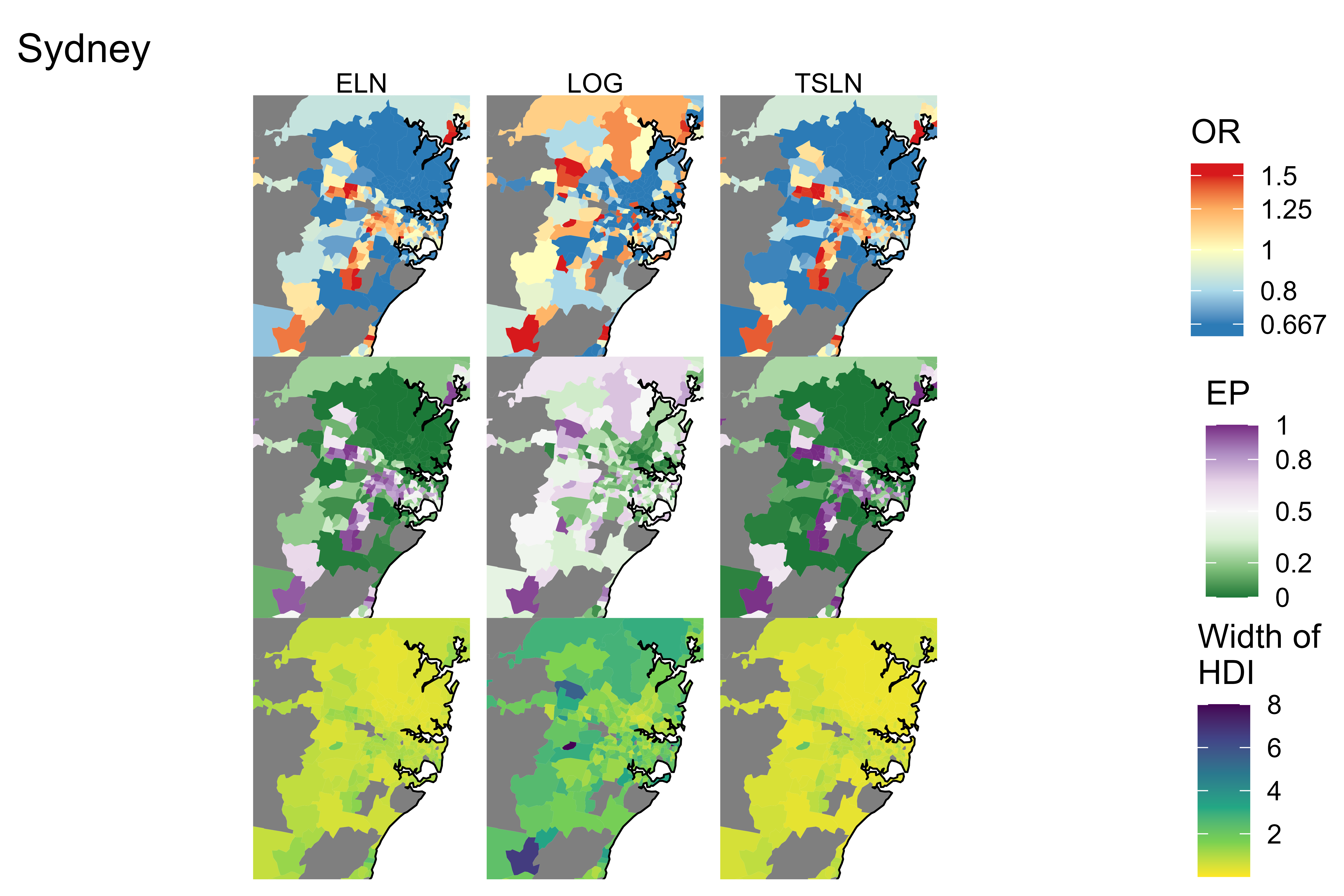}
    \caption{Choropleth maps displaying the modeled ORs for smoking prevalence in 282 SA2s in and around Sydney, Australia. For each model, we mapped the posterior medians, exceedance probabilities ($EP$s) and the width of the 95\% HDI. Note that some values are lower than the range of color scales shown --- for these values, the lowest color is used. Gray areas were excluded from estimation due to having 2016-2017 ERPs smaller or equal to 10.}
    \label{fig:map_or_Sydney}
\end{figure}

\newpage
\section{Simulation experiment for double smoothing} \label{supp:simulation}

The goal of this simulation study was to explore the relationship between the performance of the TSLN approach and the level of smoothing applied by the stage 1 model (Section 2.1.1 in the main paper). The process to simulate data was the same as that described in \cref{supp:sim_alg}. The only difference is that the vector of area level proportions ranged from $0.05$ to $0.3$ (e.g. $\mathbf{U} = \left( U_1 = 0.05,\dots, U_{100} = 0.3 \right)$).

A single synthetic census was used with 168,258 individuals and 100 areas. We repeatedly sampled from the synthetic census to obtain $D = 100$ unique samples (repetitions). Although $N_i$ and $n_i$ were fixed, the areas to be sampled and which individuals were sampled in each selected area were stochastic, resulting in different $n$ for each repetition $d$. The median sample size, $n$, and area sample size, $n_i$ were 755, and 7 respectively. Across all repetitions, the median (IQR) proportion of sampled areas that gave stable direct estimates was 0.62 (0.58, 0.65).

\subsection{Models}

To control the level of smoothing induced by the stage 1 model, we varied the fixed residual error, denoted by $\sigma_e$. Thus, the stage 1 model was

\begin{eqnarray}
    y_{ij} & \sim & \jdist{Bernoulli}{\pi_{ij}}^{\tilde{w}_{ij}} \label{eq:sim_model_s1}
    \\
    \jdist{logit}{\pi_{ij}} & = & \alpha + \mathbb{I}\lb{\text{RA}} v_i + \epsilon_{ij} \nonumber
    \\
    v_i & \sim & \jdist{N}{0, \sigma_v^2} \nonumber
    \\
    \epsilon_{ij} & \sim & \jdist{N}{0,\sigma_e^2},  \nonumber
\end{eqnarray}

\noindent where the indicator function, $\mathbb{I}\lb{\text{RA}}$, was used to include and omit the area-level random effect. For each of 

\begin{align*}
    \sigma_e &= \{0.01, 0.1, 0.25, 0.5, 0.75,
    \\
    & 1, 1.25, 1.5, 1.75, 2, 2.5, 3, 3.5\},
\end{align*}

\noindent we fit a stage 1 model with and without the area-level random effect. Thus, for each repetition, we fit 26 models. 

The stage 2 model was constant throughout this simulation study. It used $\mathbf{Z}$ as the only area level covariate.  



\subsection{Performance metrics}

To summarize the simulation results we compared the modeled prevalence estimates to the true values using mean absolute relative bias (MARB), mean relative root mean square error (MRRMSE), coverage and the width of the 95\% highest posterior density intervals (HPDIs) (Section 4.1 of the main paper). 

In addition to the area linear comparison (ALC) metric (Section 2.1.3 of the main paper) for the purposes of this simulation study we proposed an alternative metric, defined as the smoothing ratio (SR). Unlike the ALC metric, which operates at the small area level, the SR compares the observed individual-level data to the predicted probabilities. We used the posterior median of the SR, which for posterior draw $t$ was given by, 

\begin{equation}
    \text{SR}^{(t)} = 1- \frac{
    \sum_{i=1}^m \left| \frac{\sum_{j = 1}^{n_i} w_{ij} \left( y_{ij} - \pi^{(t)}_{ij} \right) }{n_i} \right|}{
    \sum_{i=1}^m \left| \frac{\sum_{j = 1}^{n_i} w_{ij} \left( y_{ij} - \hat{\mu}^D \right) }{n_i} \right|}, \label{eq:sr}
\end{equation}

\noindent where $\hat{\mu}^D$ was the overall prevalence. Higher values of both the SR and ALC were preferred. 

For each of the 100 repetitions, 13 $\sigma_e$ values, and 2 area-level random effect options (in or out) we derived the performance and smoothing metrics. Thus, the results presented here are based on 2600 pairs of performance and smoothing metrics. To assess how the performance metrics were affected by the SR and the ALC, we fit quadratic quantile regressions with the performance metric as the dependent variable and one of the SR or the ALC as the fixed and quadratic terms. In the quantile regressions, we used the 5th, 20th, 50th (median), 80th, and 95th percentiles. 

\subsection{Results}

\begin{figure}[H]
    \centering
    \includegraphics[width=\columnwidth]{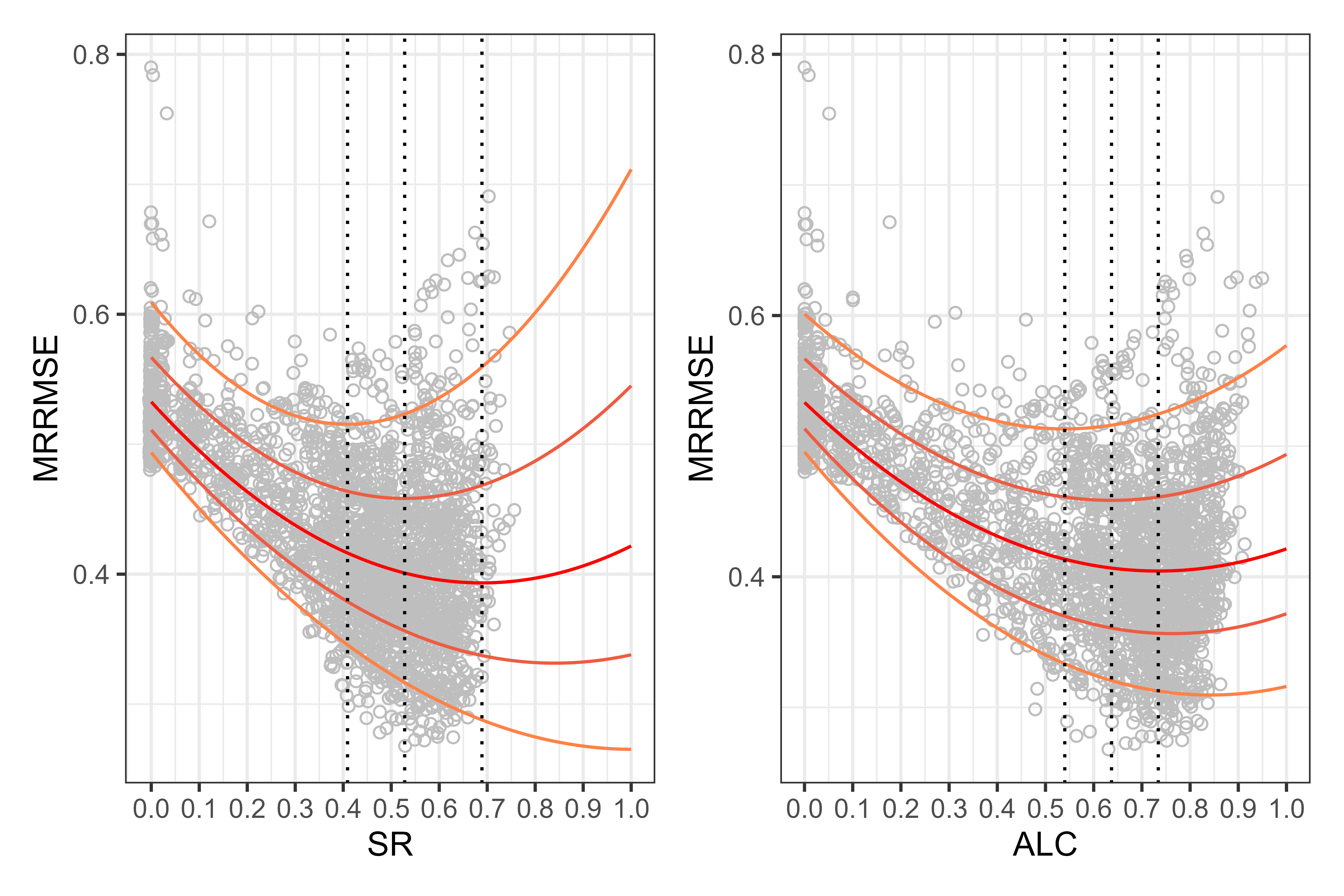}
    \caption{Scatter plot visualising the relationship between the smoothing ratio (SR) (left) and area linear comparison (ALC) (right) with the mean relative root mean squared error (MRRMSE). Grey points denote the observed MRRSME and SR or ALC pairs. The five colored lines give the fitted values from univariate quadratic quantile regression using the 5th, 20th, 50th (median), 80th, and 95th percentiles. Red denotes the median quantile fitted line. The three vertical dotted lines give the global minimums of the 50th, 80th and 95th percentile fitted lines.}
    \label{fig:MRRMSE_quantreg}
\end{figure}

\begin{figure}[H]
    \centering
    \includegraphics[width=\columnwidth]{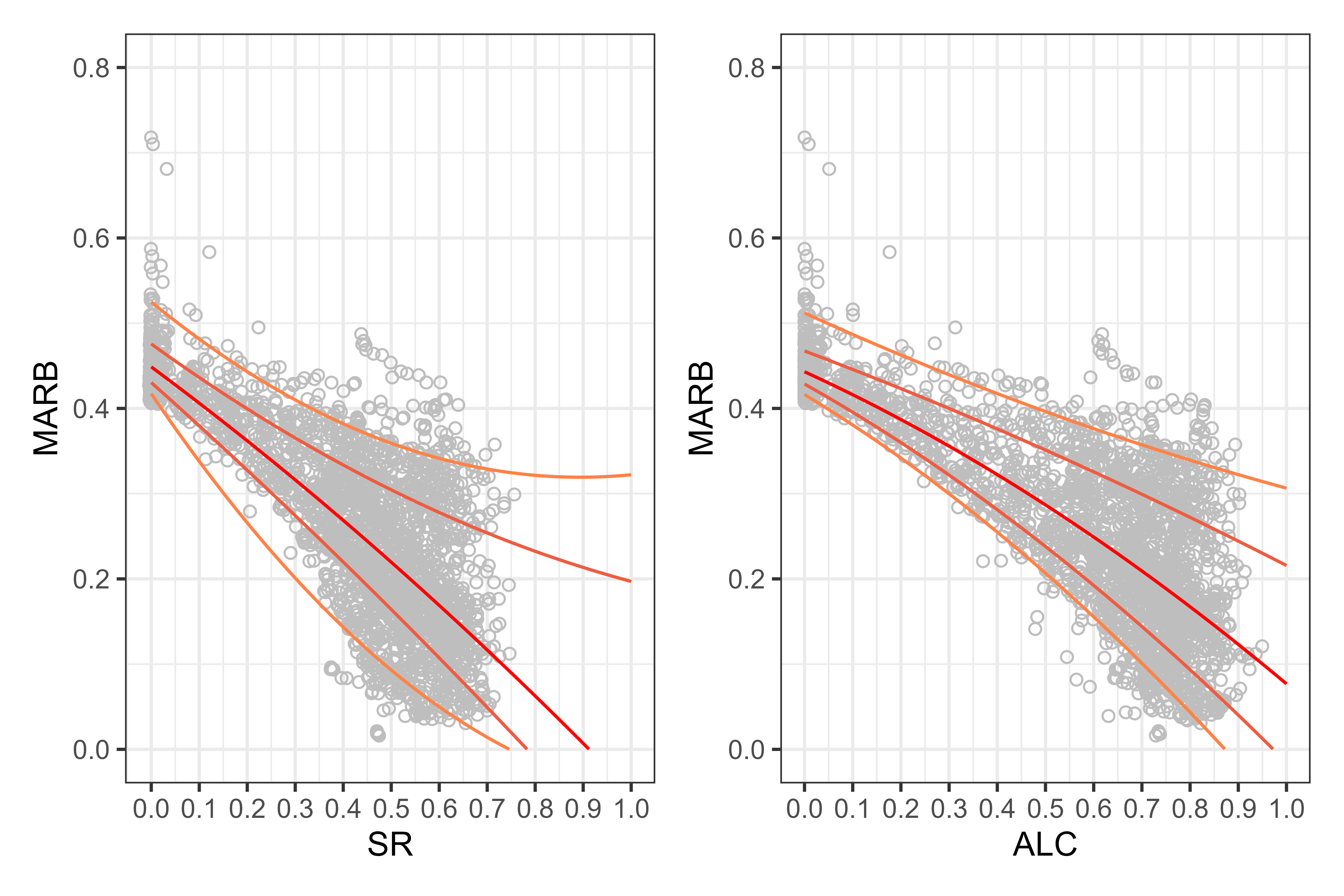}
    \caption{Scatter plot visualising the relationship between the smoothing ratio (SR) (left) and area linear comparison (ALC) (right) with the mean absolute relative bias (MARB). Grey points denote the observed MARB and SR or ALC pairs. The five colored lines give the fitted values from univariate quadratic quantile regression using the 5th, 20th, 50th (median), 80th, and 95th percentiles. Red denotes the median quantile fitted line. }
    \label{fig:MARB_quantreg}
\end{figure}

\begin{figure}[H]
    \centering
    \includegraphics[width=\columnwidth]{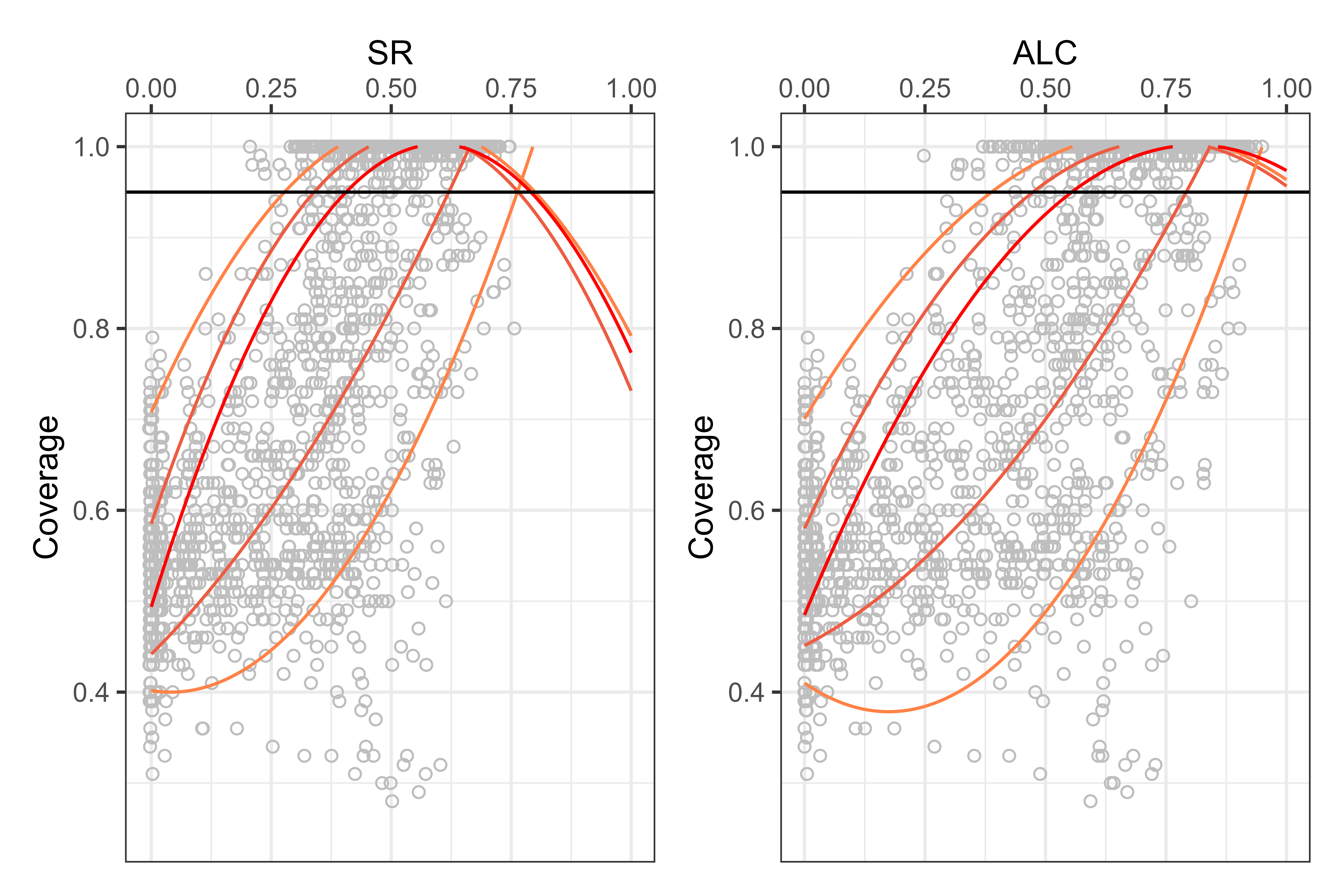}
    \caption{Scatter plot visualising the relationship between the smoothing ratio (SR) (left) and area linear comparison (ALC) (right) with coverage at the 95\% level. Grey points denote the coverage and SR or ALC pairs. The five colored lines give the fitted values from univariate quadratic quantile regression using the 5th, 20th, 50th (median), 80th, and 95th percentiles. Red denotes the median quantile fitted line. The horizontal black line gives the nomimal 95\% level.}
    \label{fig:Coverage_quantreg}
\end{figure}

\begin{figure}[H]
    \centering
    \includegraphics[width=\columnwidth]{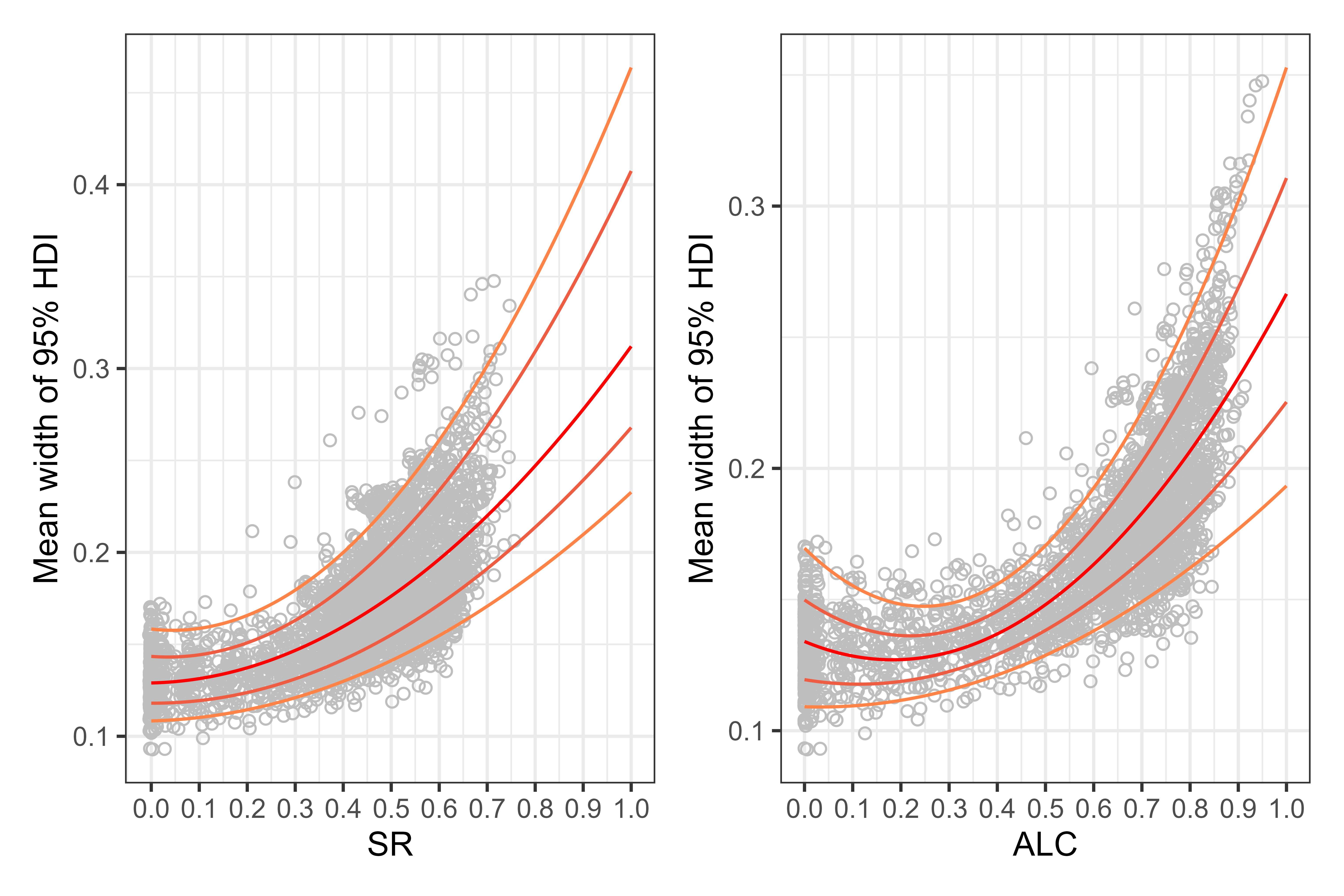}
    \caption{Scatter plot visualising the relationship between the smoothing ratio (SR) (left) and area linear comparison (ALC) (right) with the mean width of the 95\% highest posterior density intervals (HPDIs). Grey points denote the mean HPDI width and SR or ALC pairs. The five colored lines give the fitted values from univariate quadratic quantile regression. Red denotes the median quantile fitted line.}
    \label{fig:HDIsize_quantreg}
\end{figure}

\begin{figure}[H]
    \centering
    \includegraphics[width=\columnwidth]{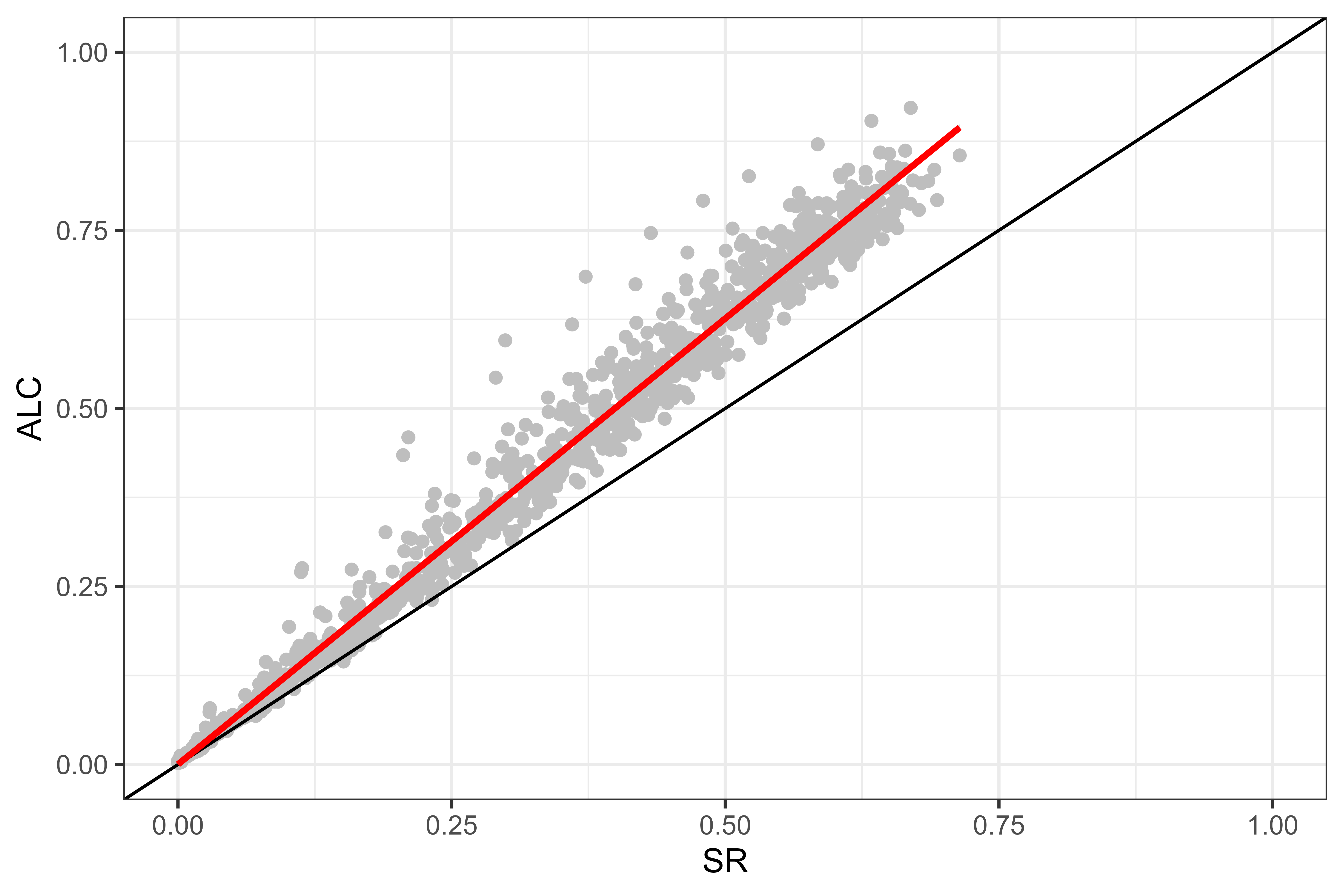}
    \caption{Scatter plot visualising the relationship between the smoothing ratio (SR) and area linear comparison (ALC). The black line gives equivalence, while the red line is a fitted linear regression line. The $R^2$ of the regression is 98\% with a 0.1 increase in SR resulting in an increase of 0.125 in the ALC. This explains why the regression line continues to drift further from the line of equivalence as the ALC and the SR increase.}
    \label{fig:ALCvsSR}
\end{figure}

\cref{fig:MRRMSE_quantreg,fig:MARB_quantreg,fig:Coverage_quantreg,fig:HDIsize_quantreg} show the simulation results, along with the fitted quantile regressions for MRRMSE, MARB, coverage and width of the HPDIs, respectively. Although the SR and the ALC appear to captured similar characteristics of smoothing, \cref{fig:ALCvsSR} shows that the ALC increases to 1 faster than the SR. Hence, the ALC was preferred as a smoothing metric since it had slightly better sensitivity. 

As expected, the TSLN approach performed better when the SR and ALC were larger. According to quantile regression on the median, by increasing the SR from 0 to 0.5, MARB and MRRMSE reduced by a factor of 2.04 and 1.32, respectively. Unlike MARB which showed a consistent downward trend (see \cref{fig:MARB_quantreg}), the quantile regression lines in \cref{fig:MRRMSE_quantreg} suggest a global minimum for MRRMSE. The minimum appears to occur between 0.4 and 0.7 for the SR and between 0.55 to 0.75 for the ALC, suggesting that increasing the metrics beyond these bounds may actually be ill-advised. Greater priority should be given to ensuring both metrics are close to the midpoint of 0.55. To further emphasize this point, \cref{table:fitted_quant_comps} illustrates how the fitted MRRMSE stagnates for ALC between 0.4 and 0.6. 

\begin{table}[H]
    \centering
    \begin{tabular}{l|lllll}
    \hline\hline
    Percentile & ALC = 0 & ALC = 0.4 & ALC = 0.5 & ALC = 0.6 & ALC = 0.7\\
    \hline
    50th & 0.53 (1.00) & 0.43 (1.24) & 0.42 (1.28) & 0.41 (1.31) & 0.40 (1.32)\\
    80th & 0.57 (1.00) & 0.47 (1.20) & 0.46 (1.22) & 0.46 (1.24) & 0.46 (1.23)\\
    95th & 0.60 (1.00) & 0.52 (1.16) & 0.51 (1.17) & 0.51 (1.17) & 0.52 (1.15)\\
    \hline\hline
    \end{tabular}
    \caption{Fitted MRRMSE according to the 50th, 80th and 95th percentile from quantile regression. Each cell gives the fitted MRRMSE for the specific quantile and ALC value. The numbers in bracket indicate the relative reduction for each row with ALC$= 0$ as the baseline.}
    \label{table:fitted_quant_comps}
\end{table}

The recommendation above is supported by \cref{fig:Coverage_quantreg} which shows that the nominal 95\% level (given by the horizontal black line) is achieved when the SR$= 0.41$ or the ALC$= 0.55$. Furthermore, \cref{fig:HDIsize_quantreg} shows that the HPDIs stay relatively constant until both the SR and ALC reach 0.5, at which point the width of the HPDIs appear to climb more steeply.  

\bibliographystyle{unsrtnat}
\bibliography{ref}  